\documentclass[useamsfonts]{pasj01}
\usepackage{amsfonts}
\usepackage{graphicx}

\Received{$\langle$reception date$\rangle$}
\Accepted{$\langle$acception date$\rangle$}
\Published{$\langle$publication date$\rangle$}

\begin{document}

\title{Photometric classification of HSC transients using machine learning}
\author{
Ichiro \textsc{Takahashi}\altaffilmark{1,2,4,*},
Nao \textsc{Suzuki}\altaffilmark{1},
Naoki \textsc{Yasuda}\altaffilmark{1},
Akisato \textsc{Kimura}\altaffilmark{3},
Naonori \textsc{Ueda}\altaffilmark{3},
Masaomi \textsc{Tanaka}\altaffilmark{4,1},
Nozomu \textsc{Tominaga}\altaffilmark{5,1},
and
Naoki \textsc{Yoshida}\altaffilmark{1,2,6,7}
}%
\altaffiltext{1}{Kavli Institute for the Physics and Mathematics of the Universe (WPI), The University of Tokyo Institutes for Advanced Study, The University of Tokyo, 5-1-5 Kashiwanoha, Kashiwa, Chiba 277-8583, Japan}
\altaffiltext{2}{CREST, JST, 4-1-8 Honcho, Kawaguchi, Saitama 332-0012, Japan}
\altaffiltext{3}{NTT Communication Science Laboratories, 2-4 Hikaridai, Seika-cho, Keihanna Science City, Kyoto 619-0237, Japan}
\altaffiltext{4}{Astronomical Institute, Tohoku University, Aoba, Sendai, Miyagi 980-8578, Japan}
\altaffiltext{5}{Department of Physics, Faculty of Science and Engineering, Konan University, 8-9-1 Okamoto, Kobe, Hyogo 658-8501, Japan}
\altaffiltext{6}{Department of Physics, Graduate School of Science, The University of Tokyo, 7-3-1 Hongo, Bunkyo-ku, Tokyo 113-0033, Japan}
\altaffiltext{7}{Institute for Physics of Intelligence, The University of Tokyo, 7-3-1 Hongo, Bunkyo-ku, Tokyo 113-0033, Japan}

\email{ichiro.takahashi@astr.tohoku.ac.jp}


\maketitle
\begin{abstract}
The advancement of technology has resulted in a rapid increase in supernova (SN) discoveries. The Subaru/Hyper Suprime-Cam (HSC) transient survey, conducted from fall 2016 through spring 2017, yielded 1824 SN candidates. This gave rise to the need for fast type classification for spectroscopic follow-up and prompted us to develop a machine learning algorithm using a deep neural network (DNN) with highway layers. 
This machine is trained by actual observed cadence and filter combinations such that we can directly input the observed data array into the machine without any interpretation. We tested our model with a dataset from the LSST classification challenge (Deep Drilling Field).
Our classifier scores an area under the curve (AUC) of 0.996 for binary classification (SN~Ia or non-SN~Ia) and 95.3\% accuracy for three-class classification (SN~Ia, SN~Ibc, or SN~II). Application of our binary classification to HSC transient data yields an AUC score of 0.925. With two weeks of HSC data since the first detection, this classifier achieves 78.1\% accuracy for binary classification, and the accuracy increases to 84.2\% with the full dataset. This paper discusses the potential use of machine learning for SN type classification purposes.
\end{abstract}

\section{Introduction}
Time domain science has become a major field of astronomical study. The discovery of the accelerating universe \citep{perlmutter99a,riess98a} evoked a series of large supernova (SN) surveys in the last few decades 
\citep{betoule14a,scolnic18a,brout19a}.    
These surveys revealed that an entirely new family of transients exists and the search for unknown populations is currently of great interest \citep{howell06a,phillips07a,quimby07b}.
For precision cosmology, it is highly important to maintain the purity of Type Ia supernovae (SNe~Ia) while performing uniform sampling of these SNe ranging from bright to faint. 
Spectroscopic follow-up has been essential to distinguish a faint SN~Ia from a Type Ib supernova (SN~Ib) and a Type Ic supernova (SN~Ic) which have similar light curve behavior \citep{scolnic14a}. 
Considering that a large amount of precious telescope time is dedicated to these follow-up programs, it is desirable to make efficient use of these telescopes.

The scale of surveys is becoming larger, and it has become impossible to trigger spectroscopic follow-up for all of the candidates in real time. It has therefore become necessary to develop a new classification scheme, and a natural path along which to proceed would be to perform photometric classification \citep{sako11a,jonesl8a}. The rise of machine learning technologies has resulted in astronomical big data commonly being analyzed by using machine learning techniques.

Neural networks have been used for photometric redshift studies from the early stages. 
Today, many Machine Learning methods are applied to the photometric redshift studies \citep{collister04a,carliles10a,pasquet19a}.
Deep neural networks (DNNs) are being used to process imaging data to identify strong lens systems \citep{petrillo17a} or for galaxy morphology classifications \citep{hausen19a}. 
Currently, machine learning is being introduced to process transient survey data for detection \citep{goldstein15a} and classification purposes \citep{charnock17a}.
A recurrent autoencoder neural network (RAENN) is introduced for photometric classification of the SN light curves \citep{villar20a}, and being applied to 2315 Pan-Starrs1 data \citep{hosseinzadeh20a}.  
In preparation for the Vera C. Rubin Observatory, a real time classifier, ALeRCE (Automatic Learning for the Rapid Classification of Events), is being developed \citep{sanchez-saez20a,forster20a} and currently applied to the Zwicky Transient Facility (ZTF) data \citep{carrasco-davis20a}.

In this paper, we introduce our attempt to apply machine learning (DNN) to the actual transient survey data.
The Hyper Suprime-Cam (HSC) \citep{miyazaki18a,Komiyama2018,kawanomoto18a,Furusawa2018}, a gigantic mosaic CCD still camera mounted on the 8.2 m Subaru Telescope, makes it possible to probe a wide field (1.77 deg$^2$ field of view) and deep space (26th mag in the $i-$band / epoch).  Our primary scientific goals are SN~Ia cosmology, Type II supernova (SN~II) cosmology, and super-luminous supernova (SLSN) studies as well as to probe unknown populations of transients.  
As reported recently \citet{yasuda19a}, more than 1800 SNe were discovered during a 6-month campaign. 
We deployed machine learning (AUC boosting) for transient detection \citep{morii16a}, where the machine determines whether a transient is ``real'' or ``bogus.''
In \citet{Kimura17}, we adopted a DNN for SN type classification from a two-dimensional image, and highway block was introduced for the optimization of layers.
This research is an extension of our previous work \citet{Kimura17} and applies a DNN to the photometric classification of transients.   
We uniquely attempt to use the observed data in a state that is as ``raw'' as possible to enable us to directly use the data as input for the machine without fitting the data or extracting characteristics.

The structure of this paper is as follows. We introduce our methods in section 2, and the data in section 3. The design of our DNN model and its application to pseudo-real LSST simulation data is described in section 4. Section 5 presents the application of our model to the actual Subaru/HSC data. We discuss the results in section 6 and conclude the paper in section 7.

\section{Methods}
\subsection{Tasks in HSC survey}
\label{sec:tasks}
The Subaru HSC Transient Survey forms part of the Subaru Strategic Project (SSP), which is a five-year program with a total of 300 dark nights \citep{aihara18a,miyazaki18a}.
The HSC-SSP Transient Survey is composed of two seasons, the first of which was executed for six consecutive months from November 2016 through April 2017.
The HSC is mounted on the prime focus for two weeks in dark time.   
Weather permitted, we aimed to observe two data points per filter per lunation.
Details of the survey strategies and the observing logs were reported in \citet{yasuda19a}, and we used the observed photometric data described in \citet{yasuda19a}.

One of our primary goals with the HSC-SSP Transient Survey is SN~Ia cosmology which aims to perform the most precise measurement of dark energy at high-redshift and test whether the dark energy varies with time \citep{linder03b}.
We have been awarded 96 orbits of Hubble Space Telescope time (WFC3 Camera) to execute precise IR photometry at the time of maximum.
Our HST program uses non-disrupted ToO, which means we are required to send the request for observation, two weeks prior to the observation.
In other words, we need to identify good candidates 2--3 weeks prior to the maximum. Although a high-redshift ($z>1$) time dilation factor helps, it is always a challenge to identify SN~Ia on the rise.

Our international collaboration team executes spectroscopic follow-up using the large telescopes in the world. 
Our target, high-redshift SN~Ia, is faint ($\sim$ 24th mag in the $i-$band) for spectroscopic identification even with the most powerful large telescopes: GMOS Gemini \citep{hook04a}, GTC OSIRIS,\footnote{GTC OSIRIS $\langle$http://www.gtc.iac.es/instruments/osiris/$\rangle$.} Keck LRIS \citep{oke95a}, VLT FORS \citep{appenzeller98a}, Subaru FOCAS \citep{kashikawa02a} and AAT AAOmega Spectrograph \citep{Saunders2004}.  
Thus, it is critical to hit the target at the time of maximum brightness either by using ToO (GTC), queue mode (VLT) or classical scheduled observation (Keck, Subaru, ATT).

A SN~Ia requires approximately 18 days from its explosion to reach its maximum in rest frame \citep{conley06a,papadogiannakis19a}. 
Given this fact, the SNe with which we are concerned is high-redshift SN~Ia ($z>1$), in which case we have approximately one month in the observed frame for a SN~Ia from the time of explosion until it reaches its maximum.
However, our task is to identify these SNe two weeks prior to the maximum, which means we have only two data points per filter. In addition to that, the sky conditions continue to change, and we may not have the data as originally planned. In reality, our identification has to proceed despite data points being missing on the rise.

In parallel to the SN~Ia cosmology program, our sister projects also need identification and classification of HSC transients.   
Specifically, the SN~II cosmology program requires timely spectroscopic follow-up to measure the expansion velocity of photosphere from the H$\beta$ line \citep{dejaeger17a}.
A SLSN is of great interest today, because it is a relatively rare event \citep{quimby11a} and its mechanism has quite a diversity \citep{galyam12a,Moriya18SLSN}, and it can be used to probe the high-redshift Universe \citep{cooke12a}. 
New types of rapid transients are also discovered by HSC, but their identities are yet to be known \citep{Tampo2020}. 
For these projects, timely spectroscopic follow-up is also critical \citep{moriya19a,curtin19a}.

An early phase SN~Ia provides us with clues on the explosion mechanism \citep{maeda18a} and progenitors \citep{cao15a}.
The advantage of HSC is its ability to survey a large volume, and in practice, it has confirmed the long-standing theoretical prediction of helium-shell detonation \citep{Jiang2017}.
Finding early phase SN~Ia is not trivial but HSC is yielding a new set of early phase SN~Ia \citep{Jiang_2020}.
Observations of early phase core-collapse SNe provide us with crucial information on the size of the progenitors \citep{thompson03a,tominaga11a} and Circumstellar Medium \citep{forster18a}.

%
%
%
%
%

\subsection{Classification method for HSC-SSP transient survey}
We designed two machine learning models with the emphasis on identifying SN~Ia, which requires a time-sensitive trigger for HST IR follow-up. 
The first model operates in binary mode and classifies whether a transient is of the SN~Ia. 
In this regard, the majority of high-redshift transients are known to be of the SN~Ia type, and our work entails searching for other unknown transients from among those labeled non-SN~Ia. 
The second model classifies a transient into one of three classes: SN~Ia, SN~Ibc, or SN~II.  
These three classes were chosen for simplicity and in fact, the majority of SNe belong to one of these three categories. 
SN~Ia is a thermonuclear explosion, and its brightness can be calibrated empirically.
SN~Ib, SN~Ic, and SN~II are all core-collapse SNe and are classified by their spectral features \citep{filippenko97a}.
The light curves of SN~Ib and SN~Ic, which are collectively referred to as SN~Ibc, are similar to those of SN~Ia and always contaminate the SN~Ia cosmology program. They are fainter than SN~Ia and redder in general. 
A major challenge of this work is to determine whether we can distinguish SN~Ibc from SN~Ia.

\section{Data}
In this section, we present the dataset we used for our study.
We first introduce our SN dataset from the HSC-SSP Transient Survey (subsection \ref{sec:hscdata}).
Then we describe the simulated photometric data to train the machine (subsection \ref{sec:training}).
Lastly, we explain the pre-processing of the above data for input into the machine (subsection \ref{sec:preproc}).

\subsection{Observed data from Subaru/HSC-SSP transient survey}
\label{sec:hscdata}
The goal of this project is to classify the light curves observed by Subaru/HSC.  
The discovery of 1824 SNe recorded during the six-month HSC-SSP Transient Survey during the period of November 2016 through April 2017 was reported and described in \citet{yasuda19a}.
The survey is composed of the Ultra-Deep and Deep layers in the COSMOS \citep{scoville07a}.
The median 5$\sigma$ limiting magnitudes per epoch are 26.4, 26.3, 26.0, 25.6, and 24.6 mag (AB) for the $g$-, $r2$-, $i2$-, $z$- and $y$-bands, respectively, for the Ultra-Deep layer. 
For the Deep layer, the depth is 0.6 mag shallower.

The SN dataset consists of time series photometric data (flux, magnitude, and their errors) in each band for each SN.
Because part of the $y$-band photometric data contains residuals due to improper background subtraction influenced by scattered light \citep{aihara18dr}, we excluded the $y$-band data from our study, considering the impact thereof on the classification performance.

The redshift information for HSC SNe is a combination of our follow-up observation results and catalogs from multiple surveys of those host galaxies.
The spectral redshifts (spec-z) are adopted from the results of the follow-up spectrum observations by AAT/AAOmega performed in 2018 and those from the DEIMOS \citep{DEIMOS2018}, FMOS-COSMOS \citep{FMOS-COSMOS2015}, C3R2 \citep{C3R2_2017}, PRIMUS \citep{PRIMUS2011} and COSMOS catalogs.
For those without spec-z, the photometric redshifts (photo-z) were adopted from the COSMOS 2015 catalog \citep{laigle16a} and those calculated from the HSC-SSP survey data \citep{HSCSSP_photo-z2018}.

\subsection{Simulated data for training}
\label{sec:training}
\begin{figure*}[htbp]
  \begin{center}
     \includegraphics[width=130mm]{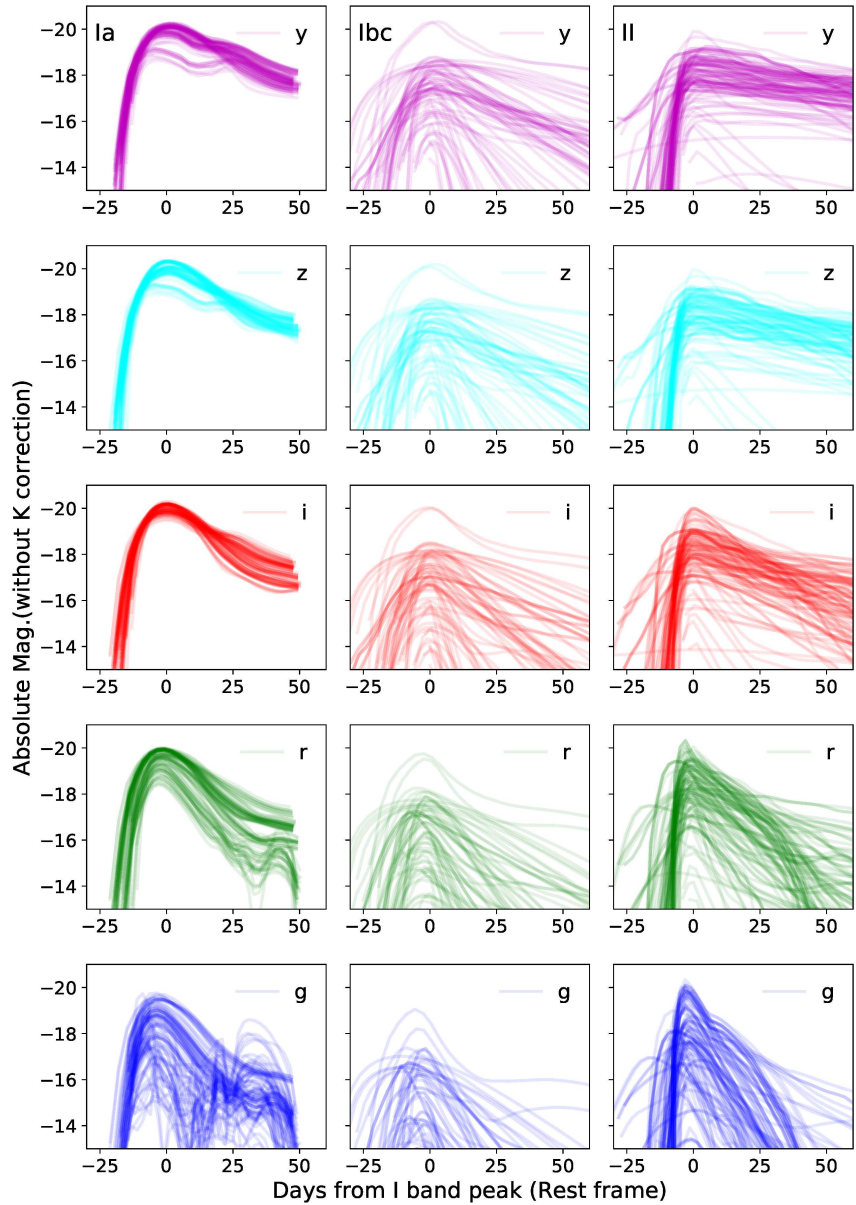}
  \end{center}
  \vspace{-6mm}
  \caption{%
  Overlay plots of simulated light curves with {\it z} between 0.1 and 1.2.
  Each panel shows the plots of SN Ia, Ibc, II data from the left, and the $g$-, $r2$-, $i2$-, $z$-, $y$-bands from the bottom.
  The variation of the curves in each panel depends on the different parameters and templates used in the simulation.
  A noise component was not added to these light curves.
  }%
  \label{fig:simLCsamples}
\end{figure*}

We are in need of simulating observed photometric data to train the machine.  
For normal SN~Ia, we used the SALT2 \citep{guy10b} model ({\it ver} 2.4) that requires two input parameters: c for color and x1 for stretch.
We adopted an asymmetric Gaussian distribution for c, and x1 from \citet{mosher14a}, and generated light curves and simulated photometric data points based on the observation schedule. 
Apart from SN~Ia, we used published spectral time series from \citet{kessler19b} which contains both observed Core-Collapse SN data and light curves from the simulation.
We combined an equal number of SN~Ia and non-SN~Ia observations to approximate the observed fractions. 
Although these fractions do not need to be exact for the purpose of our classification, it is important to avoid using an unbalanced dataset.
For three class classification, we set the ratio of SN~Ia:SN~Ibc:SN~II=10:3:7.
The redshift distribution of galaxies was taken from the COSMOS survey \citep{laigle16a}, and we distributed the simulated SNe accordingly from $z=$0.1 through $z=$2.0.
Throughout the study reported in this paper, we used $\Lambda$CDM cosmology with $\Omega_{m}$=0.3 and h=0.7.
We used the filter response including system throughput from \citet{kawanomoto18a}. 
Examples of simulated photometric data are shown in figure \ref{fig:simLCsamples}.

A complication that arises when attempting to simulate realistic data is that the machine does not accept expected errors. Alternatively, we may not have identified a good method for including errors.   
In this study, we therefore simply used brute force, namely, we placed the expected photometric error on top of the simulated data such that a simulated data point behaves similarly to one of the many realizations.
The magnitude of the error in the HSC simulation would have to consider varies from night to night because of sky conditions. 
We measured and derived the flux vs. error relationship from the actual observed data at simulating epoch and applied that relationship to the simulated photometric data.

\begin{figure}[htbp]
  \begin{center}
     \includegraphics[width=\columnwidth]{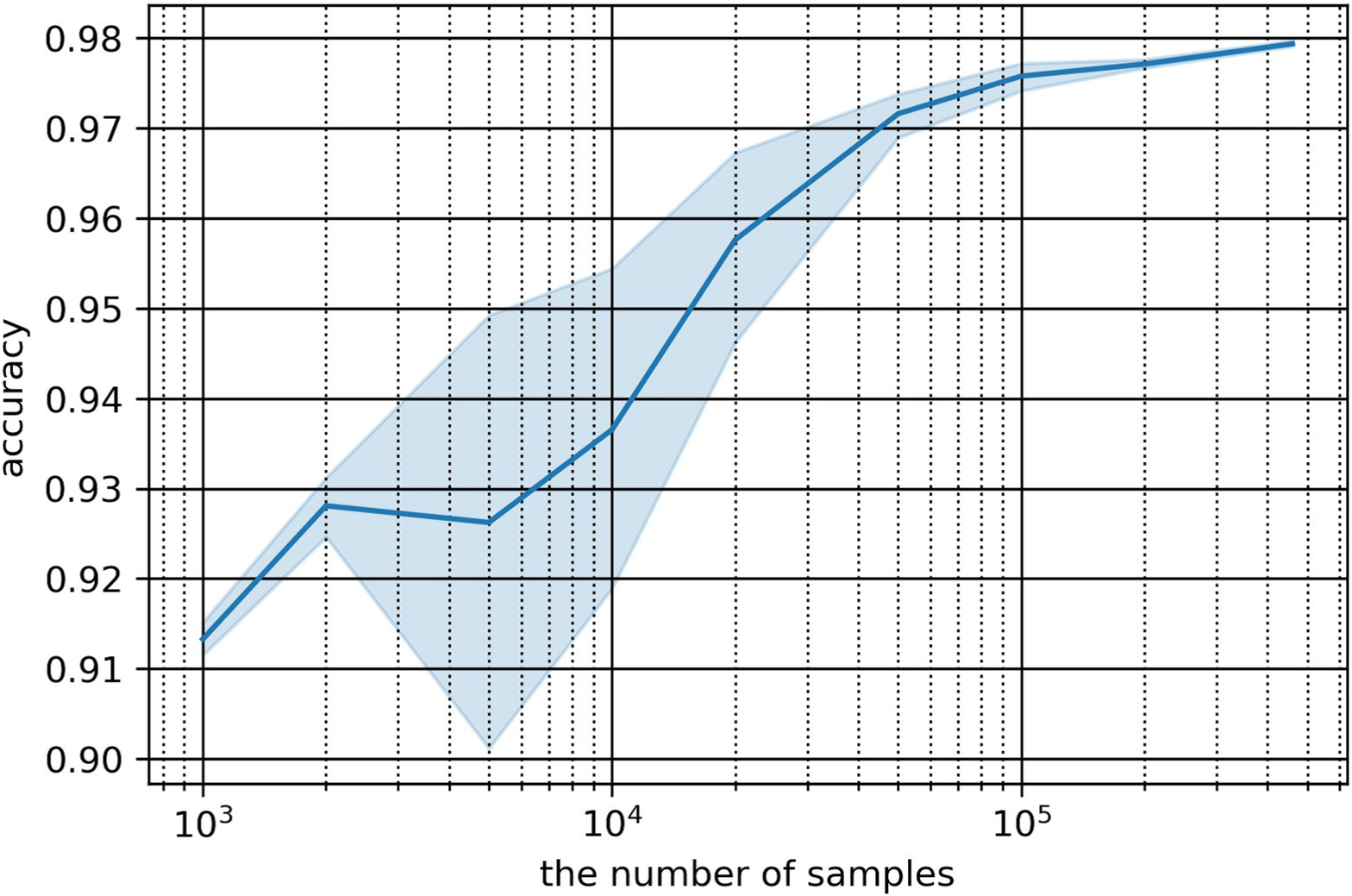}
  \end{center}
  \caption{%
Convergence test to determine the number of light curves that would need to be generated to train the machine. 
The solid line shows the mean accuracy of five classifiers. The shaded area shows the standard deviation of the classifiers, which were trained with 5-fold cross validation using the training dataset. 
The results indicate that more than 100,000 light curves would be required for training.
  }%
  \label{fig:size_convergence_test}
\end{figure}

Guided by the ``accuracy'' (subsection \ref{hyperparametersearch}), we determined the number of light curves that would be required for training.   
Based on our convergence test, we concluded that we would need to generate more than 100,000 light curves for training as shown figure \ref{fig:size_convergence_test}.
We omitted the curves with less than 3$\sigma$ detection at the maximum because our detection criterion was 5$\sigma$. 
For training, we generated 514,954 light curves for HSC observed data. 
Their final class ratio after these criteria is SN~Ia:Ibc:II=0.59:0.07:0.34, and their peak timings are randomly shifted by 450 days.
%

%
\subsection{Preprocessing of input data}
\label{sec:preproc}
Based on our pre-experiment with the simulated dataset, we found the machine to perform best by using a combination of the normalized flux ($f$) and pseudo-absolute magnitude ($M$):
\begin{equation}
    x = \left( M_1^\mathrm{abs}, \ldots, M_P^\mathrm{abs}, f_{1}^{\mathrm{scale}}, \ldots, f_{P}^{\mathrm{scale}} \right)^T,
\end{equation}
where $f_{i}^{\mathrm{scale}}$ is the $i$-th raw observed flux normalized by its maximum flux:
\begin{equation}
    f_{i}^{\mathrm{scale}} = \frac{f_i}{\max \left(f_1, \ldots, f_P \right)},    \label{eq:scaled_flux}
\end{equation}
and $M_i^\mathrm{abs}$ is the $i$-th pseudo-absolute observed magnitude.
For simplicity, we ignored K-correction and used the distance modulus (DM(z)) based on $\Lambda$CDM with the photometric redshift from its host galaxy.
\begin{eqnarray}
    M_i^\mathrm{abs} = m_i - \mathrm{DM}\left(z\right),
\end{eqnarray}
We can justify this operation because the training set and the observed dataset are processed using the same approach.
In the case of the existence of K-correction offset, both datasets would experience this in the same way.
In addition, because the observed flux could take on a negative value owing to statistical fluctuation, 
we adopt hyperbolic sine to imitate the magnitude system we use \citep{lupton99a}.  
\begin{eqnarray}
    m_i = 27.0 - \frac{2.5}{\log 10} \sinh^{-1} \frac{f_i}{2}. \label{eq:mag} 
\end{eqnarray}
In fact, combination of the flux and magnitudes is redundant, because knowledge of the one would enable us to calculate the other explicitly.   
However, based on our experiment, the score of the machine improves by using both.  
We suspected that the distribution in flux (linear) differs from that of the magnitude space (log), which provides the machine with additional information.
Thus, we used the pseudo-absolute magnitude ($M$) and normalized flux ($f$) as an input.

\section{Deep neural network classifier}
\label{sec:DNN}
With the rise of Big Data, the use of machine learning techniques has played a critical role in the analysis of astronomical data. Techniques such as random forest, support vector machine, and convolution neural network have been used for photometric data analysis \citep{pasquet19a}, galaxy classifications \citep{hausen19a}, and spectral classifications \citep{garciadias18a,muthukrishna19c,sharma20a}.

In our work, we seek to classify SNe from photometric data.
Our approach entails making use of the observed data without pre-processing or parameterization.
In this regard, we rely on deep learning to make our work possible.
We decided to test the extent to which deep learning could provide useful results without extracting features such as color, the width of the light curve, and the peak magnitude.

The fact that we went one step further by leaving the observed data as raw as possible, means that our input consists of a simple array of magnitudes. 
An attempt such as this would not have been possible ten years ago; however, owing to advancements in computing and the deep learning technique, this has become reality. 
Among the many machine learning methods, we decided to use a DNN to enable us to classify astronomical objects from the raw observed data.

\subsection{Model design}
\label{sec:model} 
In this section, we describe the design of our DNN model,\footnote{The code for our model is available at $\langle$https://github.com/ichiro-takahashi/snclass$\rangle$.} 
which accepts an array of observed magnitudes as its input and outputs the SN classification with probabilities. We adopted a highway layer (also known as a ``layer in layer'', \cite{srivastava15a}) as a core part of our network. Compared to plain DNN, the performance of a highway layer improves when the network is deep in terms of parameter optimization \citep{srivastava15b}.   

\begin{figure*}[htbp]
  \begin{center}
     \includegraphics[width=130mm]{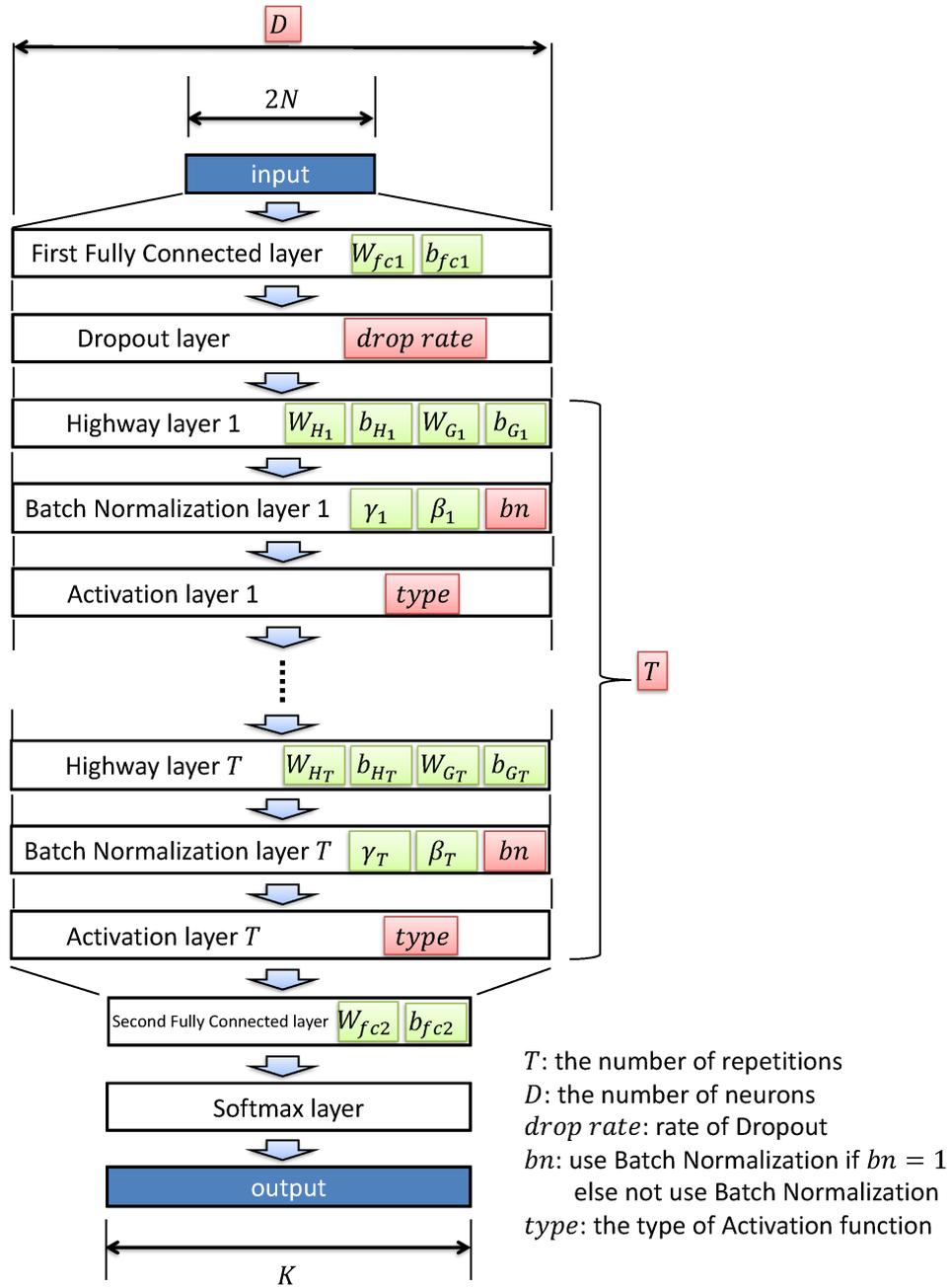}
  \end{center}
  \caption{\label{dnnmodel}
  Architecture of the deep neural network classifier. 
  The green boxes are parameters optimized by the gradient descent method during training. 
  The red boxes are hyperparameters that are optimized during the hyperparameter search. 
  The batch normalization layer has four variables ($\mu, \sigma^2, \gamma, \beta$), where $\mu$ and $\sigma^2$ are intended to learn the statistics (mean and variance) of the value through the layer, respectively, and $\gamma$ and $\beta$ are scale and shift parameters, respectively, to adjust the output. 
  Note that $\mu$ and $\sigma^2$ are not updated by gradient descent; instead, they are updated by the moving average. 
  They were omitted from the figure for simplicity.
  }%
  \label{fig:dnn_model}
\end{figure*}

Similar to other DNN models, this model proceeds through a training and validation process to optimize the parameters, and we describe the steps below. Our terminology is commonly used in the world of DNN, but as this is a new introduction to the astronomical community, we explain each step in detail. The architecture of our model is summarized in figure \ref{fig:dnn_model}.
Ultimately, each SN is assigned a probability of belonging to a certain astrophysical class, in our case, the type of SN. 

{\bf Input}: Our input is an array of magnitudes and normalized fluxes of the $i$th SN in the training dataset:
\begin{equation}
      x_i = \left( M_{i1}, M_{i2}, \ldots M_{ij} \ldots , M_{iN}, f_{i1}, f_{i2}, \ldots, f_{iN} \right)^T
\end{equation}
We do not explicitly specify the time at which or the particular filter with which the data were recorded, but this information is recorded as an order inside the array. 
The philosophy here is that the training set, which is composed of simulated data of the same array length, holds information on the filter and dates. 
For example, the $j$th magnitude in the array is data recorded on a certain date and by a certain filter. 
The combination of the date and filter is identical to those in the training set. Therefore, the $j$th component implicitly contains unique information about the dates and filters.   
Considering that the input consists of a combination of the magnitude and normalized flux, the size of our input array is $1\times2N$ per SN where $N$ is the number of data points.


{\bf First Fully Connected layer:}
We decided to make use of $D$ neurons, also known as the number of ``hidden layers,'' and $D$ is greater than the number of input components ($2N$). However, the dimension of $D$ is not known in advance, and this is one of the hyperparameters we optimize later in this section. Because the dimensionality of the input ($2N$) could differ from the number of optimized neurons $D$, we would need to adjust the number of dimensions and that is the role of this first fully connected layer $F(x)$.
\begin{equation}
    F \left(x, \left\{W_{fc1},b_{fc1}\right\}\right) = W_{fc1} x + b_{fc1} \in \mathbb{R}^D.
\end{equation}
$F(x)$ is given by a linear combination of matrix $W_{fc} \in \mathbb{R}^{D\times 2N}$ and a vector $b_{fc} \in \mathbb{R}^D$. The initial value of $W_{fc}$ is generated by Gaussian distribution and $b_{fc}$ is initialized by $\mathbf{0}$, which is a $D$-dimensional vector of which all the elements are zero. We used the Python wrapper library {\it dm-sonnet} \footnote{{\it dm-sonnet} $\langle$https://github.com/deepmind/sonnet$\rangle$.} (version 1.23) and its function {\it linear} to supply the $F(x)$ when plugging in ``$2N$'' and ``$D$.''  
Subsequently, $W_{fc}$ and $b_{fc}$ are optimized by the open source machine-learning package {\it Tensorflow} (version 1.14) \citep{Abadi2016}.  
Unless stated otherwise, we used the libraries from {\it Tensorflow}.

{\bf Dropout layer:}
To obtain a robust result, it is always best to train all of the neurons as an ensemble and avoid a situation in which one of the neurons adversely affects the result. Dropout is a process in which certain neurons are randomly dropped from training and the dropout rate can be optimized as one of the hyperparameters \citep{dropout}.

{\bf Highway layer:}
We adopted a Highway layer \citep{srivastava15a} and optimized the number of layers therein during the hyperparameter search. In theory, it would be possible to design a very deep layer with more layers than would be necessary. However, in reality, it is not trivial to optimize the number of layers.   
The depth and/or size of the layers of the DNN model are directly related to the complexity of the features used as input, and greatly affect the computational performance of the task.  
Thus, an overly deep model would complicate the learning process and cause performance degradation.
A highway layer is a technique that stabilizes the learning process by devising the network structure.
We previously used a highway layer, which we tested on 2D images, and it delivered good performance \citep{Kimura17}. Details of the use and advantages of the highway layer are provided in \citet{Kimura17}. This encouraged us to adopt a highway layer scheme for this analysis.
The output of the highway layer is calculated from the values of multiple paths.
The output, ${\bf Highway} \left(x\right)$, is formulated as
\begin{equation}
    {\bf Highway} \left(x\right) = G \left(x\right) \otimes H \left(x\right) + C \left(x\right) \otimes x \in \mathbb{R}^D,
\end{equation}
where $H$ is a nonlinear transformation layer, $G$ is the transformation gate function layer and controls the transformation of input, $C$ is the carry gate function layer, and $\otimes$ provides the element-wise product, also known as the Hadamard product.
A highway layer includes several other layers, a structure known as ``layer in layer.'' Each function is defined as follows: 
\begin{eqnarray}
    H \left(x\right) &=& \mathrm{a} \left( F \left(x, \left\{W_{H_{fc}}, b_{H_{fc}}\right\}\right) \right) \in \mathbb{R}^D, \\
    G \left(x\right) &=& \mathrm{a} \left( F \left(x, \left\{W_{G_{fc}}, b_{G_{fc}}\right\}\right) \right) \in \mathbb{R}^D, \\
    C \left(x\right) &=& \mathbf{1} - G \left(x\right) \in \mathbb{R}^D,
\end{eqnarray}
where $a$ is an activation function, namely, sigmoid.
\begin{eqnarray*}
    \mathrm{a} \left(p\right) &=& \left( \sigma\left(p_1\right),\sigma\left(p_2\right), \ldots, \sigma\left(p_D\right) \right)^T, \; p \in \mathbb{R}^D, \\
    \sigma \left(p_i\right) &=& \frac{1}{1 + e^{-p_i}},
\end{eqnarray*}
where $D$ is the number of neurons. Each element of $G(x)$ always takes a value between 0 and 1. Eventually, the Highway layer behaves as follows:
\begin{equation}
    {\bf Highway(x)}=\left\{
    \begin{array}{@{}ll@{}}
      x, & \mathrm{if} \ G \left(x\right)= \mathbf{0} \\
      H(x), & \mathrm{if} \ G \left(x\right)= \mathbf{1} 
    \end{array}\right.
\end{equation}
Along with the dimensions of the hidden layer $D$, the dropout ratio, batch normalization, and the types of activation function, the number of repetitions $T$ is one of the hyperparameters and is optimized by performing a hyperparameter search. Details are provided in subsection \ref{hyperparametersearch}.

{\bf Batch Normalization layer:}
We adopt batch normalization \citep{batch_norm} to accelerate and stabilize the optimization. 
Even if a large number of parameters need to be trained, batch normalization facilitates convergence of the training process, reduces errors on the slope when we apply entropy minimization, prevents the average and dispersion from becoming exponentially large in deep layers, and minimizes the biases on outputs  \citep{understanding_batch_norm}.
Batch normalization is effective in many cases. 
However, the performance of a model that employs both batch normalization and dropout may degrade  (\cite{dropout_and_batch_norm}).   

{\bf Activation layer:}
Each neuron is activated by using nonlinear transformation. 
Nonlinearity is an important component of DNN, because it allows a wide variety of expressions. 
Note that the majority of the layers, including fully connected layers, involve a linear transformation and, even if a number of layers were to exist, it would be equivalent to one single linear transformation. 
Thus, nonlinear transformation is essential to allow each neuron the freedom to have any necessary values.
For the first iteration, we do not know what kind of transformation is the best; thus, the transformation itself is taken as one of the hyperparameters. 
In our work, we used the functions ``tf.nn,'' in {\it Tensorflow}.

{\bf Second Fully Connected layer:}
After $T$ repetitions of the highway layer, batch normalization layer, and activation layer, it is necessary to convert the number of neurons to the number of SNe times the number of SN classes. 
This operation is opposite to that of the first fully connected layer.

{\bf Softmax layer:}
The Softmax layer normalizes the input value of this layer, which is denoted by $h \in \mathbb{R}^D$. The output value is $\hat{y} \in \mathbb{R}^D$ and each element $\hat{y}_k$ is expressed as
\begin{equation}
    \hat{y}_k = \frac{\exp \left( h_k \right)}{\sum_{k'=1}^K \exp \left( h_{k'} \right)}.
\end{equation}
The normalized value $\hat{y}$ satisfies $\hat{y}_k \geq 0$ and $\sum_k^K \hat{y}_k =1$.
We can interpret $\hat{y}_k$ as the probability that the input $x$ belongs to class $k$.
However, we note that this is a pseudo-probability and that it differs from the statistical probability.
  
\subsection{Hyperparameter search}\label{hyperparametersearch}
We perform a hyperparameter search by combining grid search and the Tree-structured Parzen Estimator (TPE) algorithm \citep{pmlr-v28-bergstra13}.
Although grid search is not suitable to search a high-dimensional space, it has the advantage of searching for multiple points in parallel.
In addition, because of the simplicity of the algorithm, it allows us to convey knowledge acquired during preliminary experiments for parameter initialization.
Meanwhile, the TPE algorithm is suitable for searching a high-dimensional space, but has the disadvantage of not knowing where to start initially.
Therefore, this time, our search was guided by the hyperparameter values that were obtained in the preliminary experiment using grid search, and these results were then used as input for the TPE algorithm.  
The ranges in which we searched for the hyperparameters are given in table \ref{tb:hp}.

\begin{figure}[htbp]
  \begin{center}
     \includegraphics[width=\columnwidth]{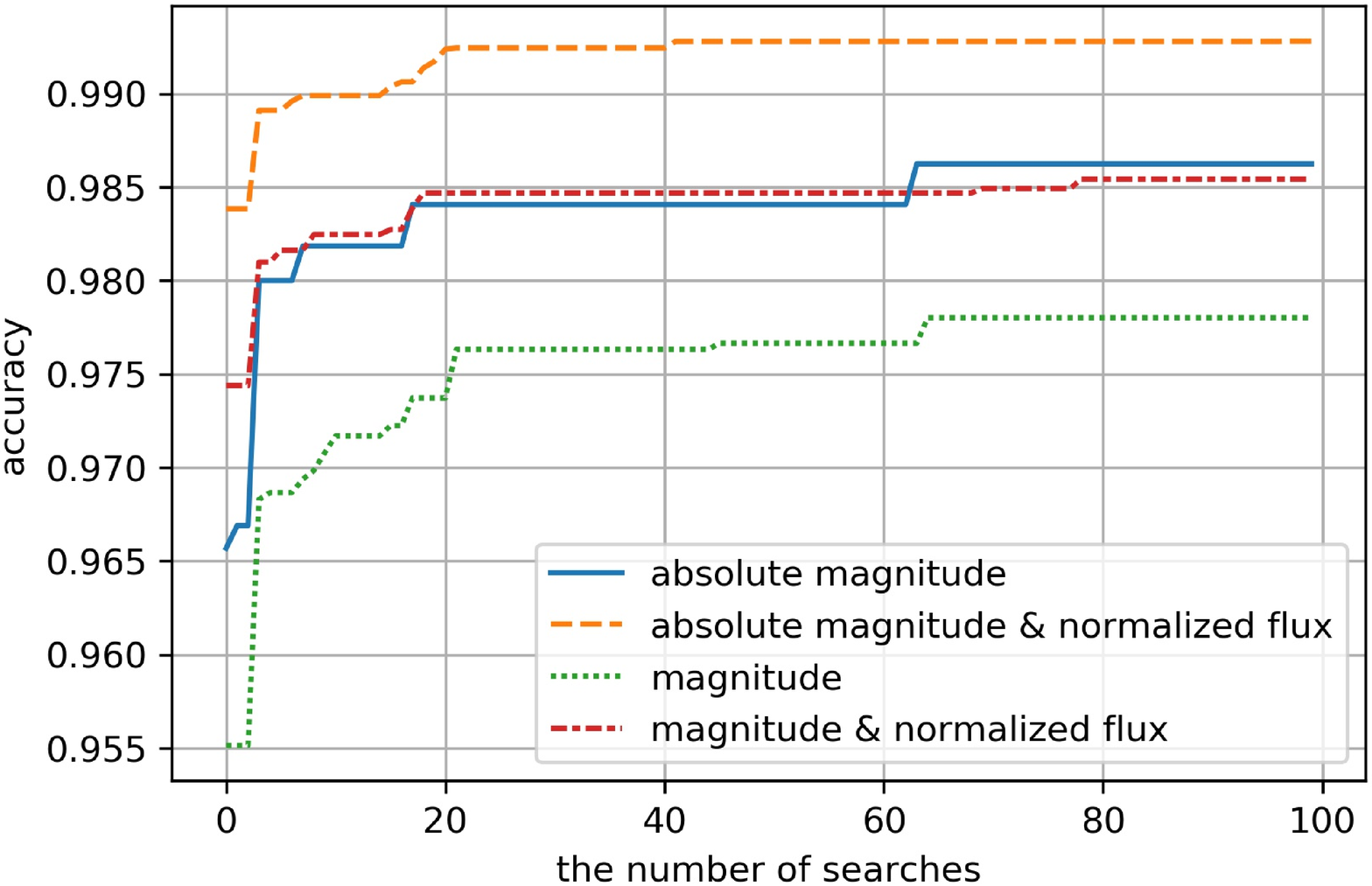}
  \end{center}
  \caption{%
  Result of iterative hyperparameter search (100 cycles) showing its convergence to its maximum performance in terms of accuracy. The task involved binary classification.
  }%
  \label{fig:hp_test}
\end{figure}

According to the usual approach, we divided the dataset into training and validation datasets.
We used the training data to optimize the DNN, and the validation data to measure the accuracy.
The hyperparameters were optimized by evaluation with the validation dataset to maximize the accuracy of this dataset.
This process was iteratively conducted 100 times to allow the accuracy to converge to its maximum (figure \ref{fig:hp_test}).

%
\begin{table}[htbp]
  \tbl{Ranges of hyperparameter search for Type Classification}{
      \begin{tabular}{lcc}
        \noalign{\vskip 2mm}
        \hline
        hyper parameter     & value (grid)  & range (TPE)\\ \hline 
        $D$                 & \{100, 300\}  & 50, \ldots, 1000   \\
        $T$                 & \{1, 3, 5\}   & 1, \ldots, 5       \\
        $bn$                & \{true\}      & \{true, false\}    \\
        $drop\mbox{ }rate$      & [5e-3, 0.035] & [5e-4, 0.25]       \\
        $type$              & \multicolumn{2}{c}{\{identity, relu, sigmoid, tanh\}} \\
        \hline
      \end{tabular}
  }\label{tb:hp}
\end{table}

We can train the DNN classifier in the same way regardless of the number of classes.
In the case of multi-type classification, the number of classes is $K = 3$ in our experiment; thus, the number of outputs of the DNN classifier is also three.
In binary classification (SN~Ia or non-SN~Ia), the number of outputs is two.

We trained the model by minimizing the cross-entropy error: 
\begin{equation}
\mathrm{CE} \left(y, \hat{y} \right) =　-\sum_{k = 1}^K y_k \log \hat{y}_k,
\end{equation}
where $y$ is the ground truth vector, which entails one-hot encoding of $K$ dimensions, and $\hat{y}$ is the DNN output vector.
We deployed the {\it Adam optimizer} \citep{Kingma2014} which uses a stochastic gradient method to optimize the model parameters.

We introduced data augmentation to prevent overfitting at the time of training.
By increasing the number of input data by using data augmentation, we prevent DNN from having to memorize the entire training dataset.
We used two data augmentation methods to augment the training dataset.
The first was to add Gaussian noise (based on the expected observed uncertainty) to the simulated flux.
The second involved the use of the mixup technique (\cite{mixup}).

Mixup generates a new virtual training dataset as follows:
\begin{eqnarray*}
    \tilde{x} &=& \lambda x_u + \left( 1-\lambda \right) x_v, \\
    \tilde{y} &=& \lambda y_u + \left( 1-\lambda \right) y_v,
\end{eqnarray*}
where $\left(x_u, y_u\right)$ and $\left(x_v, y_v\right)$ are samples drawn at random from the training dataset, $x$ is the input vector, $y$ is the one-hot label vector, and the mixing ratio $\lambda \in \left[0, 1\right]$ is drawn from a random distribution, of which the density is low near 0 and 1 and higher near 0.5. 
The datasets generated in this way are suitable to enable the DNN to learn the classification boundaries.

As described above, the hyperparameters (red boxes in figure \ref{fig:dnn_model}) of the model are optimized by maximizing the accuracy, whereas 
the model parameters (green boxes in figure \ref{fig:dnn_model}) are optimized by minimizing the cross-entropy error.
%
%

\subsection{Testing the DNN model with PLAsTiCC dataset} 
\label{sec:p}
%
Before we applied our model to the data observed by HSC, we tested it with a dataset resulting from the LSST simulated classification challenge, i.e., the PLAsTiCC dataset \citep{allam18a,malz19a}, which is composed of realistic photometric data with errors on time-variable objects.
To evaluate our model, we required a dataset with labels of true identity. 
The PLAsTiCC Deep Drilling Field (DDF) dataset contains data similar to those in the HSC-SSP Transient Survey, and we took advantage thereof.
However, we generated the training set by ourselves and selected not to use the training set provided by the PLAsTiCC team because we knew that the size of their training dataset was insufficient to achieve maximum performance (figure \ref{fig:size_convergence_test}).

The training dataset was created by using the method described in subsection \ref{sec:training}. We generated 370,345 light curves based on the filter response and photometric zero-point for LSST \citep{ivezic19a}.
These light curves are composed of the different types of SNe in the ratio SN~Ia:Ibc:II=0.60:0.06:0.34, and their peaks are randomly shifted in time. 
%
The test dataset was created by extracting 2,297 light curves from the PLAsTiCC dataset.
These light curves are labeled Ia, Ibc, or II, to identify the type of SN each curve represented. The light curves were simulated to occur in the COSMOS field.

\begin{figure}[htbp]
  \begin{center}
     \includegraphics[width=\columnwidth]{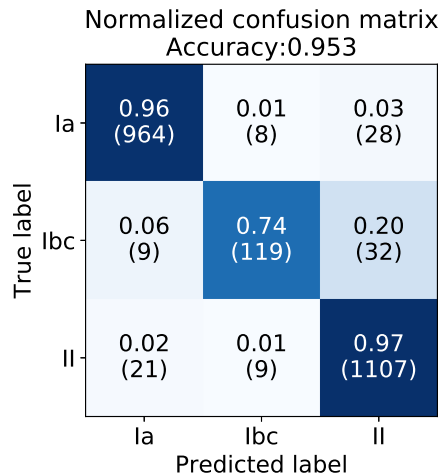}
  \end{center}
  \caption{%
  Normalized confusion matrix in the three-class classification of the PLAsTiCC dataset. The classifier received the pseudo-absolute magnitude and normalized flux as its input. The proportions in each row sum to 1. The numbers in parentheses represent the raw numbers.
  }%
  \label{fig:plasticc_3class_CM}
\end{figure}

\begin{figure}[htbp]
  \begin{center}
     \includegraphics[width=\columnwidth]{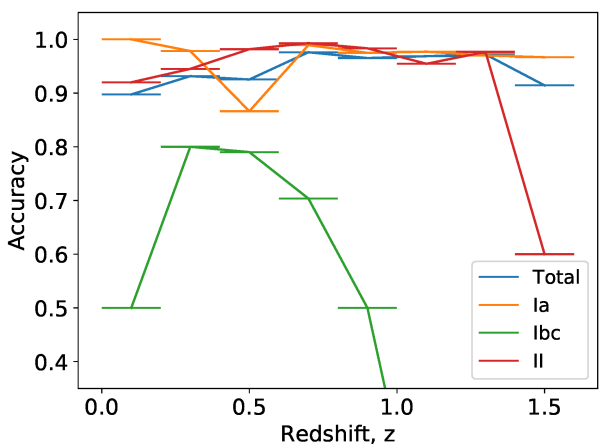}
  \end{center}
  \caption{%
  Accuracy as a function of the redshift in the three-class classification of the PLAsTiCC dataset.
  }%
  \label{fig:misclass_rate_3class}
\end{figure}

We used the area under the curve (AUC) of the receiver operating characteristic (ROC) curve, the precision-recall curve in two-class classification, and the accuracy from the confusion matrix in three-class classification as a metric to evaluate our model.
Different combinations of inputs were tested to determine which performs best when using the PLAsTiCC dataset. Our input could be a combination of the arrays of normalized flux ($f$), magnitude ($m$), or the pseudo-absolute magnitude ($M$).
Table \ref{tab:p_test} lists the AUC for two-class classification and the accuracy for three-class classification when using PLAsTiCC data, respectively.
Our investigation showed that a combination of the normalized flux ($f$) and pseudo-absolute magnitude ($M$) performs best, and, although the information is redundant, we suspect the different distribution of the data provides the machine with additional guidance.
The AUC values for the ROC curve and the precision-recall curve are 0.996 and 0.995, respectively.
Figure\ \ref{fig:plasticc_3class_CM} shows the confusion matrix for the three-class classification, with the total accuracy calculated as 95.3\%.
As is always the case in the real world, it is difficult to classify SN~Ibc, but the effect on overall accuracy is relatively small.
%
\begin{table}[htbp]
\tbl{Classification performance of each input for the PLAsTiCC dataset.}{
\begin{tabular}{cccccc}
\noalign{\vskip 2mm}
\hline
\multicolumn{3}{c}{Input\footnotemark[$*$]}   & \multicolumn{2}{c}{AUC} & Accuracy \\
\hline
$M$ & $m$ & $f$ &  ROC &  Pre.-Rec. &\\
\hline
\checkmark &            & \checkmark &    0.996 &       0.995 &     0.953\\
\checkmark &            &            &    0.995 &       0.993 &     0.952\\
           & \checkmark & \checkmark &    0.995 &       0.993 &     0.948\\
           & \checkmark &            &    0.995 &       0.991 &     0.940\\
\hline
\end{tabular}
}\label{tab:p_test}
\begin{tabnote}
\footnotemark[$*$] Input to classifier is displayed as a check mark. $M$: pseudo-absolute magnitude, $m$: magnitude, $f$: normalized flux.
\end{tabnote}
\end{table}
%
%
Table \ref{tab:snr_vs_AUC} summarizes the classification performance for each group of the test set divided according to the maximum signal-to-noise ratio of the photometric data.
It shows that the classification performance tends to improve as the maximum signal-to-noise ratio increases.
\begin{table}[htbp]
\tbl{Classification performance for the maximum signal-to-noise ratio (SNR).}{
\begin{tabular}{cccc}
\noalign{\vskip 2mm}
\hline
Max. SNR & Number & \multicolumn{2}{c}{AUC\footnotemark[$*$]} \\
\hline
         &        & ROC   & Pre.-Rec. \\
\hline
$<$ 5    & 31     & 0.975 & 0.983 \\
5 -- 10  & 736    & 0.989 & 0.983 \\   
10 -- 20 & 814    & 0.999 & 0.998 \\
$>$ 20   & 716    & 0.999 & 0.999 \\
\hline
All      & 2297   & 0.996 & 0.995 \\
\hline
\end{tabular}
}\label{tab:snr_vs_AUC}
\begin{tabnote}
\footnotemark[$*$] AUC for the best performing model ($M+f$).
\end{tabnote}
\end{table}
In the three-class classification of the PLAsTiCC dataset,
107 SNe were misclassified and have the following characteristics:
\begin{itemize}
\item 54\% (58/107) of them  were ``incomplete events'' that did not include the peak phase of the SN (the period of 10 days before and 20 days after the peak in the observed frame) in the photometric data, whereas they only constitute 38\% of all events.
\item Of the remaining misclassifications, 29\% (14/49) are on the boundary where the difference in probability between the correct class and the predicted class is less than 0.1.
\item In more than half of the remaining 35 events, SN Ibc was misclassified as either SN Ia or II.
\end{itemize}
Figure\ \ref{fig:misclass_rate_3class} shows the accuracy against redshift for the PLAsTiCC dataset.
The accuracy for SN Ibc is lower than that of the other classes; in particular, it is greatly reduced at redshifts beyond 1.0, and also decreased at redshift {\it z} of 0.1 to 0.2.
Manual verification of individual misclassifications revealed that, although certain misclassified SNe are too faint to classify even with conventional methods, a few bright SNe were also completely misclassified.

\section{Application to HSC-SSP transient survey}
\label{sec:h}
%
We applied the developed classifier to the dataset acquired during the HSC-SSP Transient Survey. This dataset includes photometric data of 1824 SNe.
As described in \citet{yasuda19a}, the survey 
was conducted in two layers with different depths and cadence, i.e., ``Deep'' and ``Ultra-Deep.''
Therefore, the number of photometric data points of an SN in each layer could be different.
Our DNN model requires exactly the same number of data points as its input; thus, we divided our dataset into five cases based on the number of photometric data points.
%
The number of SNe for each case is summarized in table\ \ref{tab:class_flag}.
For example, the number of SNe in Case 0 is 709, and they are in the Ultra-Deep field. Each SN is represented by a total of 
42 epochs of photometric data in four bands ($g$-, $r2$-, $i2$- and $z$-band).
The number of epochs and filter schedule for Case 0 SNe are summarized in table \ref{tab:HSCsurvey_schedule}.
The introduction of these five cases enabled the machine to correctly classify 1812 HSC SNe, which corresponds to 99.3\% of the 1824 SNe.
The remaining 12 SNe were excluded owing to missing data.
%
\begin{table}[htbp]
\tbl{Number of SNe for each Case.}{
\begin{tabular}{lrrrr}
\noalign{\vskip 1mm}
\hline
Case & Epoch & Number & Fraction \\
\hline
0     & 42      & 709        & 0.391 \\
1     & 26      & 646        & 0.357 \\
2     & 19      & 271        & 0.150 \\
3     & 10      & 122        & 0.067 \\
4     & 4        & 64        & 0.035 \\
\hline
\end{tabular}
}\label{tab:class_flag}
\end{table}
\begin{table}[htbp]
\tbl{Number of input epochs and the schedule for Case 0 SNe.}{
\begin{tabular}{llp{15em}}
\noalign{\vskip 2mm}
\hline
Filter & Epochs & Elapsed day\\
\hline
$g$ & 8 & 2, 40, 63, 70, 92, 119, 126, 154\\
$r2$ & 9 & 5, 32, 61, 71, 92, 103, 122, 129, 151\\
$i2$ & 13 & 2, 6, 32, 40, 61, 68, 71, 94, 101, 120, 127, 154, 155\\
$z$ & 12 & 0, 6, 30, 40, 59, 68, 90, 101, 119, 126, 151, 157\\
\hline
\end{tabular}
}\label{tab:HSCsurvey_schedule}
\end{table}
%
%
We subjected the aforementioned five cases of observed HSC data to both two-class and three-class classification.
For each of these cases, the machine needs to be trained independently with a dedicated training dataset.
Thus, the hyperparameters were optimized for each case and are reported in table \ref{tb:searched_hp_class}.
The following subsections (subsection \ref{sec:h2} and \ref{sec:h3}) describe the performance evaluation for each classification.
\begin{table*}[t]
  \tbl{Optimized hyperparameters for classification}{
      \begin{tabular}{lcccllllllllll}
      \noalign{\vskip 1mm}
\hline
      & \multicolumn{3}{l}{Input\footnotemark[$*$]}            & \multicolumn{5}{l}{Two-Class}      &  \multicolumn{5}{l}{Three-Class}   \\ \hline
      & $M$        & $m$        & $f$        & $T$ & $D$   & $drop\mbox{ }rate$ & $bn$ & $type$    & $T$ & $D$   & $drop\mbox{ }rate$ & $bn$ & $type$    \\ \hline      
Case 0& \checkmark &            & \checkmark & 5 & 178 & 9.47e-3   & 1  & sigmoid & 4 & 429 & 1.20e-3   & 0  & linear  \\
      & \checkmark &            &            & 3 & 247 & 9.68e-4   & 1  & sigmoid & 4 & 516 & 2.54e-3   & 0  & tanh    \\
      &            & \checkmark & \checkmark & 4 & 531 & 6.43e-3   & 0  & linear  & 4 & 608 & 1.72e-2   & 0  & linear  \\
      &            & \checkmark &            & 4 & 411 & 9.00e-2   & 1  & sigmoid & 4 & 838 & 1.36e-3   & 0  & tanh    \\ \hline
Case 1& \checkmark &            & \checkmark & 5 & 734 & 8.75e-4   & 0  & tanh    & 4 & 915 & 1.03e-2   & 0  & linear  \\
      & \checkmark &            &            & 2 & 389 & 7.17e-2   & 1  & sigmoid & 5 & 698 & 2.79e-2   & 0  & linear  \\
      &            & \checkmark & \checkmark & 2 & 647 & 7.92e-4   & 0  & tanh    & 4 & 540 & 1.45e-3   & 0  & linear  \\
      &            & \checkmark &            & 2 & 342 & 1.42e-2   & 1  & sigmoid & 2 & 520 & 9.99e-4   & 1  & sigmoid \\ \hline
Case 2& \checkmark &            & \checkmark & 2 & 368 & 1.79e-3   & 1  & sigmoid & 4 & 698 & 9.08e-4   & 0  & linear  \\
      & \checkmark &            &            & 4 & 920 & 1.44e-3   & 0  & sigmoid & 4 & 614 & 7.03e-3   & 0  & linear  \\
      &            & \checkmark & \checkmark & 5 & 572 & 4.27e-3   & 1  & sigmoid & 4 & 896 & 8.58e-3   & 0  & linear  \\
      &            & \checkmark &            & 4 & 640 & 1.12e-1   & 1  & sigmoid & 5 & 300 & 5.00e-1   & 1  & sigmoid \\ \hline
Case 3& \checkmark &            & \checkmark & 5 & 893 & 1.42e-3   & 1  & sigmoid & 4 & 522 & 5.02e-4   & 0  & tanh    \\
      & \checkmark &            &            & 4 & 880 & 1.98e-2   & 1  & sigmoid & 5 & 841 & 4.96e-2   & 1  & sigmoid \\
      &            & \checkmark & \checkmark & 3 & 300 & 5.00e-3   & 1  & linear  & 3 & 462 & 9.28e-4   & 1  & sigmoid \\
      &            & \checkmark &            & 3 & 930 & 9.77e-2   & 1  & sigmoid & 3 & 300 & 5.00e-3   & 1  & sigmoid \\ \hline
Case 4& \checkmark &            & \checkmark & 5 & 379 & 4.21e-3   & 1  & sigmoid & 3 & 484 & 2.13e-3   & 0  & linear  \\
      & \checkmark &            &            & 2 & 631 & 1.77e-2   & 0  & sigmoid & 4 & 243 & 3.21e-3   & 1  & sigmoid \\
      &            & \checkmark & \checkmark & 5 & 140 & 4.04e-3   & 1  & sigmoid & 5 & 389 & 1.23e-4   & 0  & tanh    \\
      &            & \checkmark &            & 4 & 567 & 5.77e-2   & 0  & sigmoid & 3 & 354 & 2.14e-1   & 1  & sigmoid \\ \hline
\end{tabular}
  }\label{tb:searched_hp_class}
\begin{tabnote}
\footnotemark[$*$] Input to classifier is displayed as a check mark. $M$: pseudo-absolute magnitude, $m$: magnitude, $f$: normalized flux.
\end{tabnote}
\end{table*}
%
%
%
\subsection{Binary classification}
\label{sec:h2}
Binary classification was performed using four versions of classifiers, which we prepared with different inputs as in the PLAsTiCC dataset. This approach allowed us to compare their performance, a summary of which is provided in table\ \ref{tab:h2_AUC}.

For the validation dataset, which is part of the simulated dataset, a higher number of input dimensions were found to improve the results, enabling any classifier to classify the data with very high AUC.
The best AUCs for all classified events are 0.993 and 0.995 for the ROC and precision-recall curve, respectively.

For the test dataset, the classification performance was verified using 1332 HSC SNe (1256 with redshift) labeled by the SALT2 light curve fitter \citep{guy2007,guy10b}, a conventional classification method.
The verification label for the HSC SNe conforms to that reported in \citet{yasuda19a}, which defines SN~Ia as SNe that satisfy all four of the following criteria for the SALT2 fitting results:
(1) color ($c$), and stretch ($x_1$) within the $3\sigma$ range of \citet{scolnickessler2016} ``All G10'' distribution, (2) absolute magnitude in B band $M_B$ brighter than $-18.5$ mag, (3) reduced $\chi ^{2}$ of less than 10, (4) number of degrees of freedom (dof) greater than or equal to five.
Other candidates that satisfy the looser set of conditions above were labeled ``Ia?.''
Specifically, the range in (1) was expanded to within 5 sigma, and the thresholds of (2) and (3) were set to $-17.5$ mag and 20, respectively.
Meanwhile, we defined non-Ia in the HSC classification as SNe that do not satisfy the conditions to be classified as ``Ia'' and ``Ia?,'' and of which the number of dof is five or more.
The number of labeled Ia, Ia?, and non-Ia are 429, 251, and 908 (410, 240, and 850, with redshift), respectively.
Apart from the above, 215 SNe with less than 5 dof were labeled as ``unclassified,'' and the remaining 21 SNe failed to be fitted.
This performance evaluation was conducted by using 428 SNe~Ia and 904 non-SNe~Ia classified by our machine.

\begin{figure*}[htbp]
    \begin{tabular}{cc}
        \begin{minipage}{0.5\hsize}
            \begin{center}
                \includegraphics[width=\columnwidth]{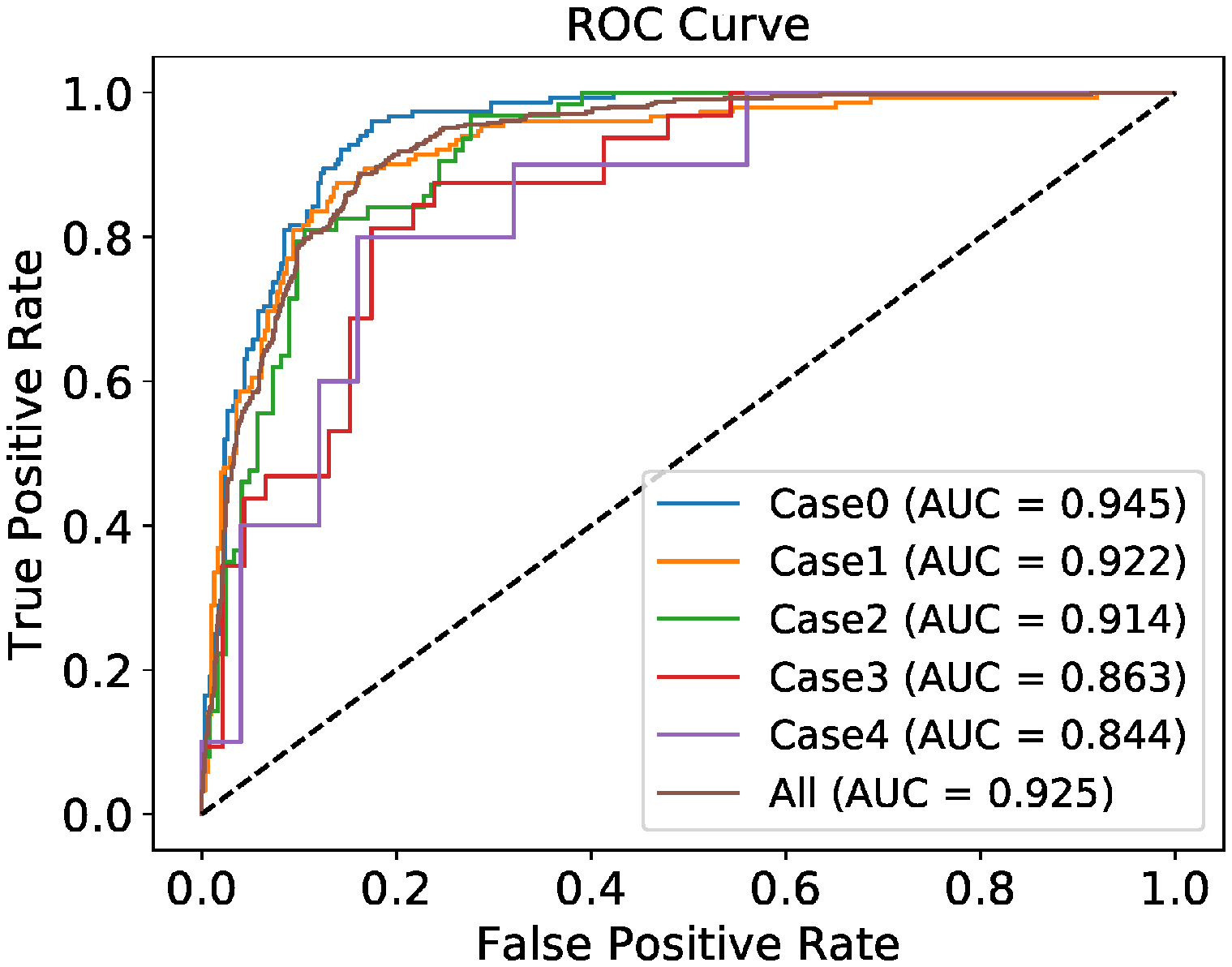}
            \end{center}
        \end{minipage}
        \begin{minipage}{0.5\hsize}
            \begin{center}
                \includegraphics[width=\columnwidth]{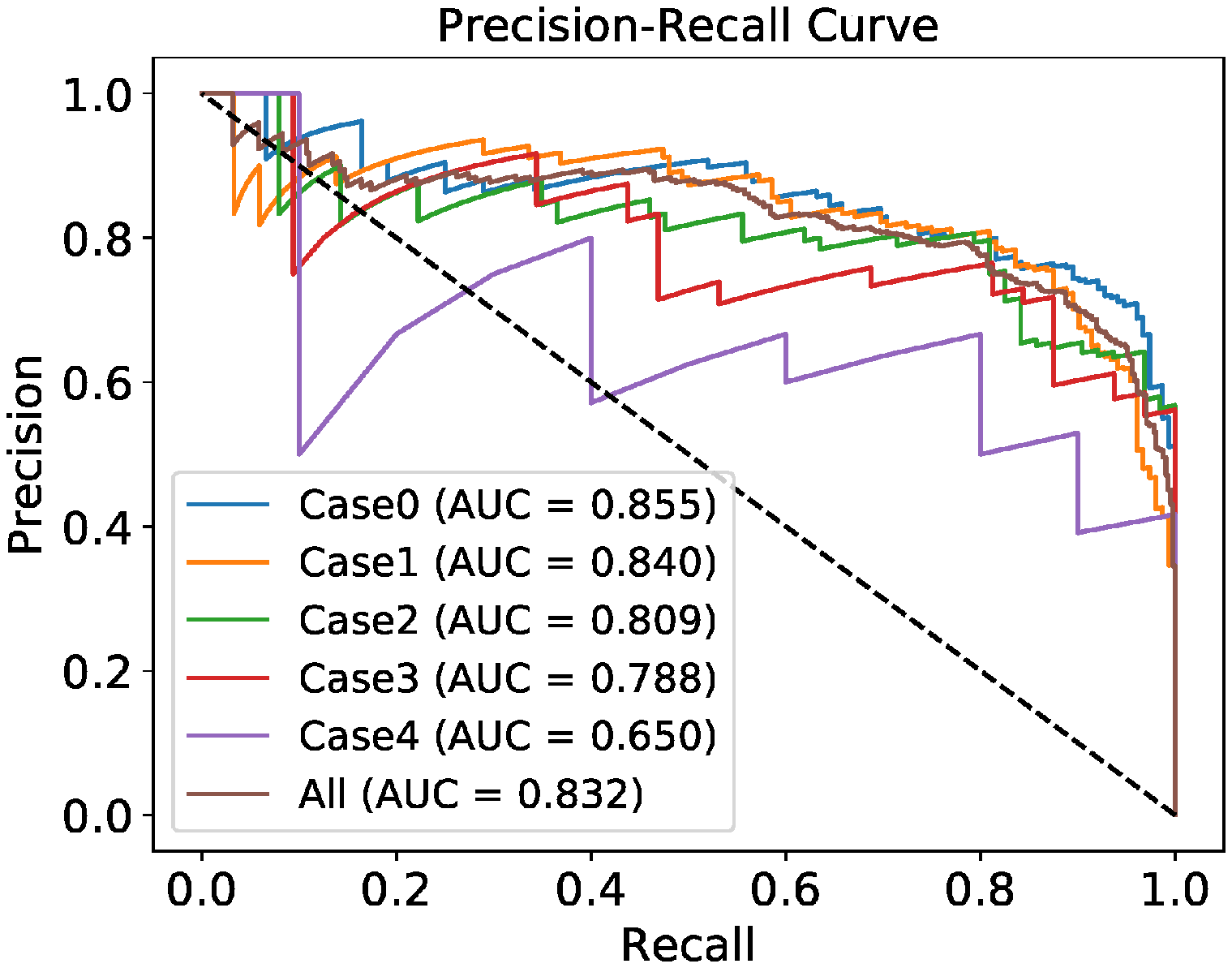}
            \end{center}
        \end{minipage}
    \end{tabular}
    \vspace{2mm}
    \caption{%
    ROC curves and precision-recall curves for the two-class classification of all labeled HSC SNe.
    The input to the classifier is the pseudo-absolute magnitude and normalized flux.
    The colored lines represent the performance for each of the five classifiers with different input cases, and that for all of their outputs.
    }
    \label{fig:h2_test_all}
\end{figure*}

\begin{figure*}[htbp]
    \begin{tabular}{cc}
        \begin{minipage}{0.5\hsize}
            \begin{center}
                \includegraphics[width=\columnwidth]{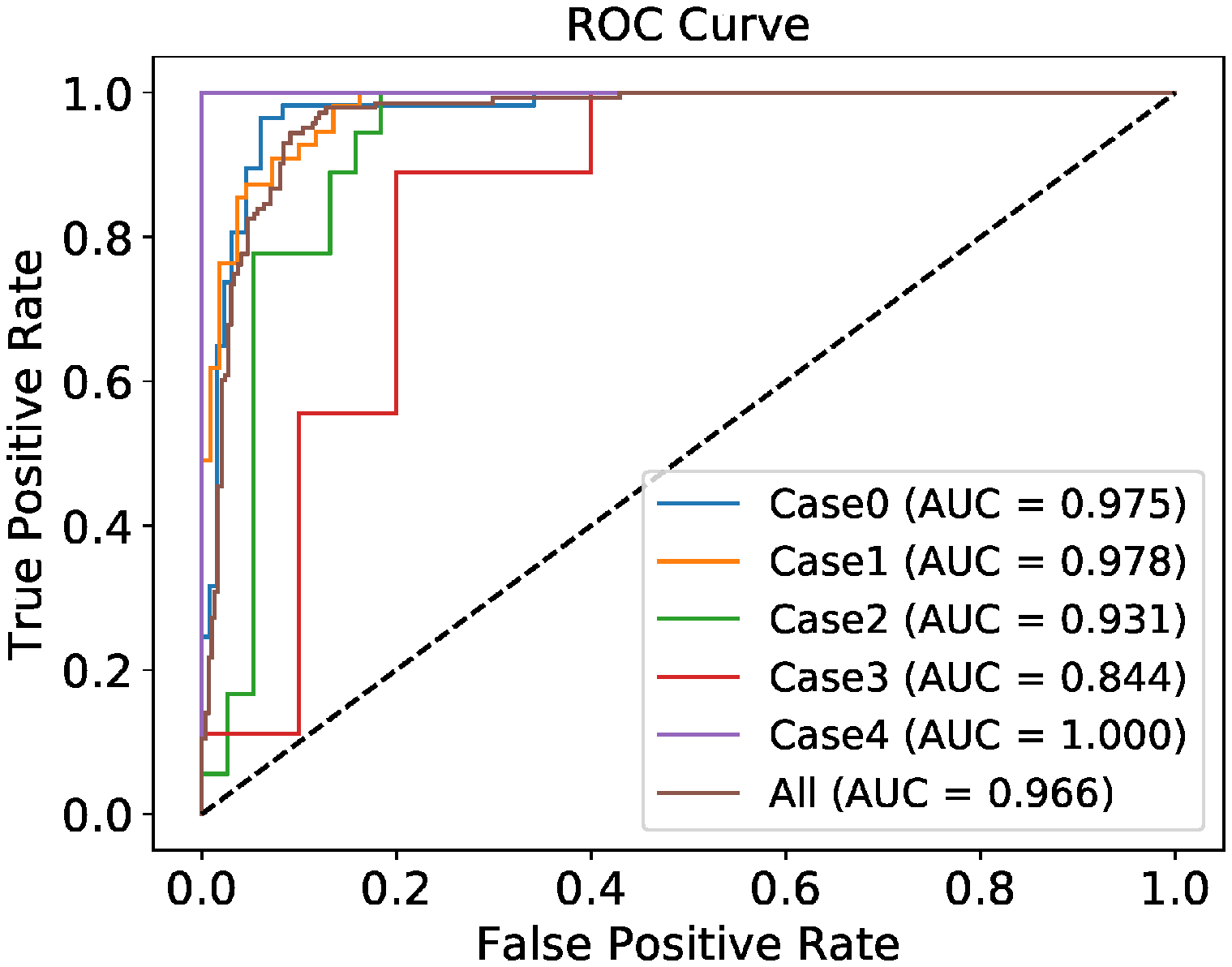}
            \end{center}
        \end{minipage}
        \begin{minipage}{0.5\hsize}
            \begin{center}
                \includegraphics[width=\columnwidth]{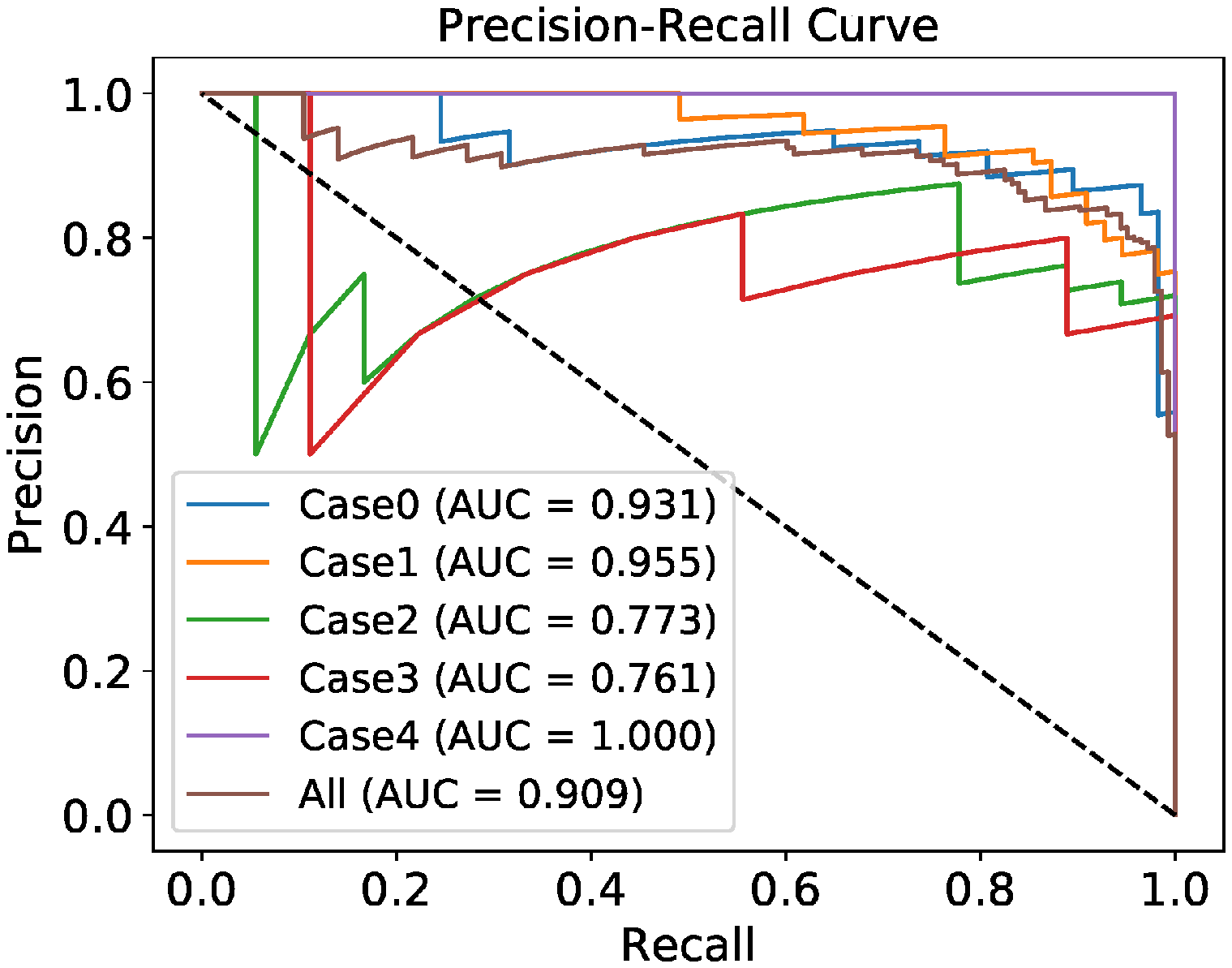}
            \end{center}
        \end{minipage}
    \end{tabular}
    \vspace{2mm}
    \caption{%
  As shown in figure \ref{fig:h2_test_all}, but for the light curve verified HSC SNe. 
}%
    \label{fig:h2_test_gold}
\end{figure*}

\begin{figure*}[htbp]
    \begin{tabular}{cc}
        \begin{minipage}{0.5\hsize}
            \begin{center}
                \includegraphics[width=\columnwidth]{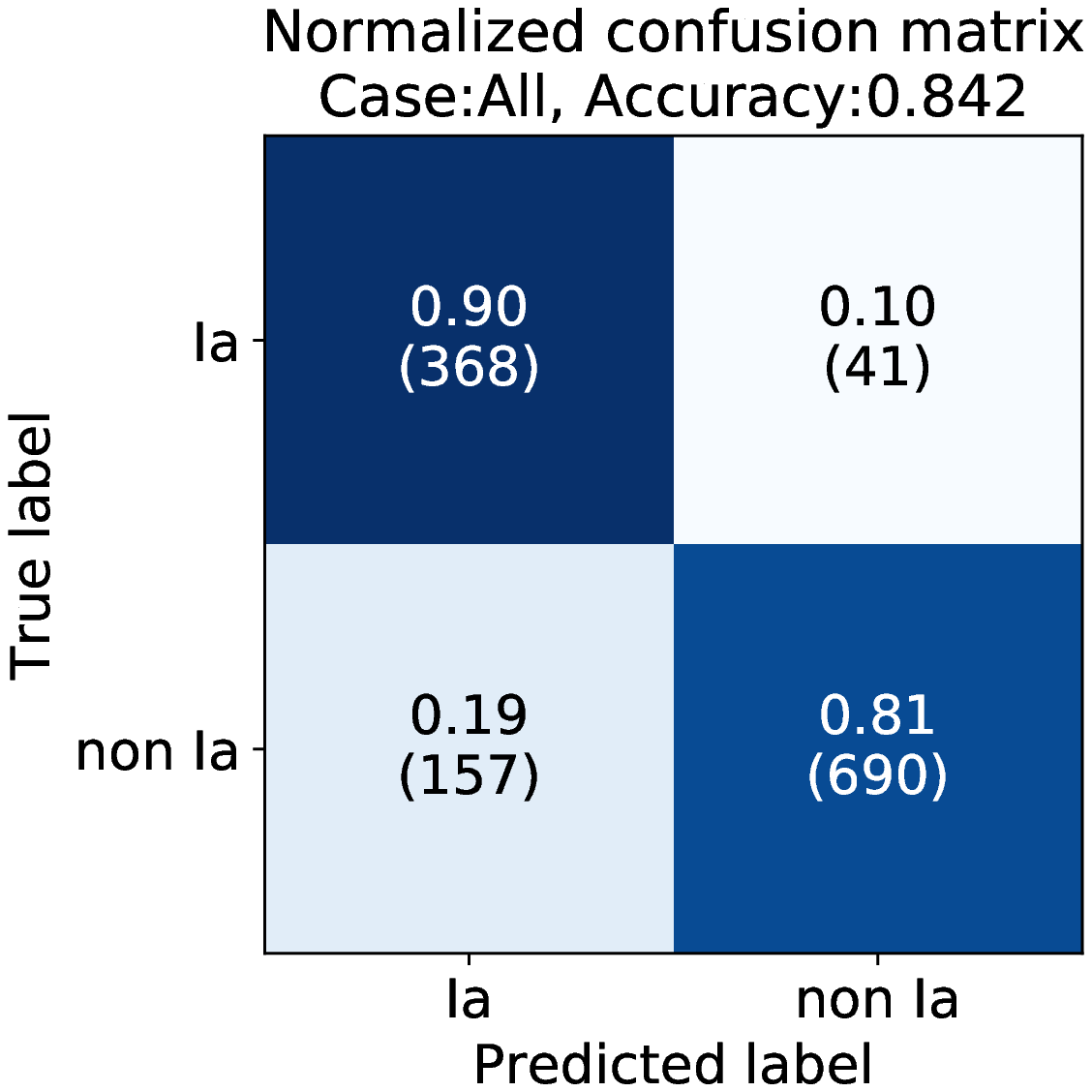}
            \end{center}
        \end{minipage}
        \begin{minipage}{0.5\hsize}
            \begin{center}
                \includegraphics[width=\columnwidth]{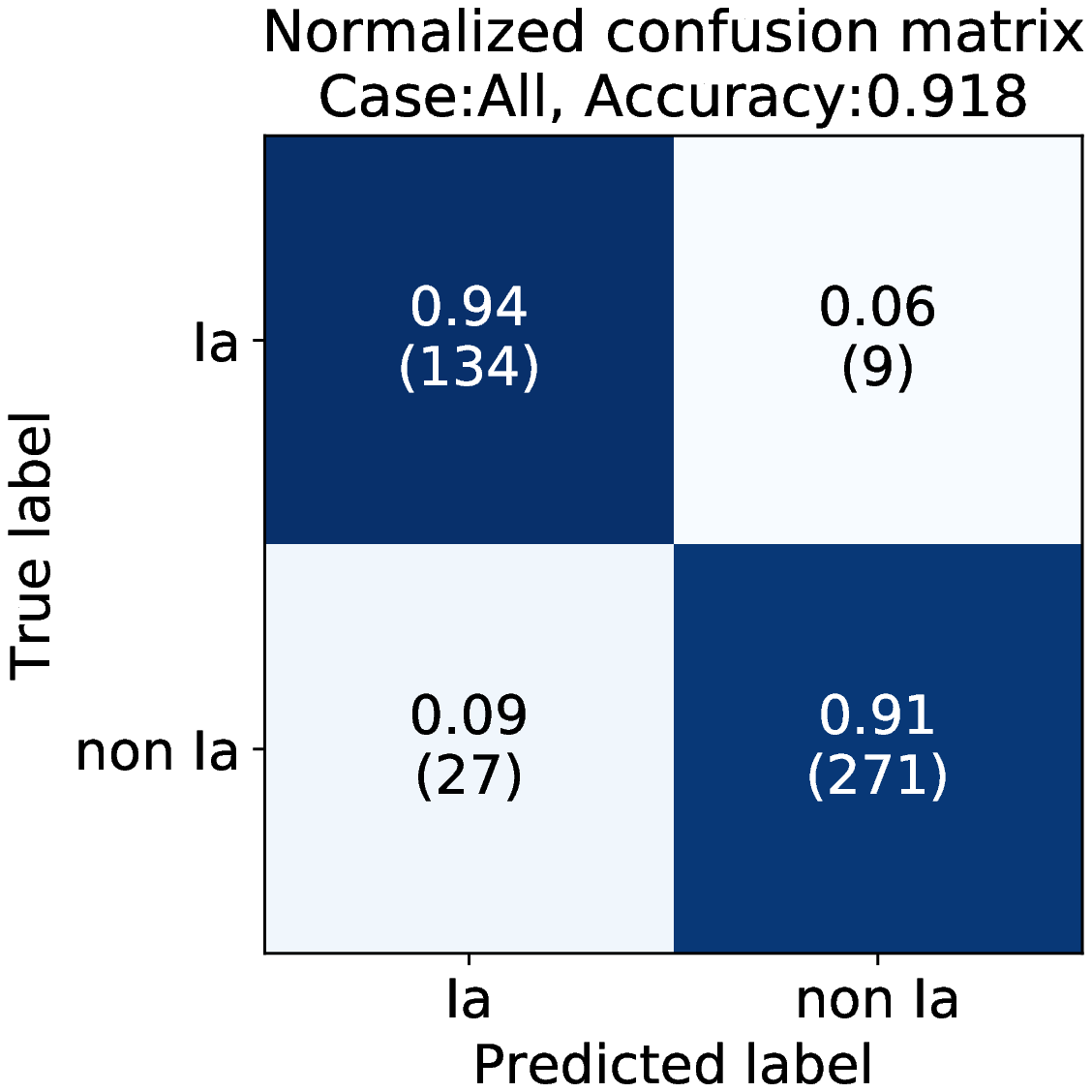}
            \end{center}
        \end{minipage}
    \end{tabular}
    \caption{%
  Normalized confusion matrices for the binary classification of 1256 labeled HSC SNe (left) and the 441 light curve verified SNe (right).
  The inputs for both classifications are the pseudo-absolute magnitude and normalized flux.
}%
    \label{fig:h2_test_CM}
\end{figure*}

We also extracted 441 ``light-curve verified SNe'' that have photometric information before and after their peak and for which spec-z are available, and verified their classification results.
Figures\ \ref{fig:h2_test_all} and \ref{fig:h2_test_gold} show the AUCs of the best classifier for all labeled HSC SNe and the light curve verified HSC SNe respectively.
The confusion matrices for each case are shown in figure\ \ref{fig:h2_test_CM}.
The best performing classifier obtained the same classification results as the conventional method for 84.2\% of 1256 labeled SNe, which is 91.8\% accurate for the 441 light curve verified SNe.
%
%
%
%
\begin{table*}[htbp]
\tbl{AUC of each input in the HSC binary classification.}{
\begin{tabular}{c|ccc|p{3em}p{1.8em}p{1.8em}p{1.8em}p{1.8em}p{1.8em}|p{3em}p{1.8em}p{1.8em}p{1.8em}p{1.8em}p{1.8em}}
\noalign{\vskip 1mm}
\hline
Dataset & \multicolumn{3}{c}{Input\footnotemark[$*$]} & \multicolumn{6}{|c|}{ROC} & \multicolumn{6}{c}{Precision-Recall} \\
\hline
 & $M$ & $m$ & $f$ & Case 0 & 1 & 2 & 3 & 4 & All & Case 0 & 1 & 2 & 3 & 4 & All \\
\hline
Validation &\checkmark &            & \checkmark &       1.000&       0.990 &       0.987 &       0.976 &       0.887 &        0.993 &          1.000 &          0.993 &          0.991 &          0.983 &          0.917 &           0.995 \\
& \checkmark &            &            &       0.999&       0.980 &       0.975 &       0.959 &       0.845 &        0.987 &          0.999 &          0.986 &          0.982 &          0.972 &          0.886 &           0.991 \\
&           & \checkmark & \checkmark &       0.999&       0.983 &       0.979 &       0.963 &       0.817 &        0.988 &          0.999 &          0.988 &          0.985 &          0.973 &          0.863 &           0.992 \\
&           & \checkmark &            &       0.996&       0.971 &       0.966 &       0.938 &       0.790 &        0.980 &          0.997 &          0.980 &          0.976 &          0.955 &          0.841 &           0.985 \\
\hline
Test& \checkmark &            & \checkmark &       0.975 &       0.978 &       0.931 &       0.844 &       1.000 &        0.966 &          0.931 &          0.955 &          0.773 &          0.761 &          1.000 &           0.909 \\
(Light curve verified)& \checkmark &            &            &       0.986 &       0.967 &       0.942 &       0.967 &       0.964 &        0.971 &          0.965 &          0.935 &          0.849 &          0.966 &          0.944 &           0.934 \\
&           & \checkmark & \checkmark &       0.947 &       0.926 &       0.887 &       0.756 &       1.000 &        0.923 &          0.836 &          0.860 &          0.803 &          0.632 &          1.000 &           0.820 \\
&           & \checkmark &            &       0.945 &       0.864 &       0.854 &       0.756 &       0.643 &        0.896 &          0.826 &          0.769 &          0.752 &          0.612 &          0.687 &           0.787 \\
\hline
Test& \checkmark &            & \checkmark &       0.945 &       0.922 &       0.914 &       0.863 &       0.844 &        0.925 &          0.855 &          0.840 &          0.809 &          0.788 &          0.650 &           0.832 \\
(All labeled)& \checkmark &            &            &       0.957 &       0.909 &       0.879 &       0.908 &       0.864 &        0.922 &          0.901 &          0.814 &          0.714 &          0.885 &          0.702 &           0.817 \\
&           & \checkmark & \checkmark &       0.915 &       0.889 &       0.885 &       0.718 &       0.711 &        0.885 &          0.780 &          0.778 &          0.768 &          0.543 &          0.363 &           0.749 \\
&           & \checkmark &            &       0.911 &       0.837 &       0.855 &       0.713 &       0.656 &        0.862 &          0.773 &          0.685 &          0.712 &          0.523 &          0.385 &           0.712 \\
\hline
\end{tabular}
}\label{tab:h2_AUC}
\begin{tabnote}
\footnotemark[$*$] Input to classifier is displayed as a check mark. $M$: pseudo-absolute magnitude, $m$: magnitude, $f$: normalized flux.
\end{tabnote}
\end{table*}
%
%
%
%
%
%
%
%
%
%
%
%

In the binary classification of 1256 labeled HSC SNe, 198 of them were misclassified.
The misclassification rate for each case is different, and tends to increase as the number of input dimensions decreases;
i.e., even though the rate is 13\% for Case 0, it is 23\% for Case 4.
As with the PLAsTiCC data, incomplete events without their peak phase constitute the majority of misclassified events in the HSC data, accounting for 47\% (93/198) of them.
The second most common cause of misclassification is an outlier value or systematic flux offset in photometric data, accounting for 34\% (67/198) of misclassifications.
Of the remaining 38 SNe, 17 are boundary events with a Ia probability of 40 to 60\%, and the remainder are events for which SALT2 fitting is ineffective.

%
%
\subsection{Multi-type classification}
\label{sec:h3}
\begin{figure}[htbp]
  \begin{center}
     \includegraphics[width=\columnwidth]{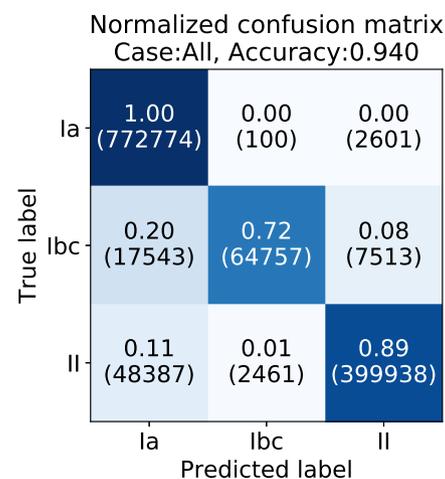}
  \end{center}
  \caption{%
  Normalized confusion matrix for validation dataset in the HSC three-class classification.
  The input is the pseudo-absolute magnitude and normalized flux.
  The proportions in each row sum to 1 (within the rounding error).
  }%
  \label{fig:h3_validation_CM}
\end{figure}

In this paper, we present the classification performance only for the validation dataset with the three-class classifier, because these three types of classification labels are not available for the HSC transients.
%
The accuracy values for each input in the three-class classification of the validation dataset are summarized in table\ \ref{tab:h3_validation}.
The best accuracy for the validation dataset is 94.0\%.
The confusion matrix of the best classifier is shown in figure \ref{fig:h3_validation_CM}.
The result represents that our classifier has a very high sensitivity toward SN~Ia, whereas it is less effective at classifying SN~Ibc.

\begin{figure}[htbp]
  \begin{center}
     \includegraphics[width=\columnwidth]{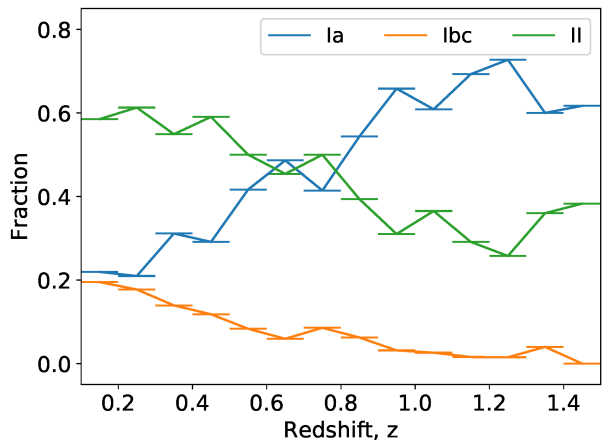}
  \end{center}
  \caption{%
  Type fractions along redshift in HSC three-class classification.
  }%
  \label{fig:hsc3_type_frac_alongz}
\end{figure}

In addition, we describe the predicted classes of actual HSC SNe classified by the three-class classifier.
Figure \ref{fig:hsc3_type_frac_alongz} shows the fractions of each type predicted by the classifier in each redshift from 0.1 to 1.5.
All of the classified HSC SNe were used to calculate the fraction.
These SN types are a combination of the outputs from the two classifiers with different inputs depending on the presence of redshift information: (1) pseudo-absolute magnitude and normalized flux, (2) magnitude and normalized flux.

%
%
%
%
\begin{table}[htbp]
\tbl{Accuracy of each input in the HSC three class classification for validation dataset.}{
\begin{tabular}{ccc|p{3em}p{1.8em}p{1.8em}p{1.8em}p{1.8em}p{1.8em}}
\noalign{\vskip 2mm} 
\hline
\multicolumn{3}{c}{Input\footnotemark[$*$]} & \multicolumn{6}{|c}{Accuracy\footnotemark[${\dagger}$]} \\
\hline
$M$ & $m$ & $f$  &  Case 0 & 1 & 2 & 3 & 4 &  All \\
\hline
\checkmark &            & \checkmark &        0.985 &        0.926 &        0.920 &        0.890 &        0.774 &         0.940 \\
\checkmark &            &            &        0.971 &        0.894 &        0.886 &        0.844 &        0.729 &         0.914 \\
           & \checkmark & \checkmark &        0.970 &        0.907 &        0.897 &        0.860 &        0.724 &         0.921 \\
           & \checkmark &            &        0.952 &        0.871 &        0.861 &        0.818 &        0.701 &         0.892 \\
\hline
\end{tabular}
}\label{tab:h3_validation}
\begin{tabnote}
\footnotemark[$*$] Input to classifier is displayed as a check mark. $M$: pseudo-absolute magnitude, $m$: magnitude, $f$: normalized flux.

\footnotemark[$\dagger$] 
Accuracy of samples extracted from each case according to the fractions in table \ref{tab:class_flag}.
\end{tabnote}
\end{table}
\subsection{Classification of HSC SNe}
We report the classification results of 1824 HSC SNe, obtained by the proposed classifiers, in e-table 1.\footnote{ E-table 1 is available on the online edition as a supplementary table. }
Part of this classification list is provided in table\ \ref{tab:h_results} as an example.
This list summarizes the probabilities predicted by the two-class and three-class classifiers for each SN, along with the redshifts of the host galaxies and the classification labels assigned on the basis of the SALT2 fitting.
The probabilities in this list are calculated from the output of the classifier with the normalized flux added to the input.
Each classification performance shown in subsections \ref{sec:h2} and \ref{sec:h3} is calculated based on the probabilities in this list.
\begin{table*}[htbp]
\tbl{Example of classification result list for HSC SNe.}{
\scriptsize
\begin{tabular}{p{4.5em}p{1.2em}p{4.0em}p{2.1em}|p{0.6em}p{1.8em}p{3.0em}|p{2.9em}|p{1.2em}p{1.2em}p{1.2em}p{0.6em}|p{2.9em}|p{1.2em}p{1.2em}p{1.2em}p{0.6em}}
\noalign{\vskip 1mm}
\hline
Name  &  Case &        {\it z} &  {\it z}\_src\footnotemark[$*$] &  \multicolumn{3}{p{6.0em}}{SALT2 fitting} &\multicolumn{5}{|p{13.5em}}{Classifier (Input\footnotemark[$\dagger$]: $M+f$)} & \multicolumn{5}{|p{12.0em}}{Classifier (Input\footnotemark[$\dagger$]: $m+f$)}\\
\hline
      &       &          &          &  dof & Type\footnotemark[$\ddagger$] & F\_cover\footnotemark[$\S$]  & 2-class &\multicolumn{4}{p{10.0em}|}{3-class}&2-class &    \multicolumn{4}{p{10.0em}}{3-class}\\
\hline
      &       &          &          &      &        &        &    Ia &    Ia &   Ibc &    II  &Type &   Ia &    Ia &   Ibc &    II &Type\\
\hline
HSC16aaau &     1 &    $0.370_{-0.072}^{+0.110}$ &         3 &    7 &    Ia? &   False &    0.556 &    0.554 &    0.022 &    0.424 &      Ia &    0.517 &    0.649 &    0.008 &    0.342 &      Ia \\
HSC16aaav &     1 &    $3.280_{-2.423}^{+0.167}$ &         4 &   17 &  nonIa &     True &    0.134 &    0.049 &    0.002 &    0.949 &      II &    0.279 &    0.356 &    0.048 &    0.596 &      II \\
HSC16aabj &     0 &    $0.361_{-0.008}^{+0.007}$ &         2 &    8 &  nonIa &   False &    0.630 &    0.667 &    0.001 &    0.331 &      Ia &    0.574 &    0.578 &    0.018 &    0.405 &      Ia \\
HSC16aabk &     1 &      -- &         0 &    9 &    Ia? &   False &      -- &      -- &      -- &      -- &     -- &    0.433 &    0.675 &    0.077 &    0.248 &      Ia \\
HSC16aabp &     1 &    $1.477_{-0.032}^{+0.037}$ &         2 &   19 &  nonIa &   False &    0.957 &    0.964 &    0.001 &    0.035 &      Ia &    0.807 &    0.871 &    0.039 &    0.090 &      Ia \\
\vdots & & & & & & & & & & & & & & & &\\
HSC17bjrb &     1 &    $0.560_{-0.036}^{+0.024}$ &         3 &    1 &  UC &   False &    0.003 &    0.007 &    0.004 &    0.989 &      II &    0.011 &    0.002 &    0.004 &    0.994 &      II \\
HSC17bjwo &     0 &    $1.449_{-0.063}^{+0.080}$ &         2 &   26 &     Ia &     True &    0.881 &    0.915 &    0.005 &    0.080 &      Ia &    0.891 &    0.935 &    0.010 &    0.055 &      Ia \\
HSC17bjya &     0 &    $1.128_{-0.000}^{+0.000}$ &         1 &   22 &  nonIa &     True &    0.130 &    0.145 &    0.039 &    0.816 &      II &    0.141 &    0.109 &    0.056 &    0.835 &      II \\
HSC17bjyn &     0 &    $0.626_{-0.000}^{+0.000}$ &         1 &   24 &     Ia &     True &    0.887 &    0.891 &    0.031 &    0.078 &      Ia &    0.965 &    0.918 &    0.007 &    0.075 &      Ia \\
HSC17bjza &     1 &    $1.350_{-0.156}^{+1.142}$ &         4 &   13 &  nonIa &     True &    0.016 &    0.041 &    0.016 &    0.943 &      II &    0.062 &    0.039 &    0.005 &    0.957 &      II \\
HSC17bkbn &     0 &    $0.863_{-0.012}^{+0.036}$ &         2 &   23 &  nonIa &     True &    0.031 &    0.025 &    0.002 &    0.973 &      II &    0.028 &    0.021 &    0.002 &    0.976 &      II \\
HSC17bkcz &     0 &    $0.795_{-0.000}^{+0.000}$ &         1 &   27 &     Ia &     True &    0.675 &    0.674 &    0.035 &    0.291 &      Ia &    0.661 &    0.789 &    0.019 &    0.191 &      Ia \\
HSC17bkef &     0 &    $2.940_{-0.087}^{+0.119}$ &         2 &    0 &   fail &    -- &    0.219 &    0.443 &    0.000 &    0.556 &      II &    0.950 &    0.947 &    0.010 &    0.043 &      Ia \\
HSC17bkem &     2 &    $0.609_{-0.000}^{+0.000}$ &         1 &   17 &     Ia &     True &    0.889 &    0.858 &    0.001 &    0.141 &      Ia &    0.901 &    0.863 &    0.023 &    0.114 &      Ia \\
HSC17bkfv &     0 &    $0.670_{-0.035}^{+0.035}$ &         3 &   23 &     Ia &     True &    0.915 &    0.906 &    0.016 &    0.078 &      Ia &    0.961 &    0.926 &    0.011 &    0.063 &      Ia \\
\vdots & & & & & & & & & & & & & & & &\\
HSC17dskd &     0 &    $0.630_{-0.000}^{+0.000}$ &         1 &    3 &  UC &   False &    0.889 &    0.863 &    0.087 &    0.050 &      Ia &    0.873 &    0.873 &    0.072 &    0.054 &      Ia \\
HSC17dsng &     0 &    $1.331_{-0.048}^{+0.048}$ &         2 &    7 &    Ia? &   False &    0.951 &    0.967 &    0.006 &    0.027 &      Ia &    0.935 &    0.895 &    0.011 &    0.094 &      Ia \\
HSC17dsoh &     0 &    $1.026_{-0.000}^{+0.000}$ &         1 &    2 &  UC &   False &    0.968 &    0.968 &    0.011 &    0.020 &      Ia &    0.911 &    0.923 &    0.022 &    0.055 &      Ia \\
HSC17dsox &     0 &    $1.137_{-0.034}^{+0.041}$ &         2 &    2 &  UC &   False &    0.708 &    0.794 &    0.019 &    0.186 &      Ia &    0.721 &    0.738 &    0.040 &    0.222 &      Ia \\
HSC17dspl &     0 &    $0.624_{-0.000}^{+0.000}$ &         1 &    9 &  nonIa &   False &    0.180 &    0.065 &    0.114 &    0.821 &      II &    0.049 &    0.103 &    0.100 &    0.797 &      II \\
\hline
\end{tabular}
}\label{tab:h_results}
\begin{tabnote}
\footnotemark[$*$] Code for redshift source.
1: spec-z, 2: COSMOS photo-z, 3: HSC photo-z Ultra-Deep, 4: HSC photo-z Deep, 0: hostless.

\footnotemark[$\dagger$] $M$: pseudo-absolute magnitude, $m$: magnitude, $f$: normalized flux.

\footnotemark[$\ddagger$] SN type labeled by SALT2 fitting, UC: unclassified.

\footnotemark[$\S$] Flag indicating whether the photometric data cover the period of 10 days before and 20 days after the peak. SNe with this flag set to False are defined as ``incomplete events.''
\end{tabnote}
\end{table*}
\subsection{Dependence on the number of epochs}
\begin{figure}[htbp]
  \begin{center}
     \includegraphics[width=\columnwidth]{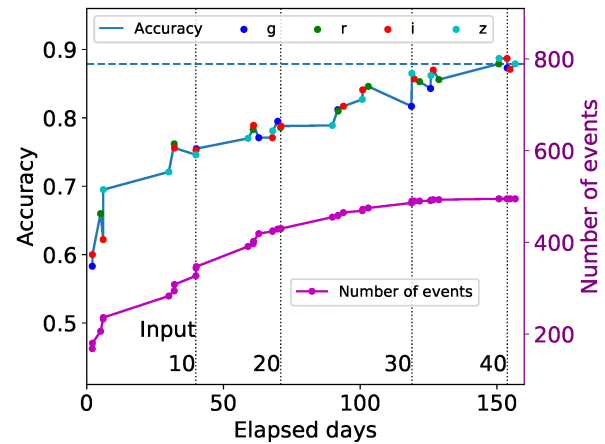}
  \end{center}
  \caption{%
  Relationship between the number of epochs and classification performance in binary classification for the Case 0 dataset. 
  The horizontal axis represents the number of elapsed days of the HSC survey, and the vertical dotted line indicates the scale of the number of photometric points that were used as input. 
  The color of each mark in accuracy indicates the band of the added photometric point. 
  The blue horizontal dotted line indicates the accuracy when using all epochs.
  }%
  \label{fig:n_observations}
\end{figure}

When using our classification method, the number of photometric data points given to the classifier increases as the survey progresses.
Therefore, we investigated the transition of performance against the number of epochs.
This was accomplished by classifying the HSC dataset by increasing the number of input data points in increments of one, and by examining the relationship between the number of epochs and the classification performance.
Binary classifiers were adopted for classification, and the accuracy calculated from each confusion matrix was used for evaluation.
Figure\ \ref{fig:n_observations} shows the transition of classification performance for the Case 0 HSC dataset along with the number of epochs.
Although the Ia accuracy is as low as 0.6 to 0.7 in the early stage of the survey with less than five epochs, it exceeds 0.8 when the number of epochs increases to 22.
The partial decrease in accuracy is thought to be due to the new SNe being found upon the addition of a photometric point.

\begin{figure*}[htbp]
    \begin{tabular}{ccc}
        \begin{minipage}{0.33\hsize}
            \begin{center}
                \includegraphics[width=\columnwidth]{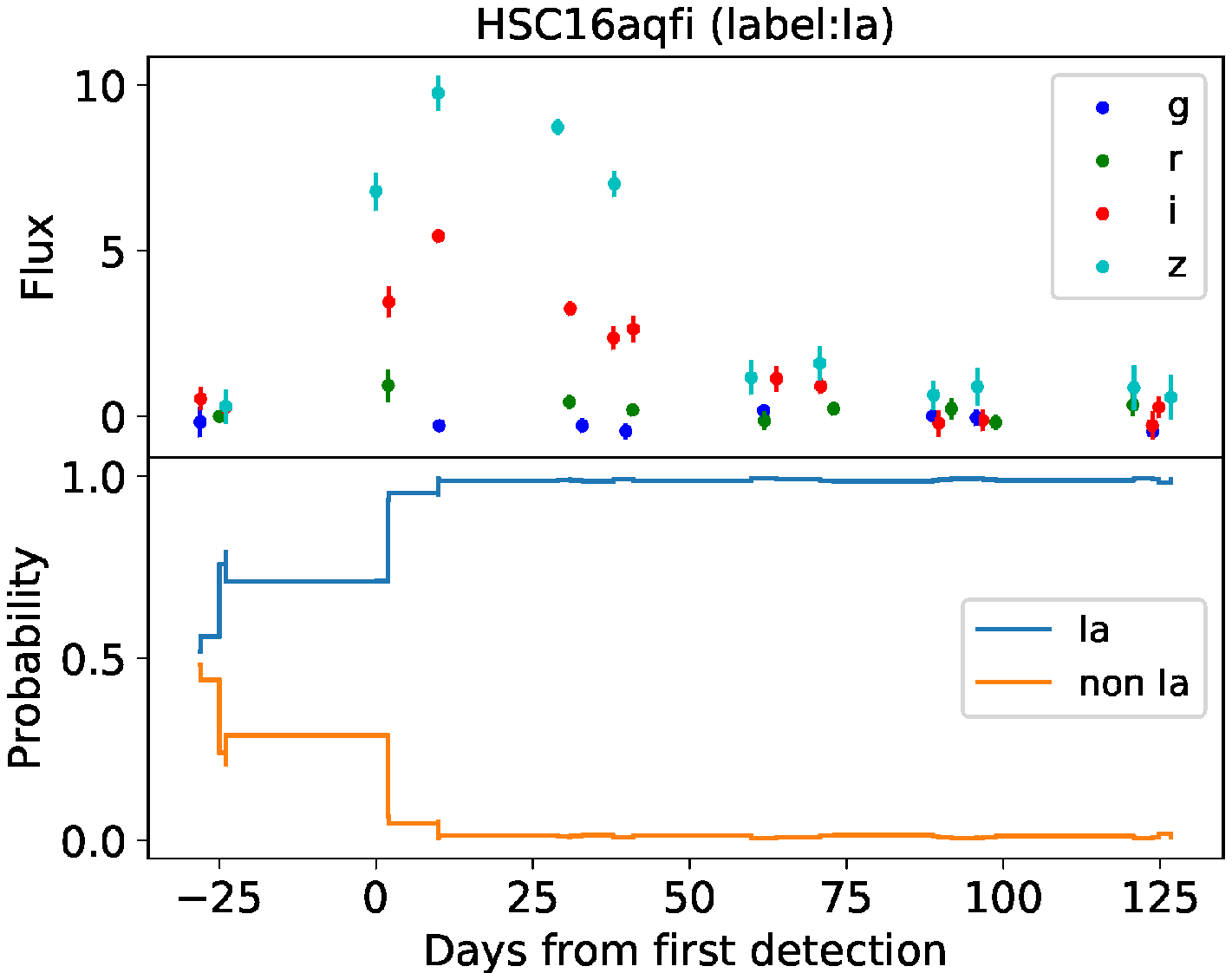}
            \end{center}
        \end{minipage}
        \begin{minipage}{0.33\hsize}
            \begin{center}
                \includegraphics[width=\columnwidth]{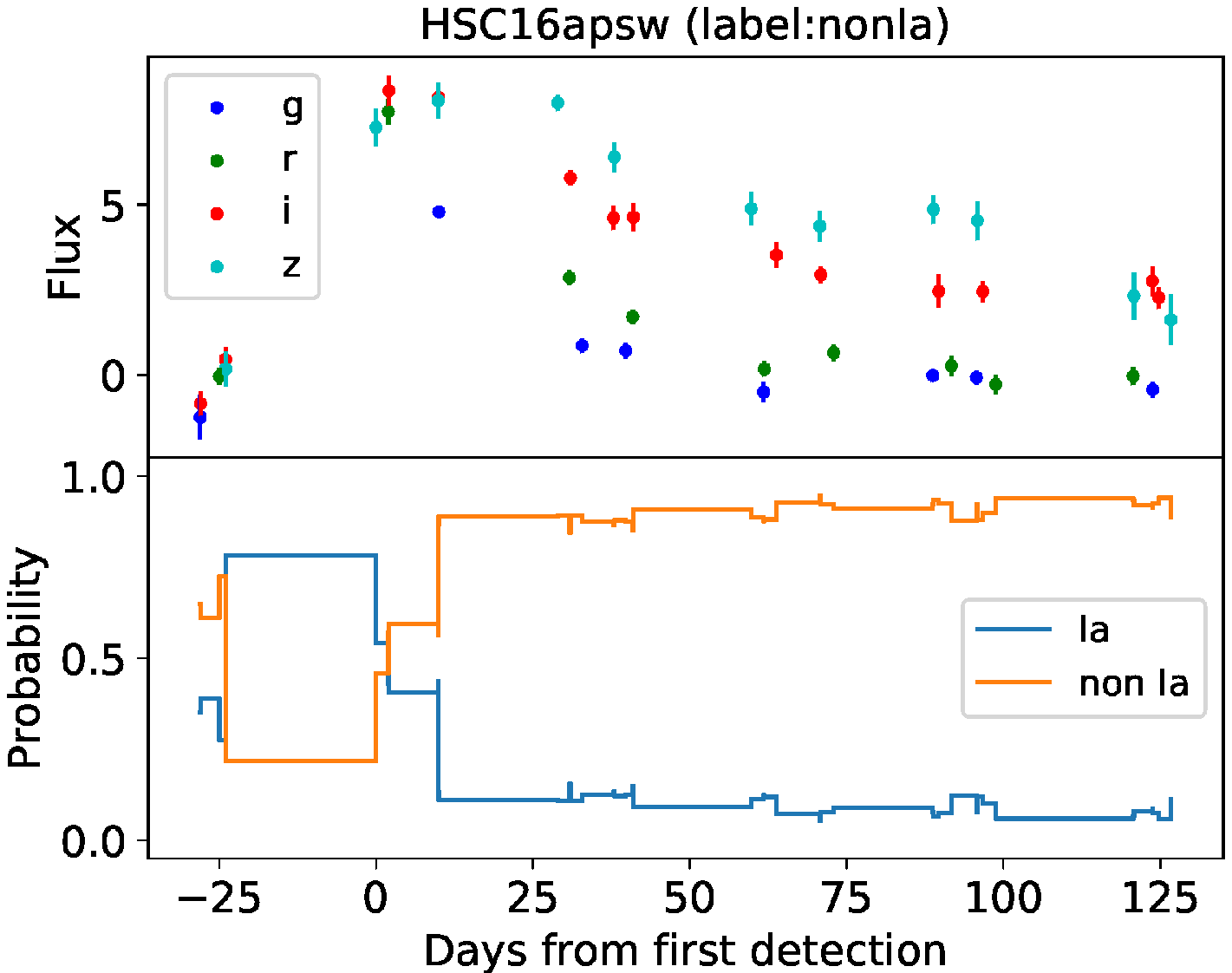}
            \end{center}
        \end{minipage}
        \begin{minipage}{0.33\hsize}
            \begin{center}
                \includegraphics[width=\columnwidth]{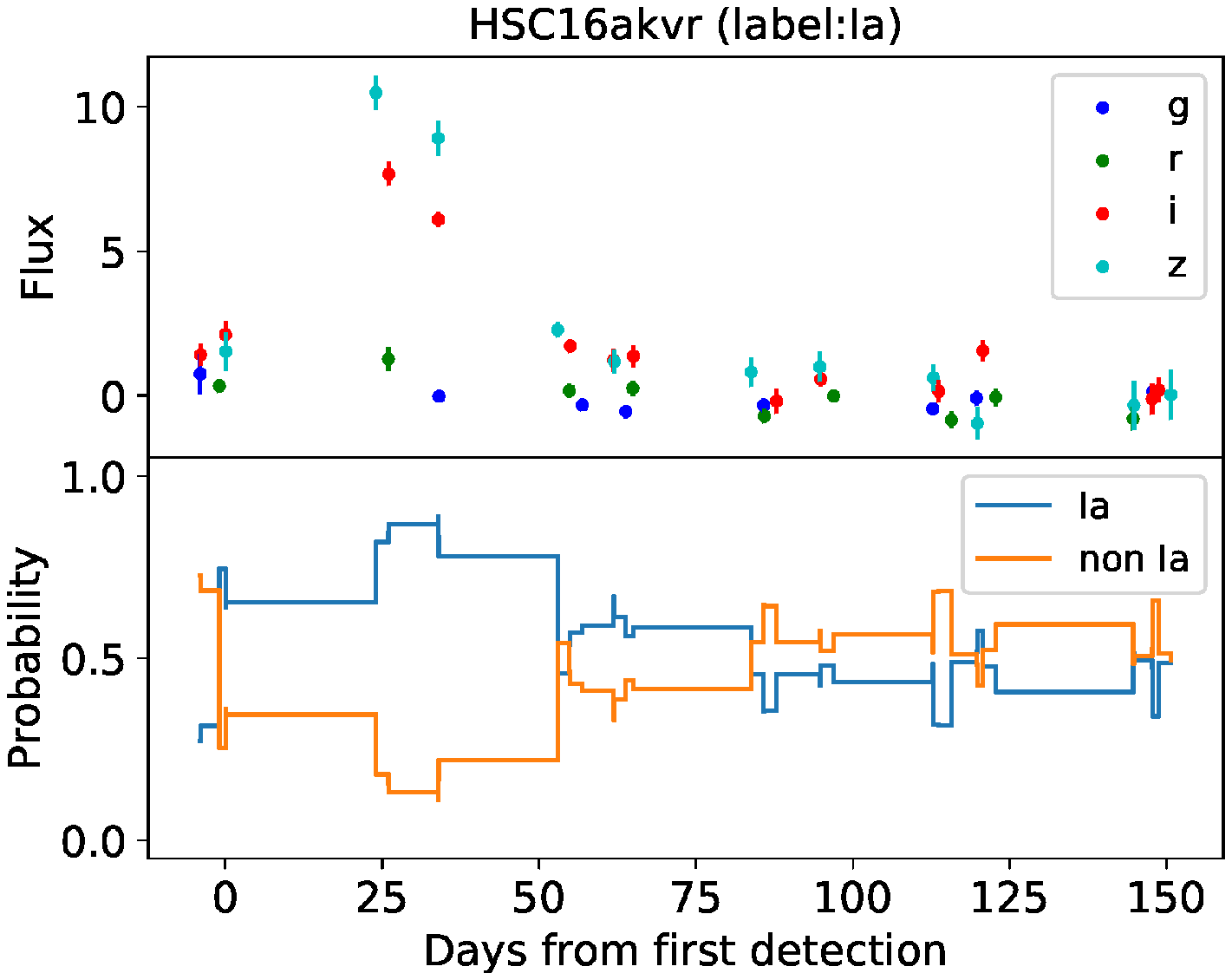}
            \end{center}
        \end{minipage}
    \end{tabular}
    \vspace{3mm}
    \caption{%
    Examples of light curves and probability transitions. The title of each plot shows the name of the SN in the HSC survey and the label classified by SALT2 fitting.
    }%
    \label{fig:lcps}
\end{figure*}

\begin{figure}[htbp]
  \begin{center}
     \includegraphics[width=\columnwidth]{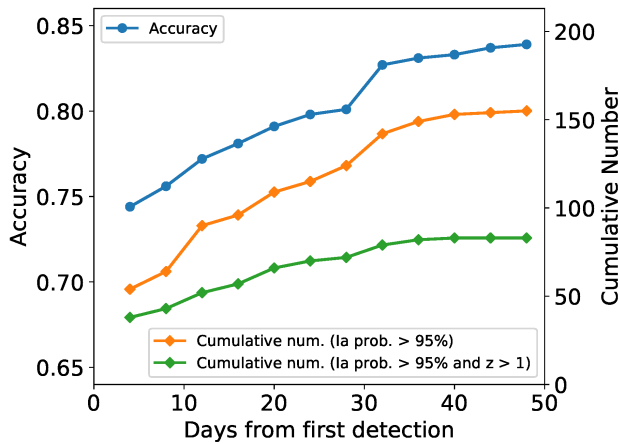}
  \end{center}
  \caption{%
  Classification accuracy and cumulative number of Type Ia SNe classified with high probability against the SN phase.
  The orange line indicates the cumulative number of SNe with Ia probability $>$ 95\%, and the green line is that for distant SNe at {\it z} $>$ 1. 
  }%
  \label{fig:n_observations_SNphase}
\end{figure}

We also investigated the classification performance during each SN phase by regrouping all events according to the length of time since the first detection.
Figure\ \ref{fig:lcps} illustrates the light curves and Ia probability transitions since the first detection of the three types of HSC SNe.
We define ``first detection'' as the first day when the SN is detected with 5$\sigma$ confidence in flux, and which is flagged as a real object by the real-bogus classifier using a convolutional neural network \citep{yasuda19a}.
The probability is updated at each new epoch.
Although the probability increases for certain events as the SN phase progresses, in the case of other events the probabilities fluctuate around 0.5 even as the observation progresses and these events cannot be clearly classified.
Figure\ \ref{fig:n_observations_SNphase} shows the accuracy of SN~Ia classification as a function of the time since the first detection. Orange and green curves show cumulative numbers of SNe~Ia.
The calculations for each performance are based on the classification results for 1161 SNe that were detected before the rising phase.
This figure presents the time span of SN photometric data that is needed for highly accurate classification using our classifier.
The classification accuracy is 78.1\% for the first two weeks of data, and after one month it increases to 82.7\%.
In addition, the number of follow-up candidates identified by the classifier can be estimated from the cumulative number in figure \ref{fig:n_observations_SNphase}.
Using data acquired within one month from the first detection, 79 SNe with {\it z} $>$ 1 could be classified with Ia probability of 95\% or more.
Because the number of these SNe is a cumulative number observed during a period of six months, dividing this by six corresponds to the number of follow-up SNe classified during a one-month survey, which is 13 events.
%

\begin{figure}[htbp]
  \begin{center}
     \includegraphics[width=\columnwidth]{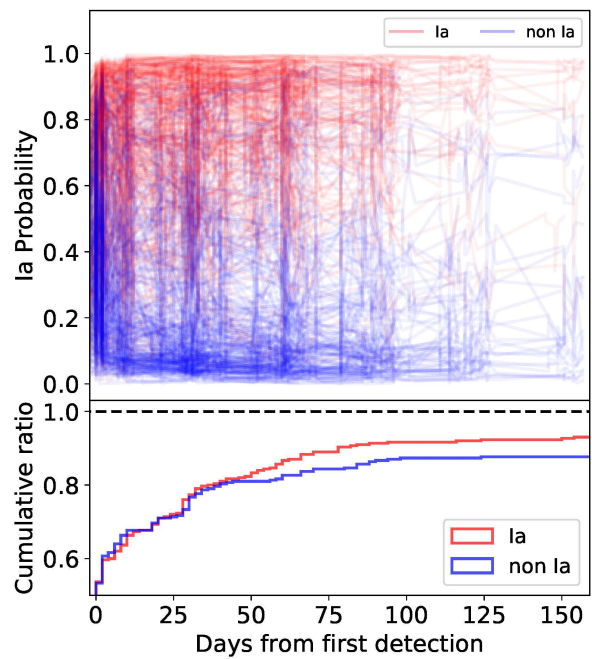}
  \end{center}
  \caption{
  Upper panel: Transition of Ia probability of SNe after first detection. Each line corresponds to one of the SNe. Different colors indicate different labels, with red being Ia and blue being non-Ia. Lower panel: Cumulative ratio for the epoch at which SNe are finally correctly classified.
  }%
  \label{fig:visualized_Ia_prob}
\end{figure}

Lastly, we studied the evolution of the SN~Ia probability along the SN phase.
Each time the number of input epochs increases, the SN~Ia probability, i.e., the output of the classifier, is updated.
In figure \ref{fig:visualized_Ia_prob} (Upper panel), we plotted the SN~Ia probability at each epoch for each of 300 SNe labeled as Ia and non-Ia based on the SALT2 fitting. The use of this result enabled us to measure the last epoch at which the correct type is classified, i.e., the epoch after which no further change in the classification occurs. The lower panel shows the cumulative ratio for the epoch. The figure shows that the classification performance improves with time and that 80\% of supernovae are correctly classified approximately 30 days after their first detection. In this figure, the initial cumulative ratio is lower than the accuracy shown in figure \ref{fig:n_observations_SNphase} because certain SNe that are initially correctly classified could ultimately be misclassified as a wrong type.

\begin{figure}[htbp]
  \begin{center}
     \includegraphics[width=\columnwidth]{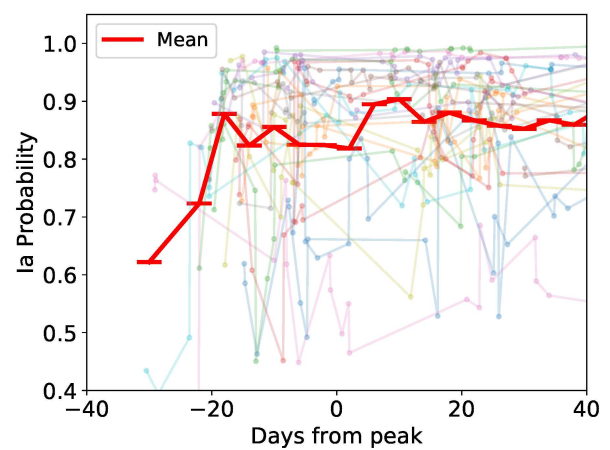}
  \end{center}
  \caption{%
  Ia probability transitions of 26 HST targets. Each of the lighter lines represents the variation for individual HST targets, and the red line is the average of these lines.
  }%
  \label{fig:HSTIaprob}
\end{figure}

%
Figure \ref{fig:HSTIaprob} shows the transitions of SN~Ia probability as a function of the number of days from the peak for 26 SNe selected as HST targets in the HSC survey, and the average of these transitions.
The SN~Ia probability for the average of the candidates is greater than 0.8 three weeks before the peak.
This means that our classifier accomplishes the task described in subsection \ref{sec:tasks} by identifying good candidate SNe even when it only has information acquired before the SN peak.
%
%
\section{Discussion}
\subsection{Factors affecting classifier performance}
%
In this study, we applied our classifier to actual HSC survey data to evaluate its classification performance.
By classifying actual data with our method, we determined that the performance difference between the validation and test sets is larger than that for the PLAsTiCC data.
Here we discuss the factors that affect the classification performance of actual data.

One possible factor is the uncertainty in the labeling of actual data using conventional methods.
In the confusion matrix for all labeled SNe in figure \ref{fig:h2_test_CM}, 47\% of the 198 misclassified SNe are incomplete events and these events only form 35\% of the 1824 HSC SNe.
The high percentage of incomplete events among the misclassified events suggests that incomplete events reduce the classification performance. 
However, the misclassification rate of incomplete events is clearly higher at 28\% (93/338) for the HSC data compared to 7\% (58/871) for PLAsTiCC data.
In addition, SALT2 fitting is equally ineffective at fitting incomplete events owing to its specification, and errors in the fitting parameters are significantly larger.
These findings suggest a degree of uncertainty in labeling with conventional classification methods.
Another source of uncertainty in labeling is the uncertainty in the redshift information used in SALT2 fitting.
For at least 16 misclassified events that were not affected by any other misclassification factors, the fitting color parameters are far outside the criteria for Ia, and the wrong redshift was probably used for fitting.
In fact, as shown in the panel on the right in figure \ref{fig:h2_test_CM}, the performance for light curve verified SNe, except for incomplete events and for events of which spec-z is not known, is superior to that of all classified events.

Another reason for the large difference in the performance of the validation and test set is the inability to perfectly simulate the observed data.
Outlier values and systematic flux offsets, which are considered to be one of the causes of misclassification as described in subsection \ref{sec:h2}, are found only in real observed data.
As described in \citet{yasuda19a}, the photometric data of the SNe in the HSC survey are measured from the difference image obtained by subtracting the reference image from the observed image.
We believe that the unsimulated residual in this subtraction created a difference between the photometric data that were used for training and observation, and increases the misclassification rate of the classifier when processing actual data.
For example, the light curve of SN HSC16akvr shown in the panel on the right in figure \ref{fig:lcps} fluctuates at its tail, and in this part of the curve the classification is not clear.
Improving the performance of the classifier would therefore necessitate the reduction or simulation of these outlier values as much as possible.
%
%
%
\subsection{Comparison with other classifiers}
\label{sec:comparison}
The direct comparison of the classification results of the classifiers is complicated by differences in methods, training datasets, or the number of inputs and outputs. 
Here, we simply compare the results obtained with the recent SN type classifier based on machine learning and our classifier with AUC of ROC in binary classification to determine whether an SN is of the type Ia.
For comparison, we use the classification results of the simulated SN light curves with redshift information from \citet{Lochner_2016}, \citet{charnock17a}, and \citet{Muthukrishna_2019}.
\citet{Lochner_2016} obtained an AUC of 0.984 by using boosted decision trees (BDTs) for classification using the SALT2 fitting parameters as input features.
\citet{charnock17a} reported an AUC of 0.986 for the classification of SNPCC data using deep recurrent neural networks (RNN) with unidirectional long short-term memory (LSTM) units.
\citet{Muthukrishna_2019} used deep RNN with Gated Recurrent Units (GRUs) to classify simulated ZTF light curves and achieved an AUC of 0.99 40 days after the trigger.
The AUC of our classifier, 0.996, is comparable to those of these recent classifiers in binary classification.

\begin{figure*}[htbp]
    \begin{tabular}{cc}
        \begin{minipage}{0.5\hsize}
            \begin{center}
                \includegraphics[width=\columnwidth]{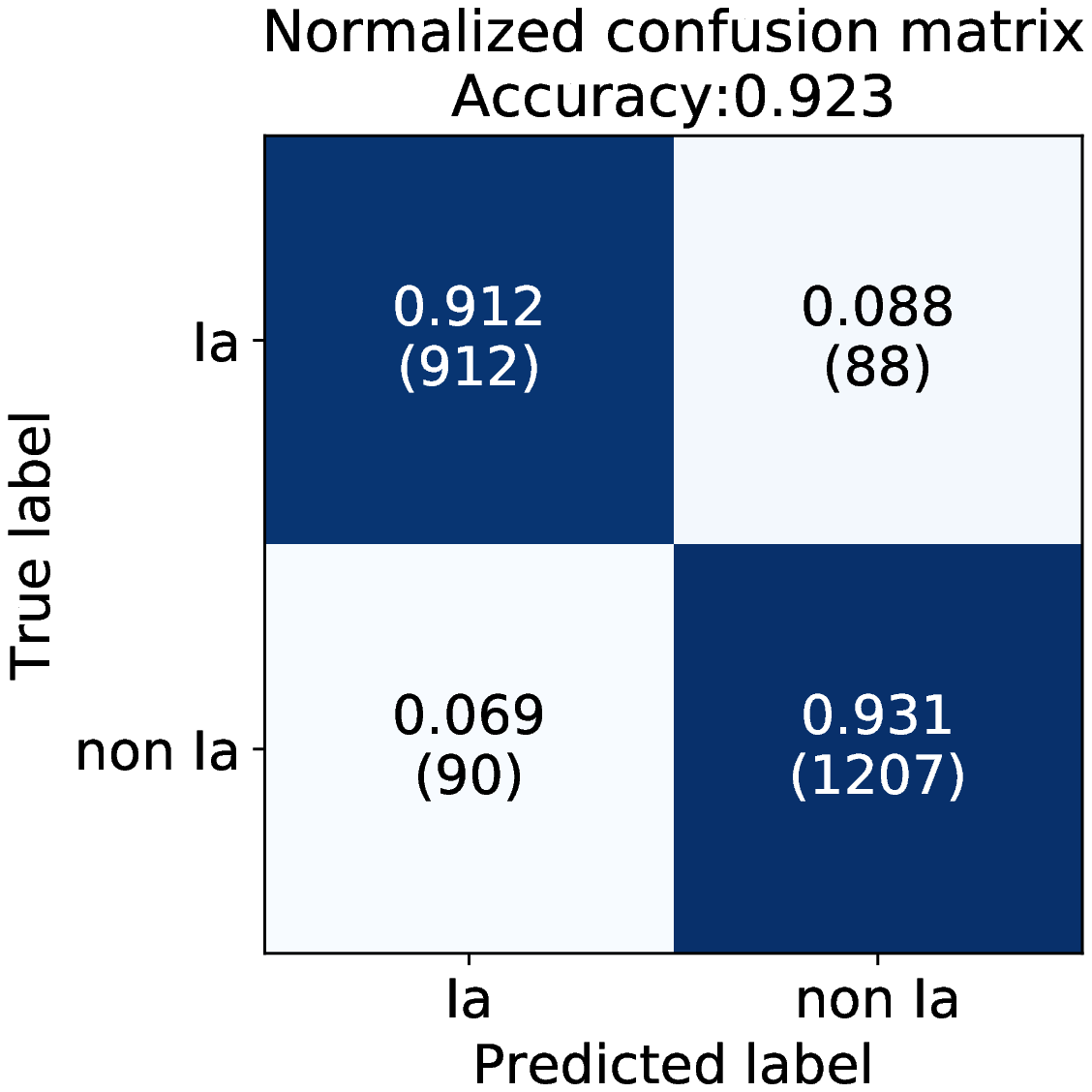}
            \end{center}
        \end{minipage}
        \begin{minipage}{0.5\hsize}
            \begin{center}
                \includegraphics[width=\columnwidth]{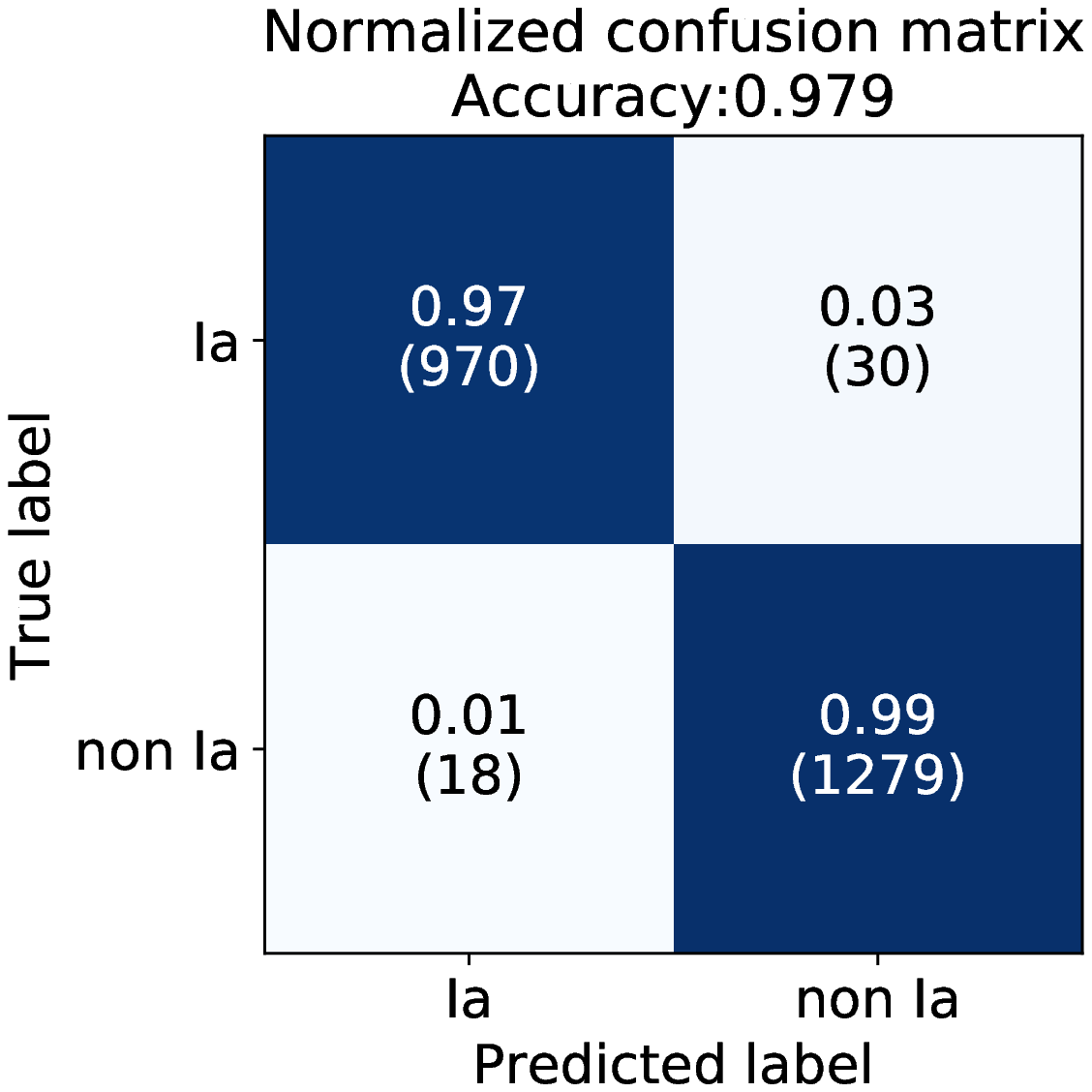}
            \end{center}
        \end{minipage}
    \end{tabular}  \caption{%
    Normalized confusion matrices of two-class classification results for the PLAsTiCC test set by \citet{boone19a} (left) and our best classifier given the pseudo-absolute magnitude and normalized flux (right).
    }%
    \label{fig:comp_plasticc_1st}
\end{figure*}

Next, we compare the results of our classifiers with those that ranked first in the PLAsTiCC Kaggle competition \citep{malz19a}.
The best classifier in the competition \citep{boone19a} was based on Light-GBM and trained with features extracted from photometric data modeled by Gaussian process regression.
The dataset that was used to train this classifier included a total of 591,410 light curves that were obtained by augmenting 100 new light curves under different observation conditions and different redshifts for each light curve of the original PLAsTiCC training set.
The classifier classifies events into 15 classes including variable objects other than SN such as microlensing events and active galactic nuclei.
We used 2,297 PLAsTiCC SN predictions, classified by both classifiers, for comparison.
Because the number of output classes is different between our classifier and the best classifier, we divided the classification results into SN~Ia and other, and performed the comparison as a binary classification.
Figure\ \ref{fig:comp_plasticc_1st} shows the confusion matrix of each classifier.
Although a strict comparison between our classifier and the best PLAsTiCC classifier is not possible because of the different training datasets and number of output classes, the capability of our classifier is comparable to that of the best PLAsTiCC classifier.
However, because our method fixes the observation schedule to the type of input, it is impossible to process all PLAsTiCC data with one classifier.
Our method is not necessarily suitable for surveys that sweep a wide area such as the LSST survey; instead, it is useful for surveys that observe the same field for a period of time, such as the HSC-SSP Transient Survey and LSST DDF.
\subsection{Data uncertainty}
Information about the flux error is also important for the classification of distant SN Ia, which is the main target of the HSC survey and has a small signal-to-noise ratio.
As mentioned in section \ref{sec:p}, the signal-to-noise ratio affects the performance of the model.
Here, we discuss the incorporation of errors for classification, including in our method.
\citet{charnock17a} added the flux error itself as part of the input vector.
In the case of feature-based classification, the flux error is used in the fitting process to calculate the features, and affects those errors.
The flux error is also used to calculate statistical features such as ``Standard Deviation / Mean,'' which is one of the features discussed by \citet{narayan18a} and \citet{Muthukrishna_2019}.
As described in subsection \ref{sec:training}, we calculate the flux error of the simulated training data based on the relationship between the flux and the flux error for each epoch of the observed data.
Then, random numbers that follow a normal distribution defined by the original fluxes and flux errors are input to the classifier as processed fluxes.

The processing of time series data with uncertainty is problematic in many application domains, not only in the field of astronomy.
For example, in the similarity matching of time series data with uncertainty,  considering continuous correlation in time series such as measuring the distance after filtering by the Uncertain Moving Average (UMA) \citep{Dallachiesa_2012} is suggested to be a promising approach.
UMA filtering uses the correlation of neighboring points and reduces the contribution of observation points with large errors.
We consider the development of a method to efficiently incorporate this error information to be necessary for improving the classification performance.
\subsection{Missing data}
Our method was developed specifically for the HSC survey, where each observed transient tends to have the same cadence by observing a certain field repeatedly.
However, this method is neither suitable for processing data with different observation schedules for each object nor for datasets from which data are missing.
In other classification methods, missing data are interpolated by linear or Gaussian processes \citep{Lochner_2016,Muthukrishna_2019}, or replaced with reference to the flux of adjacent photometric points \citep{charnock17a}.
Our approach is to handle missing data by preparing multiple classifiers.
The advantage of our classifier is that it is able to use the photometric information directly for classification with less pre-processing, compared to methods that use interpolation or replacement.
In the case of the HSC survey, most of the observation schedules of all supernovae can be roughly divided into two types that correspond to either the Ultra-Deep or Deep layers.
Therefore, we can classify 99.3\% of SNe using only five classifiers with different input schedules, including those for SNe data containing missing data (table \ref{tab:class_flag}).
%
%
%
\subsection{Type fractions of HSC SNe}
The elemental abundance of the solar system \citep{grevesse98a} originates from the cosmic history of SNe \citep{maraston05a,kobayashi00a}.  
Recent studies showed that the solar abundance pattern is observed in other systems \citep{ramirez09a} and clusters \citep{mernier18a}. 
Investigation of the origin of elements in the context of cosmic evolution is thus important \citep{fukugita04a}.

It is now well established that the star formation rate peaks at $z\sim2$, and that the chemical composition of our system is a mixture of SN~Ia and Core-Collapse SNe \citep{tsujimoto95a,kobayashi11a}.
Deriving the SN rate would require careful analysis \citep{dilday08a,brown19a,frohmaier19a} and is beyond the scope of this paper, but at least we can verify the consistency with previous work in terms of the relative fraction.
The Lick Observatory Supernova Survey \citep{li11a} reported the relative ratio of SN~Ia:Ibc:II to be 0.24:0.19:0.57, whereas we obtained 0.22:0.19:0.59 even at $z\lesssim0.2$.

Based on our survey depth, we have a complete sampling of SN~Ia up to $z\sim1.1$ although we lose the completeness of Core-Collapse SN in much lower redshift given the fact that the magnitudes of SN~Ibc and SN~II are fainter by 2$\sim$3 mag at maximum.
Increasing redshift causes the SN~II fraction to decrease whereas the SN~Ia fraction increases as shown in figure \ref{fig:hsc3_type_frac_alongz}. 
This completeness effect is simply due to the magnitude difference and does not reflect the cosmological SN rates.
By adopting the SN~Ia rate from \citet{graur14a} and the Core-Collapse SN rate from \citet{strolger15a}, we can estimate the completeness of Core-Collapse SN. 
At $z\sim0.3$, the Core-Collapse SN completeness is 78\%, and it is reduced to 49\% at $z\sim0.5$. 
The reason for the SN~II fraction not approaching zero at $z\sim1$ in figure \ref{fig:hsc3_type_frac_alongz} is that the magnitude of the dispersion of Core-Collapse SN \citep[$\sigma$ $\sim$ 1.2 mag]{li11a,kessler19b} is much larger than that of SN~Ia \citep[$\sigma$ $\sim$ 0.5 mag]{rubin15a}, and they are more abundant at $z\sim1$ by a factor of 4--5 \citep{madau98a,hounsell18a}. 
Although a more careful investigation would be necessary, we deduce the completeness of Core-Collapse SN at $z\sim1$ to be 12\%.

%
%
%
\section{Conclusions}
%
In this paper, we present a model of a classifier that classifies SN types by directly accepting photometric data as input. The classification performance of the classifier was discussed in detail.
Our DNN classifier is trained with simulated SN photometric data and was shown to classify PLAsTiCC data and actual HSC SNe data with high accuracy of 95.3\% and 84.2\%, respectively.
Our study of the number of input dimensions also indicated that our classifier can classify the HSC survey data with sufficient accuracy by even using two weeks of pre-peak data since the first detection.
Based on these results, we concluded that this classifier has sufficient classification performance for subsequent type-specific studies and for the selection of follow-up targets even in actual HSC surveys.
%
\begin{ack}
The Hyper Suprime-Cam (HSC) collaboration includes the astronomical communities of Japan, Taiwan, and Princeton University. The HSC instrumentation and software were developed by the National Astronomical Observatory of Japan (NAOJ), the Kavli Institute for the Physics and Mathematics of the Universe (Kavli IPMU), the University of Tokyo, the High Energy Accelerator Research Organization (KEK), the Academia Sinica Institute for Astronomy and Astrophysics in Taiwan (ASIAA), and Princeton University. Funding was contributed by the FIRST program from the Japanese Cabinet Office, the Ministry of Education, Culture, Sports, Science and Technology (MEXT), the Japan Society for the Promotion of Science (JSPS), the Japan Science and Technology Agency (JST), the Toray Science Foundation, NAOJ, Kavli IPMU, KEK, ASIAA, and Princeton University.

This paper makes use of software developed for the Large Synoptic Survey Telescope. We thank the LSST Project for making their code available as free software at  http://dm.lsst.org

The Pan-STARRS1 Surveys (PS1) have been made possible through contributions of the Institute for Astronomy, the University of Hawaii, the Pan-STARRS Project Office, the Max-Planck Society and its participating institutes, the Max Planck Institute for Astronomy, Heidelberg and the Max Planck Institute for Extraterrestrial Physics, Garching, The Johns Hopkins University, Durham University, the University of Edinburgh, Queen’s University Belfast, the Harvard-Smithsonian Center for Astrophysics, the Las Cumbres Observatory Global Telescope Network Incorporated, the National Central University of Taiwan, the Space Telescope Science Institute, the National Aeronautics and Space Administration under Grant No. NNX08AR22G issued through the Planetary Science Division of the NASA Science Mission Directorate, the National Science Foundation under Grant No. AST-1238877, the University of Maryland, and Eotvos Lorand University (ELTE) and the Los Alamos National Laboratory.

We thank Y. Imoto for his enormous contribution to the development of the classification model, and the data classification.
We also thank T. J. Moriya, J. Jiang and other members of the HSC transient working group for helpful discussions and comments on the manuscript.

This work is supported by JST CREST Grant Number JPMHCR1414, JST AIP Acceleration Research Grant Number JP20317829, MEXT KAKENHI Grant Numbers 18H04345 (N.Ya.), 17H06363 (M.T.), and JSPS KAKENHI Grant Numbers 18K03696, 20H04730, 20HT0063 (N.S.), 16H02183 (M.T.), 19H00694 (M.T.), 20H00179 (N.T.).

This research is based in part on data collected at the Subaru Telescope and retrieved from the HSC data archive system, which is operated by the Subaru Telescope and Astronomy Data Center at NAOJ.
\end{ack}
\appendix 
\section*{Redshift estimation}
\label{sec:est_redshift}
The HSC-SSP Transient Survey includes SNe without redshift because their host galaxies are not clearly identified. These SNe are referred to as hostless SNe.
In the redshift list of HSC SNe that was used for type classification, 6\% (108/1824) corresponds to these hostless SNe.
To be able to process these SNe with our classifier and to address the possibility of host galaxy misidentification,
we could estimate the redshift $z$ of a SN from the photometric data.

The redshift was estimated by using a model with the same structure as that in figure \ref{fig:dnn_model} except that the DNN output is a scalar, and that the model does not include a final softmax layer for redshift estimation.
The objective function to optimize is the squared error between the ground truth $z$ and the output value $\hat{y}$.
We measured the accuracy of the model with the coefficient of determination $R^2$; that is,
\begin{eqnarray*}
    R^2 = 1 - \frac{\sum_n \left| z_n - \hat{y}_n \right|^2}{\sum_n \left| z_n - \bar{z} \right|^2}, 
\end{eqnarray*}
where $z_n$ is the redshift value of the $n$-th sample, $\hat{y}_n$ is the output of the $n$-th sample, 
and $\bar{z}$ is the mean redshift of the dataset.
We performed the same hyperparameter search in the model as that described in subsection \ref{hyperparametersearch}, including data augmentation.
The accuracy of this estimator is discussed below.

\begin{figure*}[htbp]
    \begin{tabular}{cc}
        \begin{minipage}{0.5\hsize}
            \begin{center}
                \includegraphics[width=\columnwidth]{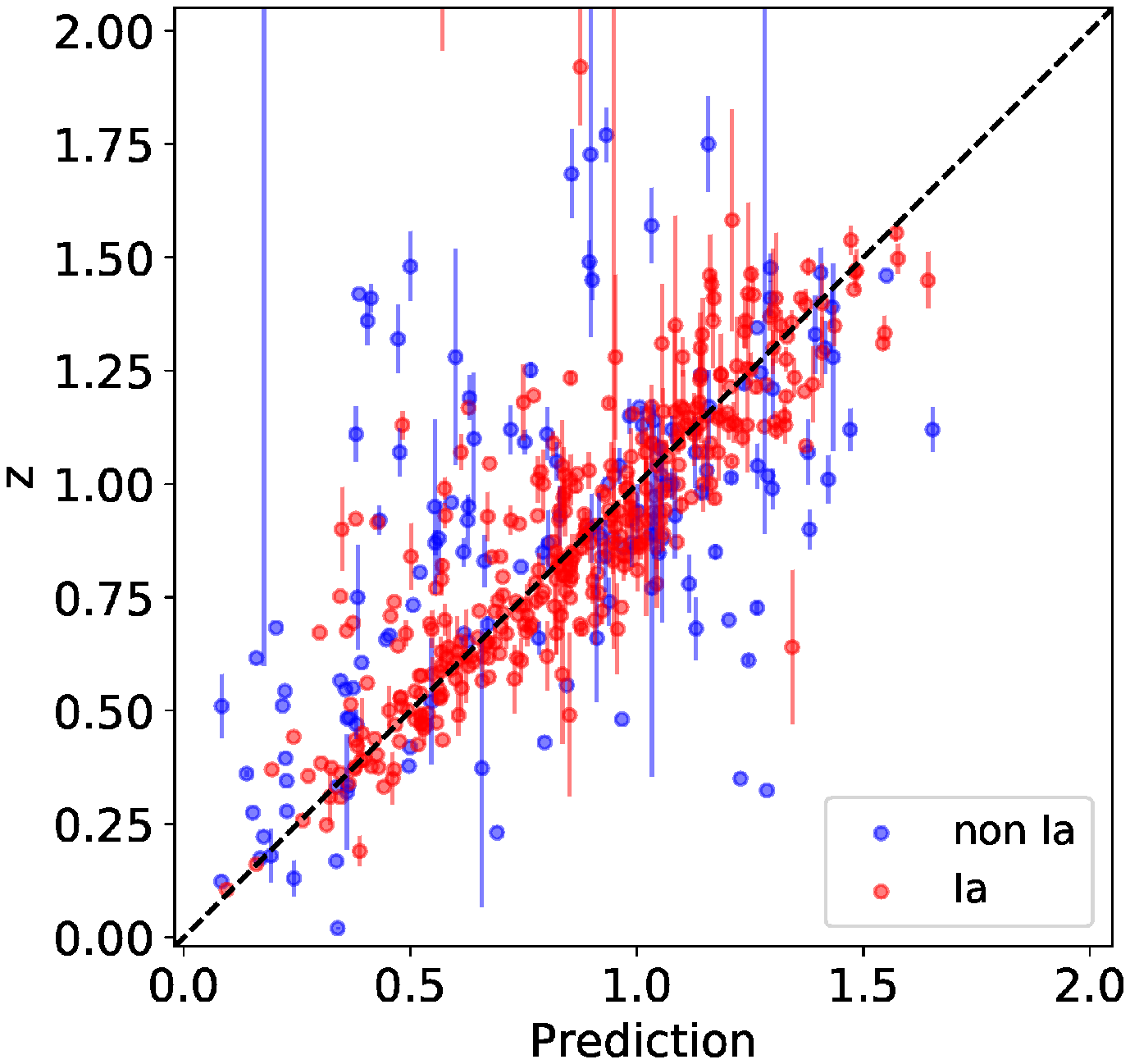}
            \end{center}
        \end{minipage}
        \begin{minipage}{0.5\hsize}
            \begin{center}
                \includegraphics[width=\columnwidth]{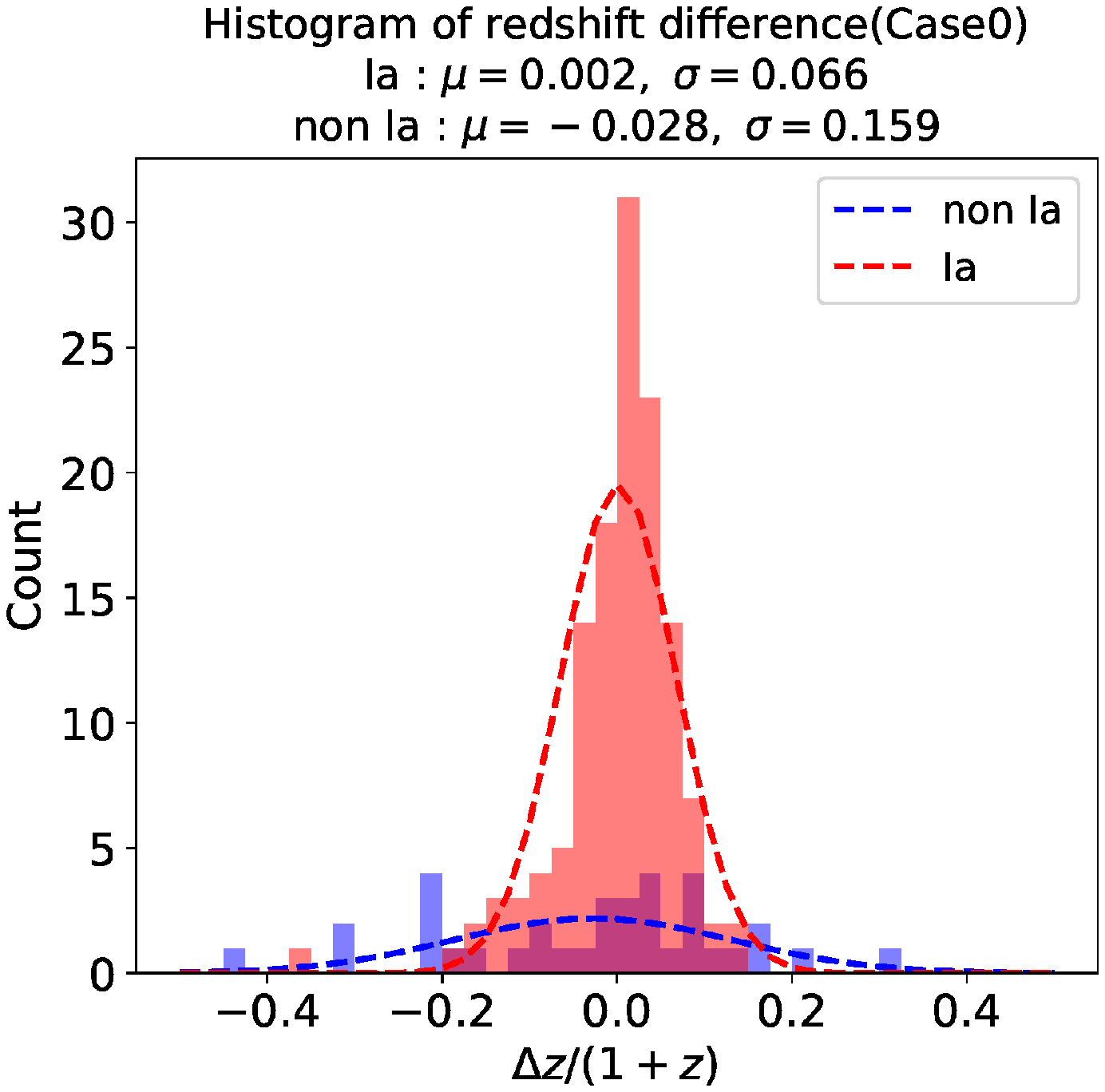}
            \end{center}
        \end{minipage}
    \end{tabular}  \caption{%
    (Left) Relation between the machine predictions and the observed redshifts for HSC SNe.
    The dashed line is the line of equality.
    The color difference represents the SALT2 fitted label.
    (Right) Normalized residual distribution for Case 0 SNe.
    }%
    \label{fig:redshift_estimation}
\end{figure*}

We applied the standard deviation of the normalized residual $(z_{\rm pred}-z_{\rm spec})/(1+z_{\rm spec})$ \citep{Salvato_2009,Salvato_2019}, used in galaxy photo-z estimation, to the performance criteria for redshift estimation.
Verified with Case 0 simulated data, the redshifts for Ia and non-Ia objects were estimated to be 0.022 and 0.076, respectively.
For the actual HSC SNe, we used the observationally obtained redshifts of host galaxies (see section \ref{sec:h}) including spec-z and photo-z for comparison with the estimates.
Figure\ \ref{fig:redshift_estimation} shows the comparison between the estimated redshifts and the observed redshifts for HSC SNe classified as Ia.
The two classes labeled by SALT2 fitting have different distributions, and the events labeled non-Ia are less accurate in terms of their redshift estimation than Ia.
The normalized residuals for each of the Ia and non-Ia Case 0 SNe labeled with SALT2 fitting are normally distributed with standard deviations of 0.066 and 0.138, respectively.
Limiting the comparison to SNe in the host galaxies with spec-z information, the standard deviations are 0.056 and 0.130, respectively.
Although this estimation accuracy is lower than the template fitting accuracy using host galaxy photometric data, it is useful not only for hostless SNe, but also to select the host galaxy by comparing the redshift estimated from the SN light curve itself with those of the galaxies.
Furthermore, this distributional difference of residuals leads to the fact that the Ia accuracy can be further improved by excluding events with large residuals when estimating the redshift of the SN and host galaxy.

%

\begin{thebibliography}{}
\expandafter\ifx\csname natexlab\endcsname\relax\def\natexlab#1{#1}\fi
\providecommand{\url}[1]{\href{#1}{#1}}

\bibitem[{{Abadi} {et~al.}(2016){Abadi}, {Agarwal}, {Barham}, {Brevdo}, {Chen},
  {Citro}, {Corrado}, {Davis}, {Dean}, {Devin}, {Ghemawat}, {Goodfellow},
  {Harp}, {Irving}, {Isard}, {Jia}, {Jozefowicz}, {Kaiser}, {Kudlur},
  {Levenberg}, {Mane}, {Monga}, {Moore}, {Murray}, {Olah}, {Schuster},
  {Shlens}, {Steiner}, {Sutskever}, {Talwar}, {Tucker}, {Vanhoucke},
  {Vasudevan}, {Viegas}, {Vinyals}, {Warden}, {Wattenberg}, {Wicke}, {Yu}, \&
  {Zheng}}]{Abadi2016}
{Abadi}, M., {et~al.} 2016, arXiv:1603.04467

\bibitem[{{Aihara} {et~al.}(2018{\natexlab{a}}){Aihara}, {Arimoto},
  {Armstrong}, {Arnouts}, {Bahcall}, {Bickerton}, {Bosch}, {Bundy}, {Capak},
  {Chan}, {Chiba}, {Coupon}, {Egami}, {Enoki}, {Finet}, {Fujimori}, {Fujimoto},
  {Furusawa}, {Furusawa}, {Goto}, {Goulding}, {Greco}, {Greene}, {Gunn},
  {Hamana}, {Harikane}, {Hashimoto}, {Hattori}, {Hayashi}, {Hayashi},
  {He{\l}miniak}, {Higuchi}, {Hikage}, {Ho}, {Hsieh}, {Huang}, {Huang},
  {Ikeda}, {Imanishi}, {Inoue}, {Iwasawa}, {Iwata}, {Jaelani}, {Jian},
  {Kamata}, {Karoji}, {Kashikawa}, {Katayama}, {Kawanomoto}, {Kayo}, {Koda},
  {Koike}, {Kojima}, {Komiyama}, {Konno}, {Koshida}, {Koyama}, {Kusakabe},
  {Leauthaud}, {Lee}, {Lin}, {Lin}, {Lupton}, {Mand elbaum}, {Matsuoka},
  {Medezinski}, {Mineo}, {Miyama}, {Miyatake}, {Miyazaki}, {Momose}, {More},
  {More}, {Moritani}, {Moriya}, {Morokuma}, {Mukae}, {Murata}, {Murayama},
  {Nagao}, {Nakata}, {Niida}, {Niikura}, {Nishizawa}, {Obuchi}, {Oguri},
  {Oishi}, {Okabe}, {Okamoto}, {Okura}, {Ono}, {Onodera}, {Onoue}, {Osato},
  {Ouchi}, {Price}, {Pyo}, {Sako}, {Sawicki}, {Shibuya}, {Shimasaku},
  {Shimono}, {Shirasaki}, {Silverman}, {Simet}, {Speagle}, {Spergel},
  {Strauss}, {Sugahara}, {Sugiyama}, {Suto}, {Suyu}, {Suzuki}, {Tait},
  {Takada}, {Takata}, {Tamura}, {Tanaka}, {Tanaka}, {Tanaka}, {Tanaka},
  {Terai}, {Terashima}, {Toba}, {Tominaga}, {Toshikawa}, {Turner}, {Uchida},
  {Uchiyama}, {Umetsu}, {Uraguchi}, {Urata}, {Usuda}, {Utsumi}, {Wang}, {Wang},
  {Wong}, {Yabe}, {Yamada}, {Yamanoi}, {Yasuda}, {Yeh}, {Yonehara}, \&
  {Yuma}}]{aihara18a}
{Aihara}, H., {et~al.} 2018{\natexlab{a}}, \pasj, 70, S4

\bibitem[{{Aihara} {et~al.}(2018{\natexlab{b}}){Aihara}, {Armstrong},
  {Bickerton}, {Bosch}, {Coupon}, {Furusawa}, {Hayashi}, {Ikeda}, {Kamata},
  {Karoji}, {Kawanomoto}, {Koike}, {Komiyama}, {Lang}, {Lupton}, {Mineo},
  {Miyatake}, {Miyazaki}, {Morokuma}, {Obuchi}, {Oishi}, {Okura}, {Price},
  {Takata}, {Tanaka}, {Tanaka}, {Tanaka}, {Uchida}, {Uraguchi}, {Utsumi},
  {Wang}, {Yamada}, {Yamanoi}, {Yasuda}, {Arimoto}, {Chiba}, {Finet},
  {Fujimori}, {Fujimoto}, {Furusawa}, {Goto}, {Goulding}, {Gunn}, {Harikane},
  {Hattori}, {Hayashi}, {He{\l}miniak}, {Higuchi}, {Hikage}, {Ho}, {Hsieh},
  {Huang}, {Huang}, {Imanishi}, {Iwata}, {Jaelani}, {Jian}, {Kashikawa},
  {Katayama}, {Kojima}, {Konno}, {Koshida}, {Kusakabe}, {Leauthaud}, {Lee},
  {Lin}, {Lin}, {Mandelbaum}, {Matsuoka}, {Medezinski}, {Miyama}, {Momose},
  {More}, {More}, {Mukae}, {Murata}, {Murayama}, {Nagao}, {Nakata}, {Niida},
  {Niikura}, {Nishizawa}, {Oguri}, {Okabe}, {Ono}, {Onodera}, {Onoue}, {Ouchi},
  {Pyo}, {Shibuya}, {Shimasaku}, {Simet}, {Speagle}, {Spergel}, {Strauss},
  {Sugahara}, {Sugiyama}, {Suto}, {Suzuki}, {Tait}, {Takada}, {Terai}, {Toba},
  {Turner}, {Uchiyama}, {Umetsu}, {Urata}, {Usuda}, {Yeh}, \&
  {Yuma}}]{aihara18dr}
---. 2018{\natexlab{b}}, \pasj, 70, S8

\bibitem[{{Appenzeller} {et~al.}(1998){Appenzeller}, {Fricke}, {F{\"u}rtig},
  {G{\"a}ssler}, {H{\"a}fner}, {Harke}, {Hess}, {Hummel}, {J{\"u}rgens},
  {Kudritzki}, {Mantel}, {Meisl}, {Muschielok}, {Nicklas}, {Rupprecht},
  {Seifert}, {Stahl}, {Szeifert}, \& {Tarantik}}]{appenzeller98a}
{Appenzeller}, I., {et~al.} 1998, The Messenger, 94, 1

\bibitem[{Bergstra {et~al.}(2013)Bergstra, Yamins, \&
  Cox}]{pmlr-v28-bergstra13}
Bergstra, J., Yamins, D., \& Cox, D. 2013, in Proceedings of Machine Learning
  Research, Vol.~28, Proceedings of the 30th International Conference on
  Machine Learning, ed. S.~Dasgupta \& D.~McAllester (Atlanta, Georgia, USA:
  PMLR), 115--123

\bibitem[{{Betoule} {et~al.}(2014){Betoule}, {Kessler}, {Guy}, {Mosher},
  {Hardin}, {Biswas}, {Astier}, {El-Hage}, {Konig}, {Kuhlmann}, {Marriner},
  {Pain}, {Regnault}, {Balland}, {Bassett}, {Brown}, {Campbell}, {Carlberg},
  {Cellier-Holzem}, {Cinabro}, {Conley}, {D'Andrea}, {DePoy}, {Doi}, {Ellis},
  {Fabbro}, {Filippenko}, {Foley}, {Frieman}, {Fouchez}, {Galbany}, {Goobar},
  {Gupta}, {Hill}, {Hlozek}, {Hogan}, {Hook}, {Howell}, {Jha}, {Le Guillou},
  {Leloudas}, {Lidman}, {Marshall}, {M{\"o}ller}, {Mour{\~a}o}, {Neveu},
  {Nichol}, {Olmstead}, {Palanque-Delabrouille}, {Perlmutter}, {Prieto},
  {Pritchet}, {Richmond}, {Riess}, {Ruhlmann-Kleider}, {Sako}, {Schahmaneche},
  {Schneider}, {Smith}, {Sollerman}, {Sullivan}, {Walton}, \&
  {Wheeler}}]{betoule14a}
{Betoule}, M., {et~al.} 2014, \aap, 568, A22

\bibitem[{Bjorck {et~al.}(2018)Bjorck, Gomes, Selman, \&
  Weinberger}]{understanding_batch_norm}
Bjorck, N., Gomes, C.~P., Selman, B., \& Weinberger, K.~Q. 2018, in Advances in
  Neural Information Processing Systems 31, ed. S.~Bengio, H.~Wallach,
  H.~Larochelle, K.~Grauman, N.~Cesa-Bianchi, \& R.~Garnett (Curran Associates,
  Inc.), 7694--7705

\bibitem[{{Boone}(2019)}]{boone19a}
{Boone}, K. 2019, \aj, 158, 257

\bibitem[{{Brout} {et~al.}(2019){Brout}, {Scolnic}, {Kessler}, {D'Andrea},
  {Davis}, {Gupta}, {Hinton}, {Kim}, {Lasker}, {Lidman}, {Macaulay},
  {M{\"o}ller}, {Nichol}, {Sako}, {Smith}, {Sullivan}, {Zhang}, {Andersen},
  {Asorey}, {Avelino}, {Bassett}, {Brown}, {Calcino}, {Carollo}, {Challis},
  {Childress}, {Clocchiatti}, {Filippenko}, {Foley}, {Galbany}, {Glazebrook},
  {Hoormann}, {Kasai}, {Kirshner}, {Kuehn}, {Kuhlmann}, {Lewis}, {Mandel},
  {March}, {Miranda}, {Morganson}, {Muthukrishna}, {Nugent}, {Palmese}, {Pan},
  {Sharp}, {Sommer}, {Swann}, {Thomas}, {Tucker}, {Uddin}, {Wester}, {Abbott},
  {Allam}, {Annis}, {Avila}, {Bechtol}, {Bernstein}, {Bertin}, {Brooks},
  {Burke}, {Carnero Rosell}, {Carrasco Kind}, {Carretero}, {Castander},
  {Cunha}, {da Costa}, {Davis}, {De Vicente}, {DePoy}, {Desai}, {Diehl},
  {Doel}, {Drlica-Wagner}, {Eifler}, {Estrada}, {Fernandez}, {Flaugher},
  {Fosalba}, {Frieman}, {Garc{\'\i}a-Bellido}, {Gruen}, {Gruendl}, {Gutierrez},
  {Hartley}, {Hollowood}, {Honscheid}, {Hoyle}, {James}, {Jarvis}, {Jeltema},
  {Krause}, {Lahav}, {Li}, {Lima}, {Maia}, {Marriner}, {Marshall}, {Martini},
  {Menanteau}, {Miller}, {Miquel}, {Ogand o}, {Plazas}, {Romer}, {Roodman},
  {Rykoff}, {Sanchez}, {Santiago}, {Scarpine}, {Schubnell}, {Serrano},
  {Sevilla-Noarbe}, {Smith}, {Soares-Santos}, {Sobreira}, {Suchyta}, {Swanson},
  {Tarle}, {Thomas}, {Troxel}, {Tucker}, {Vikram}, {Walker}, {Zhang}, \& {DES
  Collaboration}}]{brout19a}
{Brout}, D., {et~al.} 2019, \apj, 874, 150

\bibitem[{{Brown} {et~al.}(2019){Brown}, {Stanek}, {Holoien}, {Kochanek},
  {Shappee}, {Prieto}, {Dong}, {Chen}, {Thompson}, {Beacom}, {Stritzinger},
  {Bersier}, \& {Brimacombe}}]{brown19a}
{Brown}, J.~S., {et~al.} 2019, \mnras, 484, 3785

\bibitem[{{Cao} {et~al.}(2015){Cao}, {Kulkarni}, {Howell}, {Gal-Yam},
  {Kasliwal}, {Valenti}, {Johansson}, {Amanullah}, {Goobar}, {Sollerman},
  {Taddia}, {Horesh}, {Sagiv}, {Cenko}, {Nugent}, {Arcavi}, {Surace},
  {Wo{\'z}niak}, {Moody}, {Rebbapragada}, {Bue}, \& {Gehrels}}]{cao15a}
{Cao}, Y., {et~al.} 2015, \nat, 521, 328

\bibitem[{{Carliles} {et~al.}(2010){Carliles}, {Budav{\'a}ri}, {Heinis},
  {Priebe}, \& {Szalay}}]{carliles10a}
{Carliles}, S., {Budav{\'a}ri}, T., {Heinis}, S., {Priebe}, C., \& {Szalay},
  A.~S. 2010, \apj, 712, 511

\bibitem[{{Carrasco-Davis} {et~al.}(2020){Carrasco-Davis}, {Reyes},
  {Valenzuela}, {F{\"o}rster}, {Est{\'e}vez}, {Pignata}, {Bauer}, {Reyes},
  {S{\'a}nchez-S{\'a}ez}, {Cabrera-Vives}, {Eyheramendy}, {Catelan},
  {Arredondo}, {Castillo-Navarrete}, {Rodr{\'\i}guez-Mancini}, {Ruz-Mieres},
  {Moya}, {Sabatini-Gacit{\'u}a}, {Sep{\'u}lveda-Cobo}, {Mahabal},
  {Silva-Farf{\'a}n}, {Camacho-I{\~n}iquez}, \& {Galbany}}]{carrasco-davis20a}
{Carrasco-Davis}, R., {et~al.} 2020, arXiv:2008.03309

\bibitem[{{Charnock} \& {Moss}(2017)}]{charnock17a}
{Charnock}, T., \& {Moss}, A. 2017, \apjl, 837, L28

\bibitem[{{Coil} {et~al.}(2011){Coil}, {Blanton}, {Burles}, {Cool},
  {Eisenstein}, {Moustakas}, {Wong}, {Zhu}, {Aird}, {Bernstein}, {Bolton}, \&
  {Hogg}}]{PRIMUS2011}
{Coil}, A.~L., {et~al.} 2011, \apj, 741, 8

\bibitem[{{Collister} \& {Lahav}(2004)}]{collister04a}
{Collister}, A.~A., \& {Lahav}, O. 2004, \pasp, 116, 345

\bibitem[{{Conley} {et~al.}(2006){Conley}, {Howell}, {Howes}, {Sullivan},
  {Astier}, {Balam}, {Basa}, {Carlberg}, {Fouchez}, {Guy}, {Hook}, {Neill},
  {Pain}, {Perrett}, {Pritchet}, {Regnault}, {Rich}, {Taillet}, {Aubourg},
  {Bronder}, {Ellis}, {Fabbro}, {Filiol}, {Le Borgne}, {Palanque-Delabrouille},
  {Perlmutter}, \& {Ripoche}}]{conley06a}
{Conley}, A., {et~al.} 2006, \aj, 132, 1707

\bibitem[{{Cooke} {et~al.}(2012){Cooke}, {Sullivan}, {Gal-Yam}, {Barton},
  {Carlberg}, {Ryan-Weber}, {Horst}, {Omori}, \& {D{\'\i}az}}]{cooke12a}
{Cooke}, J., {et~al.} 2012, \nat, 491, 228

\bibitem[{{Curtin} {et~al.}(2019){Curtin}, {Cooke}, {Moriya}, {Tanaka},
  {Quimby}, {Bernard}, {Galbany}, {Jiang}, {Lee}, {Maeda}, {Morokuma},
  {Nomoto}, {Pignata}, {Pritchard}, {Suzuki}, {Takahashi}, {Tanaka},
  {Tominaga}, {Yamaguchi}, \& {Yasuda}}]{curtin19a}
{Curtin}, C., {et~al.} 2019, \apjs, 241, 17

\bibitem[{Dallachiesa {et~al.}(2012)Dallachiesa, Nushi, Mirylenka, \&
  Palpanas}]{Dallachiesa_2012}
Dallachiesa, M., Nushi, B., Mirylenka, K., \& Palpanas, T. 2012, Proc. VLDB
  Endow., 5, 1662^^e2^^80^^931673

\bibitem[{{de Jaeger} {et~al.}(2017){de Jaeger}, {Galbany}, {Filippenko},
  {Gonz{\'a}lez-Gait{\'a}n}, {Yasuda}, {Maeda}, {Tanaka}, {Morokuma}, {Moriya},
  {Tominaga}, {Nomoto}, {Komiyama}, {Anderson}, {Brink}, {Carlberg},
  {Folatelli}, {Hamuy}, {Pignata}, \& {Zheng}}]{dejaeger17a}
{de Jaeger}, T., {et~al.} 2017, \mnras, 472, 4233

\bibitem[{{Dilday} {et~al.}(2008){Dilday}, {Kessler}, {Frieman}, {Holtzman},
  {Marriner}, {Miknaitis}, {Nichol}, {Romani}, {Sako}, {Bassett}, {Becker},
  {Cinabro}, {DeJongh}, {Depoy}, {Doi}, {Garnavich}, {Hogan}, {Jha}, {Konishi},
  {Lampeitl}, {Marshall}, {McGinnis}, {Prieto}, {Riess}, {Richmond},
  {Schneider}, {Smith}, {Takanashi}, {Tokita}, {van der Heyden}, {Yasuda},
  {Zheng}, {Barentine}, {Brewington}, {Choi}, {Crotts}, {Dembicky}, {Harvanek},
  {Im}, {Ketzeback}, {Kleinman}, {Krzesi{\'n}ski}, {Long}, {Malanushenko},
  {Malanushenko}, {McMillan}, {Nitta}, {Pan}, {Saurage}, {Snedden}, {Watters},
  {Wheeler}, \& {York}}]{dilday08a}
{Dilday}, B., {et~al.} 2008, \apj, 682, 262

\bibitem[{{Filippenko}(1997)}]{filippenko97a}
{Filippenko}, A.~V. 1997, \araa, 35, 309

\bibitem[{{F{\"o}rster} {et~al.}(2018){F{\"o}rster}, {Moriya}, {Maureira},
  {Anderson}, {Blinnikov}, {Bufano}, {Cabrera-Vives}, {Clocchiatti}, {de
  Jaeger}, {Est{\'e}vez}, {Galbany}, {Gonz{\'a}lez-Gait{\'a}n}, {Gr{\"a}fener},
  {Hamuy}, {Hsiao}, {Huentelemu}, {Huijse}, {Kuncarayakti}, {Mart{\'\i}nez},
  {Medina}, {Olivares E.}, {Pignata}, {Razza}, {Reyes}, {San Mart{\'\i}n},
  {Smith}, {Vera}, {Vivas}, {de Ugarte Postigo}, {Yoon}, {Ashall}, {Fraser},
  {Gal-Yam}, {Kankare}, {Le Guillou}, {Mazzali}, {Walton}, \&
  {Young}}]{forster18a}
{F{\"o}rster}, F., {et~al.} 2018, Nature Astronomy, 2, 808

\bibitem[{{F{\"o}rster} {et~al.}(2020){F{\"o}rster}, {Cabrera-Vives},
  {Castillo-Navarrete}, {Est{\'e}vez}, {S{\'a}nchez-S{\'a}ez}, {Arredondo},
  {Bauer}, {Carrasco-Davis}, {Catelan}, {Elorrieta}, {Eyheramendy}, {Huijse},
  {Pignata}, {Reyes}, {Reyes}, {Rodr{\'\i}guez-Mancini}, {Ruz-Mieres},
  {Valenzuela}, {Alvarez-Maldonado}, {Astorga}, {Borissova}, {Clocchiatti}, {De
  Cicco}, {Donoso-Oliva}, {Graham}, {Kurtev}, {Mahabal}, {Maureira},
  {Molina-Ferreiro}, {Moya}, {Palma}, {P{\'e}rez-Carrasco}, {Protopapas},
  {Romero}, {Sabatini-Gacit{\'u}a}, {S{\'a}nchez}, {San Mart{\'\i}n},
  {Sep{\'u}lveda-Cobo}, {Vera}, \& {Vergara}}]{forster20a}
---. 2020, arXiv:2008.03303

\bibitem[{{Frohmaier} {et~al.}(2019){Frohmaier}, {Sullivan}, {Nugent}, {Smith},
  {Dimitriadis}, {Bloom}, {Cenko}, {Kasliwal}, {Kulkarni}, {Maguire}, {Ofek},
  {Poznanski}, \& {Quimby}}]{frohmaier19a}
{Frohmaier}, C., {et~al.} 2019, \mnras, 486, 2308

\bibitem[{{Fukugita} \& {Peebles}(2004)}]{fukugita04a}
{Fukugita}, M., \& {Peebles}, P.~J.~E. 2004, \apj, 616, 643

\bibitem[{{Furusawa} {et~al.}(2018){Furusawa}, {Koike}, {Takata}, {Okura},
  {Miyatake}, {Lupton}, {Bickerton}, {Price}, {Bosch}, {Yasuda}, {Mineo},
  {Yamada}, {Miyazaki}, {Nakata}, {Koshida}, {Komiyama}, {Utsumi},
  {Kawanomoto}, {Jeschke}, {Noumaru}, {Schubert}, {Iwata}, {Finet},
  {Fujiyoshi}, {Tajitsu}, {Terai}, \& {Lee}}]{Furusawa2018}
{Furusawa}, H., {et~al.} 2018, \pasj, 70, S3

\bibitem[{{Gal-Yam}(2012)}]{galyam12a}
{Gal-Yam}, A. 2012, Science, 337, 927

\bibitem[{{Garcia-Dias} {et~al.}(2018){Garcia-Dias}, {Allende Prieto},
  {S{\'a}nchez Almeida}, \& {Ordov{\'a}s-Pascual}}]{garciadias18a}
{Garcia-Dias}, R., {Allende Prieto}, C., {S{\'a}nchez Almeida}, J., \&
  {Ordov{\'a}s-Pascual}, I. 2018, \aap, 612, A98

\bibitem[{{Goldstein} {et~al.}(2015){Goldstein}, {D'Andrea}, {Fischer},
  {Foley}, {Gupta}, {Kessler}, {Kim}, {Nichol}, {Nugent}, {Papadopoulos},
  {Sako}, {Smith}, {Sullivan}, {Thomas}, {Wester}, {Wolf}, {Abdalla},
  {Banerji}, {Benoit-L{\'e}vy}, {Bertin}, {Brooks}, {Carnero Rosell},
  {Castander}, {da Costa}, {Covarrubias}, {DePoy}, {Desai}, {Diehl}, {Doel},
  {Eifler}, {Fausti Neto}, {Finley}, {Flaugher}, {Fosalba}, {Frieman},
  {Gerdes}, {Gruen}, {Gruendl}, {James}, {Kuehn}, {Kuropatkin}, {Lahav}, {Li},
  {Maia}, {Makler}, {March}, {Marshall}, {Martini}, {Merritt}, {Miquel},
  {Nord}, {Ogando}, {Plazas}, {Romer}, {Roodman}, {Sanchez}, {Scarpine},
  {Schubnell}, {Sevilla-Noarbe}, {Smith}, {Soares-Santos}, {Sobreira},
  {Suchyta}, {Swanson}, {Tarle}, {Thaler}, \& {Walker}}]{goldstein15a}
{Goldstein}, D.~A., {et~al.} 2015, \aj, 150, 82

\bibitem[{{Graur} {et~al.}(2014){Graur}, {Rodney}, {Maoz}, {Riess}, {Jha},
  {Postman}, {Dahlen}, {Holoien}, {McCully}, {Patel}, {Strolger},
  {Ben{\'\i}tez}, {Coe}, {Jouvel}, {Medezinski}, {Molino}, {Nonino}, {Bradley},
  {Koekemoer}, {Balestra}, {Cenko}, {Clubb}, {Dickinson}, {Filippenko},
  {Frederiksen}, {Garnavich}, {Hjorth}, {Jones}, {Leibundgut}, {Matheson},
  {Mobasher}, {Rosati}, {Silverman}, {U}, {Jedruszczuk}, {Li}, {Lin},
  {Mirmelstein}, {Neustadt}, {Ovadia}, \& {Rogers}}]{graur14a}
{Graur}, O., {et~al.} 2014, \apj, 783, 28

\bibitem[{{Grevesse} \& {Sauval}(1998)}]{grevesse98a}
{Grevesse}, N., \& {Sauval}, A.~J. 1998, \ssr, 85, 161

\bibitem[{{Guy} {et~al.}(2007){Guy}, {Astier}, {Baumont}, {Hardin}, {Pain},
  {Regnault}, {Basa}, {Carlberg}, {Conley}, {Fabbro}, {Fouchez}, {Hook},
  {Howell}, {Perrett}, {Pritchet}, {Rich}, {Sullivan}, {Antilogus}, {Aubourg},
  {Bazin}, {Bronder}, {Filiol}, {Palanque-Delabrouille}, {Ripoche}, \&
  {Ruhlmann-Kleider}}]{guy2007}
{Guy}, J., {et~al.} 2007, \aap, 466, 11

\bibitem[{{Guy} {et~al.}(2010){Guy}, {Sullivan}, {Conley}, {Regnault},
  {Astier}, {Balland}, {Basa}, {Carlberg}, {Fouchez}, {Hardin}, {Hook},
  {Howell}, {Pain}, {Palanque-Delabrouille}, {Perrett}, {Pritchet}, {Rich},
  {Ruhlmann-Kleider}, {Balam}, {Baumont}, {Ellis}, {Fabbro}, {Fakhouri},
  {Fourmanoit}, {Gonz{\'a}lez-Gait{\'a}n}, {Graham}, {Hsiao}, {Kronborg},
  {Lidman}, {Mourao}, {Perlmutter}, {Ripoche}, {Suzuki}, \& {Walker}}]{guy10b}
---. 2010, \aap, 523, A7

\bibitem[{{Hasinger} {et~al.}(2018){Hasinger}, {Capak}, {Salvato}, {Barger},
  {Cowie}, {Faisst}, {Hemmati}, {Kakazu}, {Kartaltepe}, {Masters}, {Mobasher},
  {Nayyeri}, {Sanders}, {Scoville}, {Suh}, {Steinhardt}, \&
  {Yang}}]{DEIMOS2018}
{Hasinger}, G., {et~al.} 2018, \apj, 858, 77

\bibitem[{{Hausen} \& {Robertson}(2019)}]{hausen19a}
{Hausen}, R., \& {Robertson}, B. 2019, arXiv:1906.11248

\bibitem[{{Hook} {et~al.}(2004){Hook}, {J{\o}rgensen}, {Allington-Smith},
  {Davies}, {Metcalfe}, {Murowinski}, \& {Crampton}}]{hook04a}
{Hook}, I.~M., {J{\o}rgensen}, I., {Allington-Smith}, J.~R., {Davies}, R.~L.,
  {Metcalfe}, N., {Murowinski}, R.~G., \& {Crampton}, D. 2004, \pasp, 116, 425

\bibitem[{{Hosseinzadeh} {et~al.}(2020){Hosseinzadeh}, {Dauphin}, {Villar},
  {Berger}, {Jones}, {Challis}, {Chornock}, {Drout}, {Foley}, {Kirshner},
  {Lunnan}, {Margutti}, {Milisavljevic}, {Pan}, {Rest}, {Scolnic}, {Magnier},
  {Metcalfe}, {Wainscoat}, \& {Waters}}]{hosseinzadeh20a}
{Hosseinzadeh}, G., {et~al.} 2020, arXiv:2008.0491

\bibitem[{{Hounsell} {et~al.}(2018){Hounsell}, {Scolnic}, {Foley}, {Kessler},
  {Miranda}, {Avelino}, {Bohlin}, {Filippenko}, {Frieman}, {Jha}, {Kelly},
  {Kirshner}, {Mandel}, {Rest}, {Riess}, {Rodney}, \& {Strolger}}]{hounsell18a}
{Hounsell}, R., {et~al.} 2018, \apj, 867, 23

\bibitem[{{Howell} {et~al.}(2006){Howell}, {Sullivan}, {Nugent}, {Ellis},
  {Conley}, {Le Borgne}, {Carlberg}, {Guy}, {Balam}, {Basa}, {Fouchez}, {Hook},
  {Hsiao}, {Neill}, {Pain}, {Perrett}, \& {Pritchet}}]{howell06a}
{Howell}, D.~A., {et~al.} 2006, \nat, 443, 308

\bibitem[{Ioffe \& Szegedy(2015)}]{batch_norm}
Ioffe, S., \& Szegedy, C. 2015, in Proceedings of Machine Learning Research,
  Vol.~37, Proceedings of the 32nd International Conference on Machine
  Learning, ed. F.~Bach \& D.~Blei (Lille, France: PMLR), 448--456

\bibitem[{{Ivezi{\'c}} {et~al.}(2019){Ivezi{\'c}}, {Kahn}, {Tyson}, {Abel},
  {Acosta}, {Allsman}, {Alonso}, {AlSayyad}, {Anderson}, {Andrew}, {Angel},
  {Angeli}, {Ansari}, {Antilogus}, {Araujo}, {Armstrong}, {Arndt}, {Astier},
  {Aubourg}, {Auza}, {Axelrod}, {Bard}, {Barr}, {Barrau}, {Bartlett}, {Bauer},
  {Bauman}, {Baumont}, {Bechtol}, {Bechtol}, {Becker}, {Becla}, {Beldica},
  {Bellavia}, {Bianco}, {Biswas}, {Blanc}, {Blazek}, {Bland ford}, {Bloom},
  {Bogart}, {Bond}, {Booth}, {Borgland}, {Borne}, {Bosch}, {Boutigny},
  {Brackett}, {Bradshaw}, {Brand t}, {Brown}, {Bullock}, {Burchat}, {Burke},
  {Cagnoli}, {Calabrese}, {Callahan}, {Callen}, {Carlin}, {Carlson}, {Chand
  rasekharan}, {Charles-Emerson}, {Chesley}, {Cheu}, {Chiang}, {Chiang},
  {Chirino}, {Chow}, {Ciardi}, {Claver}, {Cohen-Tanugi}, {Cockrum}, {Coles},
  {Connolly}, {Cook}, {Cooray}, {Covey}, {Cribbs}, {Cui}, {Cutri}, {Daly},
  {Daniel}, {Daruich}, {Daubard}, {Daues}, {Dawson}, {Delgado}, {Dellapenna},
  {de Peyster}, {de Val-Borro}, {Digel}, {Doherty}, {Dubois},
  {Dubois-Felsmann}, {Durech}, {Economou}, {Eifler}, {Eracleous}, {Emmons},
  {Fausti Neto}, {Ferguson}, {Figueroa}, {Fisher-Levine}, {Focke}, {Foss},
  {Frank}, {Freemon}, {Gangler}, {Gawiser}, {Geary}, {Gee}, {Geha}, {Gessner},
  {Gibson}, {Gilmore}, {Glanzman}, {Glick}, {Goldina}, {Goldstein}, {Goodenow},
  {Graham}, {Gressler}, {Gris}, {Guy}, {Guyonnet}, {Haller}, {Harris},
  {Hascall}, {Haupt}, {Hernand ez}, {Herrmann}, {Hileman}, {Hoblitt},
  {Hodgson}, {Hogan}, {Howard}, {Huang}, {Huffer}, {Ingraham}, {Innes},
  {Jacoby}, {Jain}, {Jammes}, {Jee}, {Jenness}, {Jernigan}, {Jevremovi{\'c}},
  {Johns}, {Johnson}, {Johnson}, {Jones}, {Juramy-Gilles}, {Juri{\'c}},
  {Kalirai}, {Kallivayalil}, {Kalmbach}, {Kantor}, {Karst}, {Kasliwal},
  {Kelly}, {Kessler}, {Kinnison}, {Kirkby}, {Knox}, {Kotov}, {Krabbendam},
  {Krughoff}, {Kub{\'a}nek}, {Kuczewski}, {Kulkarni}, {Ku}, {Kurita}, {Lage},
  {Lambert}, {Lange}, {Langton}, {Le Guillou}, {Levine}, {Liang}, {Lim},
  {Lintott}, {Long}, {Lopez}, {Lotz}, {Lupton}, {Lust}, {MacArthur}, {Mahabal},
  {Mand elbaum}, {Markiewicz}, {Marsh}, {Marshall}, {Marshall}, {May},
  {McKercher}, {McQueen}, {Meyers}, {Migliore}, {Miller}, {Mills}, {Miraval},
  {Moeyens}, {Moolekamp}, {Monet}, {Moniez}, {Monkewitz}, {Montgomery},
  {Morrison}, {Mueller}, {Muller}, {Mu{\~n}oz Arancibia}, {Neill}, {Newbry},
  {Nief}, {Nomerotski}, {Nordby}, {O'Connor}, {Oliver}, {Olivier}, {Olsen},
  {O'Mullane}, {Ortiz}, {Osier}, {Owen}, {Pain}, {Palecek}, {Parejko},
  {Parsons}, {Pease}, {Peterson}, {Peterson}, {Petravick}, {Libby Petrick},
  {Petry}, {Pierfederici}, {Pietrowicz}, {Pike}, {Pinto}, {Plante}, {Plate},
  {Plutchak}, {Price}, {Prouza}, {Radeka}, {Rajagopal}, {Rasmussen},
  {Regnault}, {Reil}, {Reiss}, {Reuter}, {Ridgway}, {Riot}, {Ritz}, {Robinson},
  {Roby}, {Roodman}, {Rosing}, {Roucelle}, {Rumore}, {Russo}, {Saha},
  {Sassolas}, {Schalk}, {Schellart}, {Schindler}, {Schmidt}, {Schneider},
  {Schneider}, {Schoening}, {Schumacher}, {Schwamb}, {Sebag}, {Selvy},
  {Sembroski}, {Seppala}, {Serio}, {Serrano}, {Shaw}, {Shipsey}, {Sick},
  {Silvestri}, {Slater}, {Smith}, {Smith}, {Sobhani}, {Soldahl},
  {Storrie-Lombardi}, {Stover}, {Strauss}, {Street}, {Stubbs}, {Sullivan},
  {Sweeney}, {Swinbank}, {Szalay}, {Takacs}, {Tether}, {Thaler}, {Thayer},
  {Thomas}, {Thornton}, {Thukral}, {Tice}, {Trilling}, {Turri}, {Van Berg},
  {Vanden Berk}, {Vetter}, {Virieux}, {Vucina}, {Wahl}, {Walkowicz}, {Walsh},
  {Walter}, {Wang}, {Wang}, {Warner}, {Wiecha}, {Willman}, {Winters},
  {Wittman}, {Wolff}, {Wood-Vasey}, {Wu}, {Xin}, {Yoachim}, \&
  {Zhan}}]{ivezic19a}
{Ivezi{\'c}}, {\v{Z}}., {et~al.} 2019, \apj, 873, 111

\bibitem[{{Jiang} {et~al.}(2017){Jiang}, {Doi}, {Maeda}, {Shigeyama}, {Nomoto},
  {Yasuda}, {Jha}, {Tanaka}, {Morokuma}, {Tominaga}, {Ivezi{\'c}},
  {Ruiz-Lapuente}, {Stritzinger}, {Mazzali}, {Ashall}, {Mould}, {Baade},
  {Suzuki}, {Connolly}, {Patat}, {Wang}, {Yoachim}, {Jones}, {Furusawa}, \&
  {Miyazaki}}]{Jiang2017}
{Jiang}, J.-a., {et~al.} 2017, \nat, 550, 80

\bibitem[{{Jiang} {et~al.}(2020){Jiang}, {Yasuda}, {Maeda}, {Doi}, {Shigeyama},
  {Tominaga}, {Tanaka}, {Moriya}, {Takahashi}, {Suzuki}, {Morokuma}, \&
  {Nomoto}}]{Jiang_2020}
---. 2020, \apj, 892, 25

\bibitem[{{Jones} {et~al.}(2018){Jones}, {Scolnic}, {Riess}, {Rest},
  {Kirshner}, {Berger}, {Kessler}, {Pan}, {Foley}, {Chornock}, {Ortega},
  {Challis}, {Burgett}, {Chambers}, {Draper}, {Flewelling}, {Huber}, {Kaiser},
  {Kudritzki}, {Metcalfe}, {Tonry}, {Wainscoat}, {Waters}, {Gall}, {Kotak},
  {McCrum}, {Smartt}, \& {Smith}}]{jonesl8a}
{Jones}, D.~O., {et~al.} 2018, \apj, 857, 51

\bibitem[{{Kashikawa} {et~al.}(2002){Kashikawa}, {Aoki}, {Asai}, {Ebizuka},
  {Inata}, {Iye}, {Kawabata}, {Kosugi}, {Ohyama}, {Okita}, {Ozawa}, {Saito},
  {Sasaki}, {Sekiguchi}, {Shimizu}, {Taguchi}, {Takata}, {Yadoumaru}, \&
  {Yoshida}}]{kashikawa02a}
{Kashikawa}, N., {et~al.} 2002, \pasj, 54, 819

\bibitem[{{Kawanomoto} {et~al.}(2018){Kawanomoto}, {Uraguchi}, {Komiyama},
  {Miyazaki}, {Furusawa}, {Finet}, {Hattori}, {Wang}, {Yasuda}, \&
  {Suzuki}}]{kawanomoto18a}
{Kawanomoto}, S., {et~al.} 2018, \pasj, 70, 66

\bibitem[{{Kessler} {et~al.}(2019){Kessler}, {Narayan}, {Avelino}, {Bachelet},
  {Biswas}, {Brown}, {Chernoff}, {Connolly}, {Dai}, {Daniel}, {Di Stefano},
  {Drout}, {Galbany}, {Gonz{\'a}lez-Gait{\'a}n}, {Graham}, {Hlo{\v{z}}ek},
  {Ishida}, {Guillochon}, {Jha}, {Jones}, {Mand el}, {Muthukrishna}, {O'Grady},
  {Peters}, {Pierel}, {Ponder}, {Pr{\v{s}}a}, {Rodney}, {Villar}, {LSST Dark
  Energy Science Collaboration}, \& {Transient and Variable Stars Science
  Collaboration}}]{kessler19b}
{Kessler}, R., {et~al.} 2019, \pasp, 131, 094501

\bibitem[{Kimura {et~al.}(2017)Kimura, Takahashi, Tanaka, Yasuda, Ueda, \&
  Yoshida}]{Kimura17}
Kimura, A., Takahashi, I., Tanaka, M., Yasuda, N., Ueda, N., \& Yoshida, N.
  2017, in 37th {IEEE} International Conference on Distributed Computing
  Systems Workshops, {ICDCS} Workshops 2017, Atlanta, GA, USA, June 5-8, 2017,
  ed. A.~Musaev, J.~E. Ferreira, \& T.~Higashino ({IEEE} Computer Society),
  354--359

\bibitem[{{Kingma} \& {Ba}(2014)}]{Kingma2014}
{Kingma}, D.~P., \& {Ba}, J. 2014, arXiv:1412.6980

\bibitem[{{Kobayashi} \& {Nakasato}(2011)}]{kobayashi11a}
{Kobayashi}, C., \& {Nakasato}, N. 2011, \apj, 729, 16

\bibitem[{{Kobayashi} {et~al.}(2000){Kobayashi}, {Tsujimoto}, \&
  {Nomoto}}]{kobayashi00a}
{Kobayashi}, C., {Tsujimoto}, T., \& {Nomoto}, K. 2000, \apj, 539, 26

\bibitem[{{Komiyama} {et~al.}(2018){Komiyama}, {Obuchi}, {Nakaya}, {Kamata},
  {Kawanomoto}, {Utsumi}, {Miyazaki}, {Uraguchi}, {Furusawa}, {Morokuma},
  {Uchida}, {Miyatake}, {Mineo}, {Fujimori}, {Aihara}, {Karoji}, {Gunn}, \&
  {Wang}}]{Komiyama2018}
{Komiyama}, Y., {et~al.} 2018, \pasj, 70, S2

\bibitem[{{Laigle} {et~al.}(2016){Laigle}, {McCracken}, {Ilbert}, {Hsieh},
  {Davidzon}, {Capak}, {Hasinger}, {Silverman}, {Pichon}, {Coupon}, {Aussel},
  {Le Borgne}, {Caputi}, {Cassata}, {Chang}, {Civano}, {Dunlop}, {Fynbo},
  {Kartaltepe}, {Koekemoer}, {Le F{\`e}vre}, {Le Floc'h}, {Leauthaud}, {Lilly},
  {Lin}, {Marchesi}, {Milvang-Jensen}, {Salvato}, {Sanders}, {Scoville},
  {Smolcic}, {Stockmann}, {Taniguchi}, {Tasca}, {Toft}, {Vaccari}, \&
  {Zabl}}]{laigle16a}
{Laigle}, C., {et~al.} 2016, \apjs, 224, 24

\bibitem[{{Li} {et~al.}(2011){Li}, {Leaman}, {Chornock}, {Filippenko},
  {Poznanski}, {Ganeshalingam}, {Wang}, {Modjaz}, {Jha}, {Foley}, \&
  {Smith}}]{li11a}
{Li}, W., {et~al.} 2011, \mnras, 412, 1441

\bibitem[{{Li} {et~al.}(2019){Li}, {Chen}, {Hu}, \&
  {Yang}}]{dropout_and_batch_norm}
{Li}, X., {Chen}, S., {Hu}, X., \& {Yang}, J. 2019, in 2019 IEEE/CVF Conference
  on Computer Vision and Pattern Recognition (CVPR), 2677--2685

\bibitem[{{Linder}(2003)}]{linder03b}
{Linder}, E.~V. 2003, Physical Review Letters, 90, 091301

\bibitem[{{Lochner} {et~al.}(2016){Lochner}, {McEwen}, {Peiris}, {Lahav}, \&
  {Winter}}]{Lochner_2016}
{Lochner}, M., {McEwen}, J.~D., {Peiris}, H.~V., {Lahav}, O., \& {Winter},
  M.~K. 2016, \apjs, 225, 31

\bibitem[{{Lupton} {et~al.}(1999){Lupton}, {Gunn}, \& {Szalay}}]{lupton99a}
{Lupton}, R.~H., {Gunn}, J.~E., \& {Szalay}, A.~S. 1999, \aj, 118, 1406

\bibitem[{{Madau} {et~al.}(1998){Madau}, {della Valle}, \&
  {Panagia}}]{madau98a}
{Madau}, P., {della Valle}, M., \& {Panagia}, N. 1998, \mnras, 297, L17

\bibitem[{{Maeda} {et~al.}(2018){Maeda}, {Jiang}, {Shigeyama}, \&
  {Doi}}]{maeda18a}
{Maeda}, K., {Jiang}, J.-a., {Shigeyama}, T., \& {Doi}, M. 2018, \apj, 861, 78

\bibitem[{{Malz} {et~al.}(2019){Malz}, {Hlo{\v{z}}ek}, {Allam}, {Bahmanyar},
  {Biswas}, {Dai}, {Galbany}, {Ishida}, {Jha}, {Jones}, {Kessler}, {Lochner},
  {Mahabal}, {Mand el}, {Mart{\'\i}nez-Galarza}, {McEwen}, {Muthukrishna},
  {Narayan}, {Peiris}, {Peters}, {Ponder}, {Setzer}, {(the LSST Dark Energy
  Science Collaboration}, {LSST Transients}, \& {Variable Stars Science
  Collaboration}}]{malz19a}
{Malz}, A.~I., {et~al.} 2019, \aj, 158, 171

\bibitem[{{Maraston}(2005)}]{maraston05a}
{Maraston}, C. 2005, \mnras, 362, 799

\bibitem[{{Masters} {et~al.}(2017){Masters}, {Stern}, {Cohen}, {Capak},
  {Rhodes}, {Castander}, \& {Paltani}}]{C3R2_2017}
{Masters}, D.~C., {Stern}, D.~K., {Cohen}, J.~G., {Capak}, P.~L., {Rhodes},
  J.~D., {Castander}, F.~J., \& {Paltani}, S. 2017, \apj, 841, 111

\bibitem[{{Mernier} {et~al.}(2018){Mernier}, {Biffi}, {Yamaguchi}, {Medvedev},
  {Simionescu}, {Ettori}, {Werner}, {Kaastra}, {de Plaa}, \& {Gu}}]{mernier18a}
{Mernier}, F., {et~al.} 2018, \ssr, 214, 129

\bibitem[{{Miyazaki} {et~al.}(2018){Miyazaki}, {Komiyama}, {Kawanomoto}, {Doi},
  {Furusawa}, {Hamana}, {Hayashi}, {Ikeda}, {Kamata}, {Karoji}, {Koike},
  {Kurakami}, {Miyama}, {Morokuma}, {Nakata}, {Namikawa}, {Nakaya}, {Nariai},
  {Obuchi}, {Oishi}, {Okada}, {Okura}, {Tait}, {Takata}, {Tanaka}, {Tanaka},
  {Terai}, {Tomono}, {Uraguchi}, {Usuda}, {Utsumi}, {Yamada}, {Yamanoi},
  {Aihara}, {Fujimori}, {Mineo}, {Miyatake}, {Oguri}, {Uchida}, {Tanaka},
  {Yasuda}, {Takada}, {Murayama}, {Nishizawa}, {Sugiyama}, {Chiba}, {Futamase},
  {Wang}, {Chen}, {Ho}, {Liaw}, {Chiu}, {Ho}, {Lai}, {Lee}, {Jeng}, {Iwamura},
  {Armstrong}, {Bickerton}, {Bosch}, {Gunn}, {Lupton}, {Loomis}, {Price},
  {Smith}, {Strauss}, {Turner}, {Suzuki}, {Miyazaki}, {Muramatsu}, {Yamamoto},
  {Endo}, {Ezaki}, {Ito}, {Kawaguchi}, {Sofuku}, {Taniike}, {Akutsu}, {Dojo},
  {Kasumi}, {Matsuda}, {Imoto}, {Miwa}, {Suzuki}, {Takeshi}, \&
  {Yokota}}]{miyazaki18a}
{Miyazaki}, S., {et~al.} 2018, \pasj, 70, S1

\bibitem[{{Morii} {et~al.}(2016){Morii}, {Ikeda}, {Tominaga}, {Tanaka},
  {Morokuma}, {Ishiguro}, {Yamato}, {Ueda}, {Suzuki}, {Yasuda}, \&
  {Yoshida}}]{morii16a}
{Morii}, M., {et~al.} 2016, \pasj, 68, 104

\bibitem[{{Moriya} {et~al.}(2018){Moriya}, {Sorokina}, \&
  {Chevalier}}]{Moriya18SLSN}
{Moriya}, T.~J., {Sorokina}, E.~I., \& {Chevalier}, R.~A. 2018, \ssr, 214, 59

\bibitem[{{Moriya} {et~al.}(2019){Moriya}, {Tanaka}, {Yasuda}, {Jiang}, {Lee},
  {Maeda}, {Morokuma}, {Nomoto}, {Quimby}, {Suzuki}, {Takahashi}, {Tanaka},
  {Tominaga}, {Yamaguchi}, {Bernard}, {Cooke}, {Curtin}, {Galbany},
  {Gonz{\'a}lez-Gait{\'a}n}, {Pignata}, {Pritchard}, {Komiyama}, \&
  {Lupton}}]{moriya19a}
{Moriya}, T.~J., {et~al.} 2019, \apjs, 241, 16

\bibitem[{{Mosher} {et~al.}(2014){Mosher}, {Guy}, {Kessler}, {Astier},
  {Marriner}, {Betoule}, {Sako}, {El-Hage}, {Biswas}, {Pain}, {Kuhlmann},
  {Regnault}, {Frieman}, \& {Schneider}}]{mosher14a}
{Mosher}, J., {et~al.} 2014, \apj, 793, 16

\bibitem[{{Muthukrishna} {et~al.}(2019{\natexlab{a}}){Muthukrishna}, {Narayan},
  {Mandel}, {Biswas}, \& {Hlo{\v{z}}ek}}]{Muthukrishna_2019}
{Muthukrishna}, D., {Narayan}, G., {Mandel}, K.~S., {Biswas}, R., \&
  {Hlo{\v{z}}ek}, R. 2019{\natexlab{a}}, \pasp, 131, 118002

\bibitem[{{Muthukrishna} {et~al.}(2019{\natexlab{b}}){Muthukrishna},
  {Parkinson}, \& {Tucker}}]{muthukrishna19c}
{Muthukrishna}, D., {Parkinson}, D., \& {Tucker}, B.~E. 2019{\natexlab{b}},
  \apj, 885, 85

\bibitem[{{Narayan} {et~al.}(2018){Narayan}, {Zaidi}, {Soraisam}, {Wang},
  {Lochner}, {Matheson}, {Saha}, {Yang}, {Zhao}, {Kececioglu}, {Scheidegger},
  {Snodgrass}, {Axelrod}, {Jenness}, {Maier}, {Ridgway}, {Seaman}, {Evans},
  {Singh}, {Taylor}, {Toeniskoetter}, {Welch}, {Zhu}, \& {ANTARES
  Collaboration}}]{narayan18a}
{Narayan}, G., {et~al.} 2018, \apjs, 236, 9

\bibitem[{{Oke} {et~al.}(1995){Oke}, {Cohen}, {Carr}, {Cromer}, {Dingizian},
  {Harris}, {Labrecque}, {Lucinio}, {Schaal}, {Epps}, \& {Miller}}]{oke95a}
{Oke}, J.~B., {et~al.} 1995, \pasp, 107, 375

\bibitem[{{Papadogiannakis} {et~al.}(2019){Papadogiannakis}, {Goobar},
  {Amanullah}, {Bulla}, {Dhawan}, {Doran}, {Feindt}, {Ferretti}, {Hangard},
  {Howell}, {Johansson}, {Kasliwal}, {Laher}, {Masci}, {Nyholm}, {Ofek},
  {Sollerman}, \& {Yan}}]{papadogiannakis19a}
{Papadogiannakis}, S., {et~al.} 2019, \mnras, 483, 5045

\bibitem[{{Pasquet} {et~al.}(2019){Pasquet}, {Bertin}, {Treyer}, {Arnouts}, \&
  {Fouchez}}]{pasquet19a}
{Pasquet}, J., {Bertin}, E., {Treyer}, M., {Arnouts}, S., \& {Fouchez}, D.
  2019, \aap, 621, A26

\bibitem[{{Perlmutter} {et~al.}(1999){Perlmutter}, {Aldering}, {Goldhaber},
  {Knop}, {Nugent}, {Castro}, {Deustua}, {Fabbro}, {Goobar}, {Groom}, {Hook},
  {Kim}, {Kim}, {Lee}, {Nunes}, {Pain}, {Pennypacker}, {Quimby}, {Lidman},
  {Ellis}, {Irwin}, {McMahon}, {Ruiz-Lapuente}, {Walton}, {Schaefer}, {Boyle},
  {Filippenko}, {Matheson}, {Fruchter}, {Panagia}, {Newberg}, {Couch}, \& {The
  Supernova Cosmology Project}}]{perlmutter99a}
{Perlmutter}, S., {et~al.} 1999, \apj, 517, 565

\bibitem[{{Petrillo} {et~al.}(2017){Petrillo}, {Tortora}, {Chatterjee},
  {Vernardos}, {Koopmans}, {Verdoes Kleijn}, {Napolitano}, {Covone},
  {Schneider}, {Grado}, \& {McFarland}}]{petrillo17a}
{Petrillo}, C.~E., {et~al.} 2017, \mnras, 472, 1129

\bibitem[{{Phillips} {et~al.}(2007){Phillips}, {Li}, {Frieman}, {Blinnikov},
  {DePoy}, {Prieto}, {Milne}, {Contreras}, {Folatelli}, {Morrell}, {Hamuy},
  {Suntzeff}, {Roth}, {Gonz{\'a}lez}, {Krzeminski}, {Filippenko}, {Freedman},
  {Chornock}, {Jha}, {Madore}, {Persson}, {Burns}, {Wyatt}, {Murphy}, {Foley},
  {Ganeshalingam}, {Serduke}, {Krisciunas}, {Bassett}, {Becker}, {Dilday},
  {Eastman}, {Garnavich}, {Holtzman}, {Kessler}, {Lampeitl}, {Marriner},
  {Frank}, {Marshall}, {Miknaitis}, {Sako}, {Schneider}, {van der Heyden}, \&
  {Yasuda}}]{phillips07a}
{Phillips}, M.~M., {et~al.} 2007, \pasp, 119, 360

\bibitem[{{Quimby} {et~al.}(2007){Quimby}, {Aldering}, {Wheeler},
  {H{\"o}flich}, {Akerlof}, \& {Rykoff}}]{quimby07b}
{Quimby}, R.~M., {Aldering}, G., {Wheeler}, J.~C., {H{\"o}flich}, P.,
  {Akerlof}, C.~W., \& {Rykoff}, E.~S. 2007, \apjl, 668, L99

\bibitem[{{Quimby} {et~al.}(2011){Quimby}, {Kulkarni}, {Kasliwal}, {Gal-Yam},
  {Arcavi}, {Sullivan}, {Nugent}, {Thomas}, {Howell}, {Nakar}, {Bildsten},
  {Theissen}, {Law}, {Dekany}, {Rahmer}, {Hale}, {Smith}, {Ofek}, {Zolkower},
  {Velur}, {Walters}, {Henning}, {Bui}, {McKenna}, {Poznanski}, {Cenko}, \&
  {Levitan}}]{quimby11a}
{Quimby}, R.~M., {et~al.} 2011, \nat, 474, 487

\bibitem[{{Ram{\'\i}rez} {et~al.}(2009){Ram{\'\i}rez}, {Mel{\'e}ndez}, \&
  {Asplund}}]{ramirez09a}
{Ram{\'\i}rez}, I., {Mel{\'e}ndez}, J., \& {Asplund}, M. 2009, \aap, 508, L17

\bibitem[{{Riess} {et~al.}(1998){Riess}, {Filippenko}, {Challis},
  {Clocchiatti}, {Diercks}, {Garnavich}, {Gilliland}, {Hogan}, {Jha},
  {Kirshner}, {Leibundgut}, {Phillips}, {Reiss}, {Schmidt}, {Schommer},
  {Smith}, {Spyromilio}, {Stubbs}, {Suntzeff}, \& {Tonry}}]{riess98a}
{Riess}, A.~G., {et~al.} 1998, \aj, 116, 1009

\bibitem[{{Rubin} {et~al.}(2015){Rubin}, {Aldering}, {Barbary}, {Boone},
  {Chappell}, {Currie}, {Deustua}, {Fagrelius}, {Fruchter}, {Hayden}, {Lidman},
  {Nordin}, {Perlmutter}, {Saunders}, {Sofiatti}, \& {Supernova Cosmology
  Project}}]{rubin15a}
{Rubin}, D., {et~al.} 2015, \apj, 813, 137

\bibitem[{{Sako} {et~al.}(2011){Sako}, {Bassett}, {Connolly}, {Dilday},
  {Cambell}, {Frieman}, {Gladney}, {Kessler}, {Lampeitl}, {Marriner}, {Miquel},
  {Nichol}, {Schneider}, {Smith}, \& {Sollerman}}]{sako11a}
{Sako}, M., {et~al.} 2011, \apj, 738, 162

\bibitem[{{Salvato} {et~al.}(2019){Salvato}, {Ilbert}, \&
  {Hoyle}}]{Salvato_2019}
{Salvato}, M., {Ilbert}, O., \& {Hoyle}, B. 2019, Nature Astronomy, 3, 212

\bibitem[{{Salvato} {et~al.}(2009){Salvato}, {Hasinger}, {Ilbert}, {Zamorani},
  {Brusa}, {Scoville}, {Rau}, {Capak}, {Arnouts}, {Aussel}, {Bolzonella},
  {Buongiorno}, {Cappelluti}, {Caputi}, {Civano}, {Cook}, {Elvis}, {Gilli},
  {Jahnke}, {Kartaltepe}, {Impey}, {Lamareille}, {Le Floc'h}, {Lilly},
  {Mainieri}, {McCarthy}, {McCracken}, {Mignoli}, {Mobasher}, {Murayama},
  {Sasaki}, {Sanders}, {Schiminovich}, {Shioya}, {Shopbell}, {Silverman},
  {Smol{\v{c}}i{\'c}}, {Surace}, {Taniguchi}, {Thompson}, {Trump}, {Urry}, \&
  {Zamojski}}]{Salvato_2009}
{Salvato}, M., {et~al.} 2009, \apj, 690, 1250

\bibitem[{{S{\'a}nchez-S{\'a}ez} {et~al.}(2020){S{\'a}nchez-S{\'a}ez}, {Reyes},
  {Valenzuela}, {F{\"o}rster}, {Eyheramendy}, {Elorrieta}, {Bauer},
  {Cabrera-Vives}, {Est{\'e}vez}, {Catelan}, {Pignata}, {Huijse}, {De Cicco},
  {Ar{\'e}valo}, {Carrasco-Davis}, {Abril}, {Kurtev}, {Borissova}, {Arredondo},
  {Castillo-Navarrete}, {Rodriguez}, {Ruz-Mieres}, {Moya},
  {Sabatini-Gacit{\'u}a}, \& {Sep{\'u}lveda-Cobo}}]{sanchez-saez20a}
{S{\'a}nchez-S{\'a}ez}, P., {et~al.} 2020, arXiv:2008.0331

\bibitem[{{Saunders} {et~al.}(2004){Saunders}, {Bridges}, {Gillingham},
  {Haynes}, {Smith}, {Whittard}, {Churilov}, {Lankshear}, {Croom}, {Jones}, \&
  {Boshuizen}}]{Saunders2004}
{Saunders}, W., {et~al.} 2004, in Society of Photo-Optical Instrumentation
  Engineers (SPIE) Conference Series, Vol. 5492, Ground-based Instrumentation
  for Astronomy, ed. A.~F.~M. {Moorwood} \& M.~{Iye}, 389--400

\bibitem[{{Scolnic} \& {Kessler}(2016)}]{scolnickessler2016}
{Scolnic}, D., \& {Kessler}, R. 2016, \apjl, 822, L35

\bibitem[{{Scolnic} {et~al.}(2014){Scolnic}, {Rest}, {Riess}, {Huber}, {Foley},
  {Brout}, {Chornock}, {Narayan}, {Tonry}, {Berger}, {Soderberg}, {Stubbs},
  {Kirshner}, {Rodney}, {Smartt}, {Schlafly}, {Botticella}, {Challis},
  {Czekala}, {Drout}, {Hudson}, {Kotak}, {Leibler}, {Lunnan}, {Marion},
  {McCrum}, {Milisavljevic}, {Pastorello}, {Sand ers}, {Smith}, {Stafford},
  {Thilker}, {Valenti}, {Wood-Vasey}, {Zheng}, {Burgett}, {Chambers},
  {Denneau}, {Draper}, {Flewelling}, {Hodapp}, {Kaiser}, {Kudritzki},
  {Magnier}, {Metcalfe}, {Price}, {Sweeney}, {Wainscoat}, \&
  {Waters}}]{scolnic14a}
{Scolnic}, D., {et~al.} 2014, \apj, 795, 45

\bibitem[{{Scolnic} {et~al.}(2018){Scolnic}, {Jones}, {Rest}, {Pan},
  {Chornock}, {Foley}, {Huber}, {Kessler}, {Narayan}, {Riess}, {Rodney},
  {Berger}, {Brout}, {Challis}, {Drout}, {Finkbeiner}, {Lunnan}, {Kirshner},
  {Sand ers}, {Schlafly}, {Smartt}, {Stubbs}, {Tonry}, {Wood-Vasey}, {Foley},
  {Hand}, {Johnson}, {Burgett}, {Chambers}, {Draper}, {Hodapp}, {Kaiser},
  {Kudritzki}, {Magnier}, {Metcalfe}, {Bresolin}, {Gall}, {Kotak}, {McCrum}, \&
  {Smith}}]{scolnic18a}
{Scolnic}, D.~M., {et~al.} 2018, \apj, 859, 101

\bibitem[{{Scoville} {et~al.}(2007){Scoville}, {Aussel}, {Brusa}, {Capak},
  {Carollo}, {Elvis}, {Giavalisco}, {Guzzo}, {Hasinger}, {Impey}, {Kneib},
  {LeFevre}, {Lilly}, {Mobasher}, {Renzini}, {Rich}, {Sanders}, {Schinnerer},
  {Schminovich}, {Shopbell}, {Taniguchi}, \& {Tyson}}]{scoville07a}
{Scoville}, N., {et~al.} 2007, \apjs, 172, 1

\bibitem[{{Sharma} {et~al.}(2020){Sharma}, {Kembhavi}, {Kembhavi}, {Sivarani},
  {Abraham}, \& {Vaghmare}}]{sharma20a}
{Sharma}, K., {Kembhavi}, A., {Kembhavi}, A., {Sivarani}, T., {Abraham}, S., \&
  {Vaghmare}, K. 2020, \mnras, 491, 2280

\bibitem[{{Silverman} {et~al.}(2015){Silverman}, {Kashino}, {Sanders},
  {Kartaltepe}, {Arimoto}, {Renzini}, {Rodighiero}, {Daddi}, {Zahid}, {Nagao},
  {Kewley}, {Lilly}, {Sugiyama}, {Baronchelli}, {Capak}, {Carollo}, {Chu},
  {Hasinger}, {Ilbert}, {Juneau}, {Kajisawa}, {Koekemoer}, {Kovac}, {Le
  F{\`e}vre}, {Masters}, {McCracken}, {Onodera}, {Schulze}, {Scoville},
  {Strazzullo}, \& {Taniguchi}}]{FMOS-COSMOS2015}
{Silverman}, J.~D., {et~al.} 2015, \apjs, 220, 12

\bibitem[{Srivastava {et~al.}(2014)Srivastava, Hinton, Krizhevsky, Sutskever,
  \& Salakhutdinov}]{dropout}
Srivastava, N., Hinton, G., Krizhevsky, A., Sutskever, I., \& Salakhutdinov, R.
  2014, Journal of Machine Learning Research, 15, 1929

\bibitem[{{Srivastava} {et~al.}(2015{\natexlab{a}}){Srivastava}, {Greff}, \&
  {Schmidhuber}}]{srivastava15a}
{Srivastava}, R.~K., {Greff}, K., \& {Schmidhuber}, J. 2015{\natexlab{a}},
  arXiv:1505.00387

\bibitem[{{Srivastava} {et~al.}(2015{\natexlab{b}}){Srivastava}, {Greff}, \&
  {Schmidhuber}}]{srivastava15b}
---. 2015{\natexlab{b}}, arXiv:1507.06228

\bibitem[{{Strolger} {et~al.}(2015){Strolger}, {Dahlen}, {Rodney}, {Graur},
  {Riess}, {McCully}, {Ravindranath}, {Mobasher}, \& {Shahady}}]{strolger15a}
{Strolger}, L.-G., {et~al.} 2015, \apj, 813, 93

\bibitem[{{Tampo} {et~al.}(2020){Tampo}, {Tanaka}, {Maeda}, {Yasuda},
  {Tominaga}, {Jiang}, {Moriya}, {Morokuma}, {Suzuki}, {Takahashi}, {Kokubo},
  \& {Kawana}}]{Tampo2020}
{Tampo}, Y., {et~al.} 2020, \apj, 894, 27

\bibitem[{{Tanaka} {et~al.}(2018){Tanaka}, {Coupon}, {Hsieh}, {Mineo},
  {Nishizawa}, {Speagle}, {Furusawa}, {Miyazaki}, \&
  {Murayama}}]{HSCSSP_photo-z2018}
{Tanaka}, M., {et~al.} 2018, \pasj, 70, S9

\bibitem[{{The PLAsTiCC team} {et~al.}(2018){The PLAsTiCC team}, {Allam},
  {Bahmanyar}, {Biswas}, {Dai}, {Galbany}, {Hlo{\v{z}}ek}, {Ishida}, {Jha},
  {Jones}, {Kessler}, {Lochner}, {Mahabal}, {Malz}, {Mand el},
  {Mart{\'\i}nez-Galarza}, {McEwen}, {Muthukrishna}, {Narayan}, {Peiris},
  {Peters}, {Ponder}, {Setzer}, {The LSST Dark Energy Science Collaboration},
  {LSST Transients}, \& {Variable Stars Science Collaboration}}]{allam18a}
{The PLAsTiCC team}, {et~al.} 2018, arXiv:1810.00001

\bibitem[{{Thompson} {et~al.}(2003){Thompson}, {Burrows}, \&
  {Pinto}}]{thompson03a}
{Thompson}, T.~A., {Burrows}, A., \& {Pinto}, P.~A. 2003, \apj, 592, 434

\bibitem[{{Tominaga} {et~al.}(2011){Tominaga}, {Morokuma}, {Blinnikov},
  {Baklanov}, {Sorokina}, \& {Nomoto}}]{tominaga11a}
{Tominaga}, N., {Morokuma}, T., {Blinnikov}, S.~I., {Baklanov}, P., {Sorokina},
  E.~I., \& {Nomoto}, K. 2011, \apjs, 193, 20

\bibitem[{{Tsujimoto} {et~al.}(1995){Tsujimoto}, {Nomoto}, {Yoshii},
  {Hashimoto}, {Yanagida}, \& {Thielemann}}]{tsujimoto95a}
{Tsujimoto}, T., {Nomoto}, K., {Yoshii}, Y., {Hashimoto}, M., {Yanagida}, S.,
  \& {Thielemann}, F.~K. 1995, \mnras, 277, 945

\bibitem[{{Villar} {et~al.}(2020){Villar}, {Hosseinzadeh}, {Berger},
  {Ntampaka}, {Jones}, {Challis}, {Chornock}, {Drout}, {Foley}, {Kirshner},
  {Lunnan}, {Margutti}, {Milisavljevic}, {Sanders}, {Pan}, {Rest}, {Scolnic},
  {Magnier}, {Metcalfe}, {Wainscoat}, \& {Waters}}]{villar20a}
{Villar}, V.~A., {et~al.} 2020, arXiv:2008.04921

\bibitem[{{Yasuda} {et~al.}(2019){Yasuda}, {Tanaka}, {Tominaga}, {Jiang},
  {Moriya}, {Morokuma}, {Suzuki}, {Takahashi}, {Yamaguchi}, {Maeda}, {Sako},
  {Ikeda}, {Kimura}, {Morii}, {Ueda}, {Yoshida}, {Lee}, {Suyu}, {Komiyama},
  {Regnault}, \& {Rubin}}]{yasuda19a}
{Yasuda}, N., {et~al.} 2019, \pasj, 71, 74

\bibitem[{Zhang {et~al.}(2017)Zhang, Ciss{\'{e}}, Dauphin, \&
  Lopez{-}Paz}]{mixup}
Zhang, H., Ciss{\'{e}}, M., Dauphin, Y.~N., \& Lopez{-}Paz, D. 2017,
  arXiv:1710.09412

\end{thebibliography}
%
 \newcommand{\noop}[1]{}

\end{document}